\documentclass[reprint,twocolumn,superscriptaddress,reprint,nofootinbib,floatfix,letterpaper]{revtex4-1}

\pdfoutput=1

\newcommand\cfig[1]{\vtop{\vskip-0.15in\hbox{#1}}}
\newcommand\xcite[1]{\cite{#1}}

\begin{document}

\begin{titlepage}

\title{Zoo of quantum-topological phases of matter}

\author{Xiao-Gang Wen}
\affiliation{Department of Physics, Massachusetts Institute of
Technology, Cambridge, Massachusetts 02139, USA}

\begin{abstract} 
What are topological phases of matter?  First, they are phases of matter at
zero temperature.  Second, they have a non-zero energy gap for the excitations
above the ground state.  Third, they are disordered liquids that seem have no
feature.  But those disordered liquids actually can have rich patterns of
many-body entanglement representing new kinds of order.  This paper will give a
simple introduction and a brief survey of topological phases of matter.  We
will first discuss topological phases that have topological order (\ie with
long range entanglement). Then we will cover topological phases that have no
topological order (\ie with only short-range entanglement).

\end{abstract}

\pacs{}

\maketitle

\end{titlepage}

{\small \setcounter{tocdepth}{1} \tableofcontents }

\section{Orders and symmetries} \label{deforder}

Condensed matter physics is a branch of science that study various properties
of all kinds of materials.  Usually for each kind of materials, we need a
different theory (or model) to explain its properties.  After seeing many
different type of theories/models for condensed matter systems, a common theme
among those theories start to emerge.  The common theme is the \emph{principle
of emergence}, which states that the properties of a material are mainly
determined by how particles are organized in the material.  This is quite
different from the point of view that the properties of a material should be
determined by the components that form the material.  In fact, all the
materials are made of same three components: electrons, protons and neutrons.
So we cannot use the richness of the components to understand the richness of
the materials. The various properties of different materials originate from
various ways in which the particles are organized.  The organizations of the
particles are called orders.  Different orders lead to different phases of
matter, which in turn leads to different properties of materials.

Therefore, according to the principle of emergence, the key to understand a
material is to understand how electrons, protons and neutrons are organized in
the material.  Based on a deep insight into phase and phase transition,
\textcite{L3726} developed a general theory of orders as well as transitions
between different phases of matter\xcite{L3726,GL5064,LanL58}. He pointed out
that the reason that different phases (or orders) are different is because they
have different symmetries.  A phase transition is simply a transition that
changes the symmetry.  Introducing order parameters that transform
non-trivially under the symmetry transformations, \textcite{GL5064} developed
the standard theory for phases and phase transitions, where
different phases of matter are classified by a pair of groups
$(G_\Psi \subset G_H)$. Here $G_H$ is the symmetry group of the system
and $G_\Psi$ the unbroken symmetry group of the equilibrium state.

Landau's theory is very successful. Using symmetry and the related group
theory, we can classify all of the 230 different kinds of crystals that can
exist in three dimensions.  By determining how symmetry changes across a
continuous phase transition, we can obtain the critical properties of the phase
transition.  The symmetry breaking also provides the origin of many gapless
excitations, such as phonons, spin waves, etc., which determine the low-energy
properties of many systems \cite{N6080,G6154}. Many of the properties of those
excitations, including their gaplessness, are directly determined by the
symmetry.

As Landau's symmetry-breaking theory has such a broad and fundamental impact on
our understanding of matter, it became a corner-stone of condensed matter
theory. The picture painted by Landau's theory is so satisfactory that one
starts to have a feeling that we understand, at least in principle, all kinds
of orders that matter can have.  One feels that we start to see the beginning
of the end of condensed matter theory. 

\section{New world of condensed matter physics} 
\label{newworld}

However, through the researches in last 30 years, a different picture starts to
emerge. It appears that what we have seen is just the end of beginning.  There
is a whole new world ahead of us waiting to be explored.  A peek into the new
world is offered by the discovery of fractional quantum Hall (FQH) effect
\cite{TSG8259}. Another peek is offered by the discovery of high $T_c$
superconductors \cite{BM8689}. Both phenomena are completely beyond the
paradigm of Landau's symmetry breaking theory. Rapid and exciting developments
in FQH effect and in high $T_c$ superconductivity resulted in many new ideas
and new concepts.  Looking back at those new developments, it becomes more and
more clear that, in last 30 years, we were actually witnessing an emergence of
a new theme in condensed matter physics. The new theme is associated with new
kinds of orders, new states of matter and new class of materials beyond
Landau's symmetry breaking theory.  This is an exciting time for condensed
matter physics.  The new paradigm may even have an impact in our understanding
of fundamental questions of nature -- the emergence of elementary particles and
the four fundamental interactions, which leads to an unification of matter and
quantum information.\footnote{See \textcite{FNN8035} and \textcite{BA8880} for emergence of
gauge interactions, \textcite{W0202,W0303a,LWqed} for unification of gauge
interactions and Fermi statistics, and \textcite{W1301,YBX1451,YX14124784} for
emergence of chiral fermions.}

The emergent new field of quantum-topological matter has developed very fast.
Many new terms are introduced, but some of them can be very
confusing:
\begin{enumerate}
\item[\textbf{Pt.1:}] 
Some \emph{Haldane phases} are \emph{topological}, while some other Haldane
phases are not topological.  Although, the Haldane phase for spin-1 chain is
topological, it is actually a product state with no \emph{topological order}.

\item[\textbf{Pt.2:}] 
\emph{Topological insulators} and \emph{topological superconductors} (\ie with
$T^2=(-)^{N_F}$ time-reversal symmetry and weak interactions) has no
\emph{topological order}.  It is wrong to characterize topological insulators
as insulators with conducting surface.

\item[\textbf{Pt.3:}] 
What is the difference between \emph{quantum spin Hall state} and \emph{spin quantum Hall state}? Are they \emph{topological insulator}?

\item[\textbf{Pt.4:}] 
``\emph{SPT state}'' is the abbreviation for both \emph{symmetry protected trivial state} and
\emph{symmetry protected topological state}. The two mean the same.

\item[\textbf{Pt.5:}] 
3+1D \emph{textbook $s$-wave superconductors}  have no topological order, while
3+1D \emph{real-life $s$-wave superconductors} have a \emph{$Z_2$-topological
order}.

\item[\textbf{Pt.6:}] 
2+1D \emph{$p+\ii p$ fermion paired state} and \emph{integer quantum Hall
states} (IQH) do not have any fractionalized topological excitations.  Some
people regard them as \emph{long-range entangled} (\ie \emph{topologically
ordered}) state while others regard them as \emph{short-range entangled} state.

\item[\textbf{Pt.7:}] 
What are the difference between \emph{Chern insulator}, \emph{quantum anomalous
Hall state}, and \emph{integer quantum Hall state}?  What are the difference
between \emph{fractionalized topological insulator} and \emph{topological
order}?

\item[\textbf{Pt.8:}] 
There is a very active search for \emph{Majorana fermions} with
\emph{non-abelian statistics}.  But should  Majorana \emph{fermion} be a
fermion that carries Fermi statistics?  Is Majorana fermion the
\emph{Bogoliubov quasiparticle} in a superconductor?

\end{enumerate}
In this paper, we will try to clarify some of those notions.  


\section{Topologically ordered phases}

\subsection{Chiral spin liquids and topological order}

After the discovery of high $T_c$ superconductors in 1986 by \textcite{BM8689},
some theorists believed that quantum spin liquids play a key role in
understanding high $T_c$ superconductors \cite{A8796}.  This is because spin
liquid can leads to a so called spin-charge separation: an electron
disintegrates into two quasiparticles -- a spinon (spin-1/2 charge-0 ) and a
holon (spin-0 charge-$e$).  Since holon is not fermion, its condensation can
leads to superconductivity - a novel mechanism of high $T_c$ superconductors.
Thus many people started to construct and study various spin
liquids.\footnote{See \textcite{BZA8773,AM8874,RK8876,AZH8845,DFM8826}}

However, despite the success of Landau symmetry-breaking theory in describing
all kind of states, the theory cannot explain and does not even allow the
existence of spin liquids (with odd number of electrons per unit cell). This
leads many theorists to doubt the very existence of spin liquids.  In early
proposals of spin liquid, the spinons are gapless and are confined at long
distance by the emergent gauge field \cite{BA8880}, adding support to the
opinion that the spin liquid is just a fiction and does not actually
exist.\footnote{Now we realized that even those gapless spin liquid can exist
as algebraic spin liquid without quasiparticles
\cite{CMM0159,RW0171,RW0201,FHM0353,HSF0437,SBS0407}.}

In 1987, \textcite{KL8795} introduced a special kind of spin liquids -- chiral
spin liquid --  in an attempt to explain high temperature superconductivity.
In contrast to many other proposed spin liquids at that time, the chiral spin
liquid was shown to have deconfined spinons (as well as deconfined holons) and
correspond to a stable zero-temperature phase.\footnote{Recently, chiral spin
liquid was shown to exist in Heisenberg model on Kagome lattice with
$J_1$-$J_2$-$J_3$ coupling \cite{HC14072740,GS14121571}.}  At first, not
believing Landau symmetry-breaking theory fails to describe spin liquids,
people still wanted to use symmetry breaking to characterize the chiral spin
liquid. They identified the chiral spin liquid as a state that breaks the time
reversal and parity symmetries, but not the spin rotation and translation
symmetries \cite{WWZcsp}. 

However, \textcite{Wtop} quickly realized that there are many different chiral
spin liquids (with different spinon statistics and spin Hall conductances) that
have exactly the same symmetry. So symmetry alone is not enough to characterize
different chiral spin liquids. This means that the chiral spin liquids contain
a new kind of order that is beyond symmetry description . This new kind of
order was named \textbf{topological order}.

Just like any concepts in physics, the concept of topological order is also
required to be defined via measurable quantities, which are called
\textbf{topological invariants}. The first discovered topological invariants
\cite{Wrig} that define topological order  were (1) the \emph{robust ground
state} degeneracy on torus and other closed space manifolds (\ie with no
boundary), (2) the non-abelian geometric phases (the \emph{modular matrices})
of the degenerate ground states, (3) the \emph{chiral central charge} $c$ of
the edge states.\footnote{The central charge $c$ of the edge states is related
to a gravitational response of the system described by a gravitational
Chern-Simons 3-form $\om_3$: ${\cal L}= \frac{2\pi c}{24} \om_3$, where $\dd
\om_3=p_1$ is the first Pontryagin class
\cite{AG14013703,GF14106812,BR150204126}.  $c$ can be measured via the thermal
Hall conductivity $K_H=c\frac{\pi k_B^2}{6\hbar}T$ \cite{KF9732}.}  It was
conjectured that those macroscopic topological invariants, or more generally,
``the total gauge structures (the Abelian one plus the non-Abelian one) on the
moduli spaces of the models defined on generic Riemann surfaces $\Si_g$
completely characterize (or classify) the topological orders in 1+2
dimensions'' \cite{Wrig}.  

Microscopically, topological order is a property of a local quantum system
whose total Hilbert space have a tensor product decomposition $
\cH^\mathrm{tot}= \bigotimes_i \cH_i$, where $\cH_i$ is the Hilbert space on
each \emph{site}. Such a tensor product decomposition is a part of the definition
of a local system, which also satisfies the condition of short-range
interaction between \emph{sites}.  Relative to such a tensor product
decomposition, a \emph{product state} is defined to be a state of the form
$|\Psi\>= \bigotimes_i |\Psi_i\>$, where $|\Psi_i\> \in \cH_i$.  In this paper,
only the tensor products of on-site states, $|\Psi_i\>$, are called product
states. With such a definition of local quantum systems, \emph{topological
order is defined to describe gapped quantum-liquids\footnote{ \textcite{ZW1490}
and \textcite{SM1403} introduced the notion of \emph{gapped quantum-liquids} to
describe a simple kind of gapped states: the states that can enlarge themselves
by dissolving product states.  Only gapped quantum-liquids have quantum field
theory descriptions at long distances. 3D gapped states obtained by stacking 2D
quantum Hall states and \textcite{H11011962} cubic code are examples of gapped
non-quantum-liquids.} that cannot be deformed into a product state without
gap-closing phase transitions}.  Such quantum liquids are said to have
\textbf{long-range entanglement} \cite{KP0604,LW0605,CGW1038}.  Long-range
entanglement is the microscopic origin of topological order.  A gapped state
that can be deformed into a product state smoothly is \textbf{short-range
entangled} and has no topological order. In particular, a product state has no
topological order.

One may wonder: why do we need such a complicated way to characterize
topological order. Is the quantized Hall conductance a more direct and simpler
way to characterize topological order, at least for quantum Hall states (see
Sec. \ref{secFQH})?  In fact, quantized Hall conductance is due to a combined
effect of $U(1)$ symmetry (\ie particle-number conservation) and topological
order (\ie long-range entanglement).  If we break the $U(1)$ symmetry,
quantum Hall states still have topological order, even though the  Hall
conductance is no longer well defined. How to characterize topological order in
such a situation?  The above characterization based on ground state degeneracy
and non-abelian geometric phases does not require symmetries and provides
\emph{a complete characterize of topological orders in 2-dimensions}.

We like to mention that the term ``topological'' in topological order and in
topological insulators/superconductors has totally different meanings.  In
topological order, the term is motivated by the low energy effective theory of
the chiral spin liquids, which is a $U(1)$ Chern-Simons theory -- a topological
quantum field theory \cite{W8951}. Here, ``topological'' really means
long-range entangled, which is a property of many-body wave functions. We may
call it \textbf{quantum topology}.  While in topological
insulators/superconductors, the term corresponds to \textbf{classical topology}
which is a property of continuous manifold, related to the difference between
sphere and torus. The vortex in superfluid, the Chern number, and the $Z_2$
index in topological insulators/superconductors belong to classical topology,
which represent a very different phenomenon. In fact, ``topological'' in
topological insulators/superconductors really means ``symmetry protected'' (see
Sec. \ref{secSPT}).

\subsection{Quantum Hall states}
\label{secFQH}

However, soon after the proposal of chiral spin liquid, experiments indicated
that high-temperature superconductors do not break the time reversal and parity
symmetries and chiral spin liquids do not describe high-temperature
superconductors \cite{LSL9239}. Thus the concept of topological order became a
concept with no experimental realization. 

But long before the discovery of high $T_c$ superconductors, \textcite{TSG8259}
discovered FQH effect, such as the filling fraction
$\nu=1/m$ \textcite{L8395} state 
\begin{align}
 \Psi_{\nu=1/m}(\{z_i\})=\prod (z_i-z_j)^m \ee^{-\frac14 \sum |z_i|^2}
\end{align}
where $z_i=x_i+\ii y_i$.  People realized that the FQH states are new states of
matter.  At first, influenced by the previous success of Landau's symmetry
breaking theory, people used order parameters and long range correlations to
describe the FQH states \cite{GM8752,ZHK8982,R8986},  which result in the
Ginzburg-Landau-Chern-Simons effective theory of quantum Hall states.  But in
quantum Hall states, there is no off-diagonal long range order in any local
operators, and thinking about it can mislead some people to wrong directions,
such as looking for Josephson effect in quantum Hall states.

If we concentrate on physical measurable quantities,
we will see that all those different FQH states have exactly the same symmetry
and conclude that we cannot use Landau symmetry-breaking theory and local order
parameters to describe different orders in FQH states.  
In fact, just like chiral spin
liquids, FQH states also contain new kind of orders beyond Landau's symmetry
breaking theory. Different FQH states are also described by different
topological orders \cite{WNtop}. The better way to see the essence
of FQH states is via topological invariants such as robust ground state
degeneracy and modular matrices, as well as the non-trivial edge states
\cite{H8285,Wcll}.  Thus the concept of topological order does have
experimental realizations in FQH systems.

One of the most striking properties of FQH states is their fractionalized
excitations, that can carry fractional statistics
\cite{H8483,ASW8422}\footnote{The possibility of fractional statistics in 2+1D
was pointed out by \textcite{LM7701} and \textcite{W8257}. The relation to
braid group was discussed by \textcite{W8413}.} and, if particle number
conserves, fractional charges \cite{TSG8259,L8395}\footnote{Fractional charge
has been directly observed via quantum shot noise in tunneling current
\cite{dRH9762}}. 

We know that a point-like excitation above the ground state is something that
can be trapped by a local change of the Hamiltonian near a spatial point $\v
x$. But some times, the local change of the ground state near $\v x$ cannot be
created by local operators.  In this case, we refer the corresponding local
change of the ground state as a \textbf{topological excitations}.  It is those
topological excitations that can carry fractional statistics/charge.

We note that the presence of any topological excitations imply a presence of
topological order in the ground state.  But the reverse is not true, the
absence of any topological excitations may not imply the absence of topological
order in the ground state.  In fact, the $E_8$ bosonic state and the IQH states
are states with topological order but no topological excitations.  

Regarding to \textbf{Pt.6} in Sec. \ref{newworld}, some people define those
states with no topological excitations as short-range entangled \cite{K11sre}.
However, since those states have non-zero \emph{chiral central charges} $c$ for
the edge states, they cannot smoothly change to product state without phase
transition.  Thus, they are topologically ordered states distinct from the
trivial product states.  Those topological orders with no topological
excitations are called \textbf{invertible topological orders}
\footnote{ For every invertible topological order $\cC$, there exist another
topological order $\cD$ -- the inverse, such that stacking $\cC$ and $\cD$ on
top of each other give us a gapped state that have no topological order, \ie
belong to the phase of product states.}, and some people refer them as
long-rang entangled \cite{CGW1038}.  Regarding to \textbf{Pt.7}, \emph{IQH
state} \cite{KDP8094}, \emph{Chern insulator} \cite{H7639,TKN8205}, 
\emph{quantum anomalous Hall state} \cite{H8815},
are just different names for the same fermionic
invertible topological order with integer \emph{chiral central charge} $c$.
Also, \emph{fractionalized topological insulator} is same as \emph{topological
order}, but may have an additional time reversal symmetry.

\subsection{Non-abelian Quantum Hall states}

In addition to the Laughlin states, more exotic non-abelian FQH states were
proposed in 1991 by two independent works.  \textcite{Wnab} pointed out that
the FQH states described by wave functions
\begin{align}
 \Psi_{\nu=\frac{n}{m}}(\{z_i\})&=[\chi_n(\{z_i\})]^m, 
\nonumber\\
\text { or }
 \Psi_{\nu=\frac{n}{m+n}}(\{z_i\})&=\chi_1(\{z_i\})[\chi_n(\{z_i\})]^m 
\end{align}
have topological excitations with \textbf{non-abelian statistics}\footnote{
\textcite{W8413} has setup a general theory and braid group for quantum
statistics in two dimensions, and \textcite{GMS8503} pointed out that such a
setup contains non-abelian representations of braid group, which correspond to
non-abelian statistics. More complete description of non-abelian statistics are
given by \textcite{W8951} and \textcite{K062}.} of type $SU(n)_m$ (which is
denoted as $A(n-1)_m$ in https://www.math.ksu.edu/$\sim$gerald/voas/)
\cite{LW170107820}. This result was obtained via the low energy $SU(m)_n$
effective Chern-Simons theory of the above states, plus the level-rank duality.
Here $\chi_n$ is the fermion wave function of $n$-filled Landau levels.  We
note that the $\nu=1/3$ Laughlin state is given by
\begin{align}
 \Psi_{\nu=1/3}(\{z_i\})&=[\chi_1(\{z_i\})]^3. 
\end{align}
So $[\chi_n(\{z_i\})]^m$ and $\chi_1(\{z_i\})[\chi_n(\{z_i\})]^m$ are
generalizations of the Laughlin state \cite{J9153}.  They both have non-trivial
edge states described by $U(1)\times SU(n)_m$ Kac-Moody current algebra
\cite{BW9215}.  

In the same year, \textcite{MR9162} proposed that the FQH state described by
Pfaffian wave function 
\begin{align}
 \Psi_{\nu=1/2}=
\mathrm{Pf}\Big[ \frac{1}{z_i-z_j} \Big]
\ee^{-\frac14 \sum |z_i|^2}
\prod (z_i-z_j)^2  .
\end{align}
has excitations with non-abelian statistics of Ising-type (or $SU(2)_2$-type).
Its edge states were studied numerically \cite{Wnabhalf} and were found to be
described by a $c=1$ chiral-boson conformal field theory (CFT) plus a $c=1/2$
Majorana fermion CFT.  Such a result about the edge states supports the
proposal that the Pfaffian state is non-abelian, since the edge for abelian FQH
states always have integer chiral central charge $c$. Later, the non-abelian
statistics in Pfaffian wave function was also confirmed by its low energy
effective $SO(5)$ level 1 Chern-Simons theory \cite{W9927} (denoted as $B2_1$
in https://www.math.ksu.edu/$\sim$gerald/voas/), as well as a plasma analogue
calculation \cite{BN10085194}.

It is possible that the $SU(2)_2$-type of non-abelian state is
realized by $\nu=5/2$ fractional quantum Hall samples
\cite{WES8776,DM08020930,RMM0899}.

\subsection{Superconducting states (with dynamical electromagnetism)}

It is interesting to point out that long before the discovery of  FQH states,
Onnes discovered superconductor in 1911 \cite{O1122}. The Ginzburg-Landau
theory for symmetry breaking phases is largely developed to explain
superconductivity.  However, the superconducting order, that motivates the
Ginzburg-Landau theory for symmetry breaking, itself is not a symmetry breaking
order.  Superconducting order (in real life with dynamical $U(1)$ gauge field)
is an order that is beyond Landau symmetry breaking theory.  Superconducting
order (in real life) is an topological order (or more precisely a \textbf{$Z_2$
topological order} or $Z_2$ gauge theory) \cite{W9141,HOS0497}. The real-life
superconductor has string-like topological excitation that can be trapped by
modifying Hamiltonian along a loop.  Such a string-like topological excitation
is the $\frac{hc}{2e}$-flux loop, since the electromagnetic $U(1)$ gauge field
is dynamical.  The presence of string-like topological excitation indicate the
superconductor has a topological order.  The textbook
superconductors usually do not contain the dynamical $U(1)$ gauge field, and
do not contain string-like topological excitation that can be trapped by
modifying Hamiltonian along a loop. This explains \textbf{Pt.5} in Sec.
\ref{newworld}.

It is quite amazing that the experimental discovery of superconducting order
did not lead to a theory of topological order. But instead, it led to a theory
of symmetry breaking order, that fails to describe superconducting order
itself.

\begin{figure}[tb]
\centerline{
\includegraphics[height=.9in]{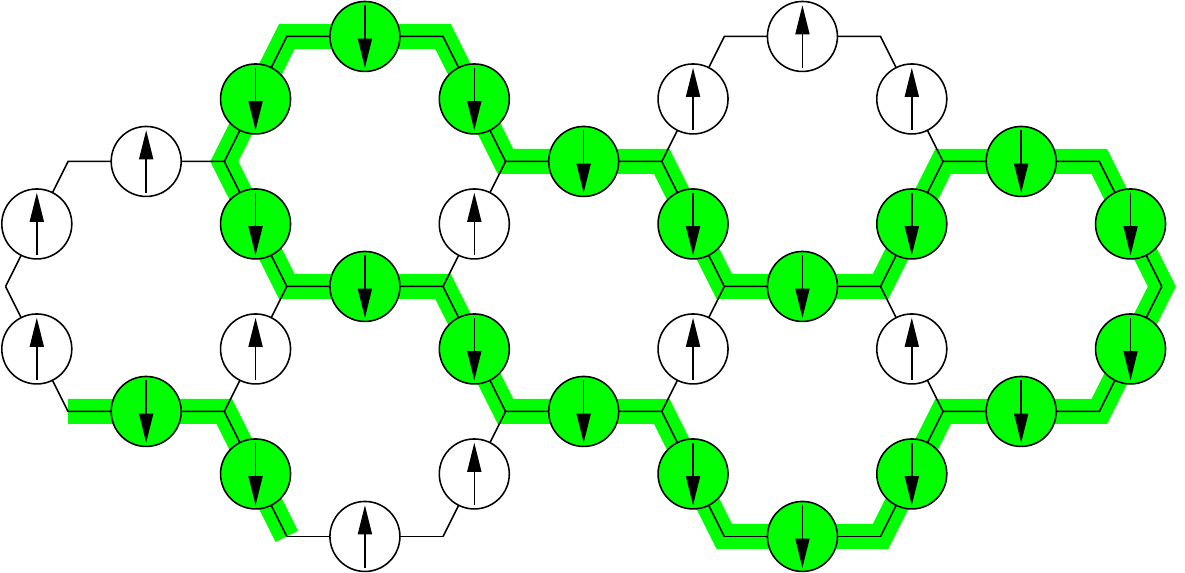}
}
\caption{
The strings in a spin-1/2 model.
In the background of up-spins, the down-spins form closed strings.
}
\label{strspin}
\end{figure}

\subsection{$Z_2$-spin liquid in 2+1D}

Since chiral spin liquid breaks the time reversal symmetry, while high $T_c$
superconductors do not break the time reversal symmetry. So chiral spin liquid
does not appear in  high $T_c$ superconductors.  This motivated people to look
for other spin liquids with deconfined spinons and holons that do not break
time reversal symmetry. This leads to the theoretical discovery of 2+1D
\textbf{$Z_2$-spin liquid} \cite{RS9173,Wsrvb} described by effective $Z_2$
gauge theory \cite{K7959} (\ie has a $Z_2$-topological order).  The
construction can be easily generalized to obtain 3+1D $Z_2$-spin liquid, which
will have a $Z_2$ topological order identical to an $s$-wave superconductor
discussed above.  Later, an exact soluble toric code model was constructed to
realize the $Z_2$ topological order \cite{K032}.  Since then, the
$Z_2$-topological order is also referred as ``toric code''.

The $Z_2$-spin liquid of spin-1/2 on Kagome lattice may be realized by
Herbertsmithite\cite{HMS0704}, as suggested by recent experiments by
\textcite{FL151102174,HL151206807}.  The early numerical calculation of
\textcite{YHW1173} suggested the spin-1/2 Heisenberg model on Kagome lattice is
gapped, but details of the results are inconsistent with $Z_2$-topological
order, which led people to suspect that the model is gapless.  A more recent
numerical calculation suggests the model to have a $Z_2$-spin liquid ground
state with long correlation length (10 unit cell length) \cite{MW160609639},
while several other calculations suggest gapless $U(1)$ spin liquid ground
states \cite{JR161002024,LX161004727,HP161106238}.  More experimental and
theoretical studies are needed to settle the issue.

\subsection{Quantum liquids of non-oriented strings}

If we do not require spin rotation symmetry,  one can use string liquid to
construct a state with $Z_2$-topological order \cite{K032}. String liquids are
long-range entangled (hence topologically ordered).  We will see how long-range
entanglement in topological order leads to fractional statistics and
topological degeneracy.  

\subsubsection{Local ``dancing'' rules in string liquids}

Given a spin-1/2 system, if we pick a particular spin-up spin-down
configuration, we will get a product state.  To construct a highly entangled
state, one may consider a equal-weight superposition of all spin-up spin-down
configurations. But this does not work. We get a product state with all spins
in $x$-direction.  So one idea to get a highly entangled state is to a partial
sum.  For example, we can view up-spins as background and lines of down-spins
as the strings (see Fig. \ref{strspin}). The simplest  topologically ordered
state in such a spin-1/2 system is given by the equal-weight superposition of all
closed strings:\cite{K032} $|\Phi_{Z_2}\>=\sum_\text{all closed strings} \left
|\cfig{\includegraphics[height=0.2in]{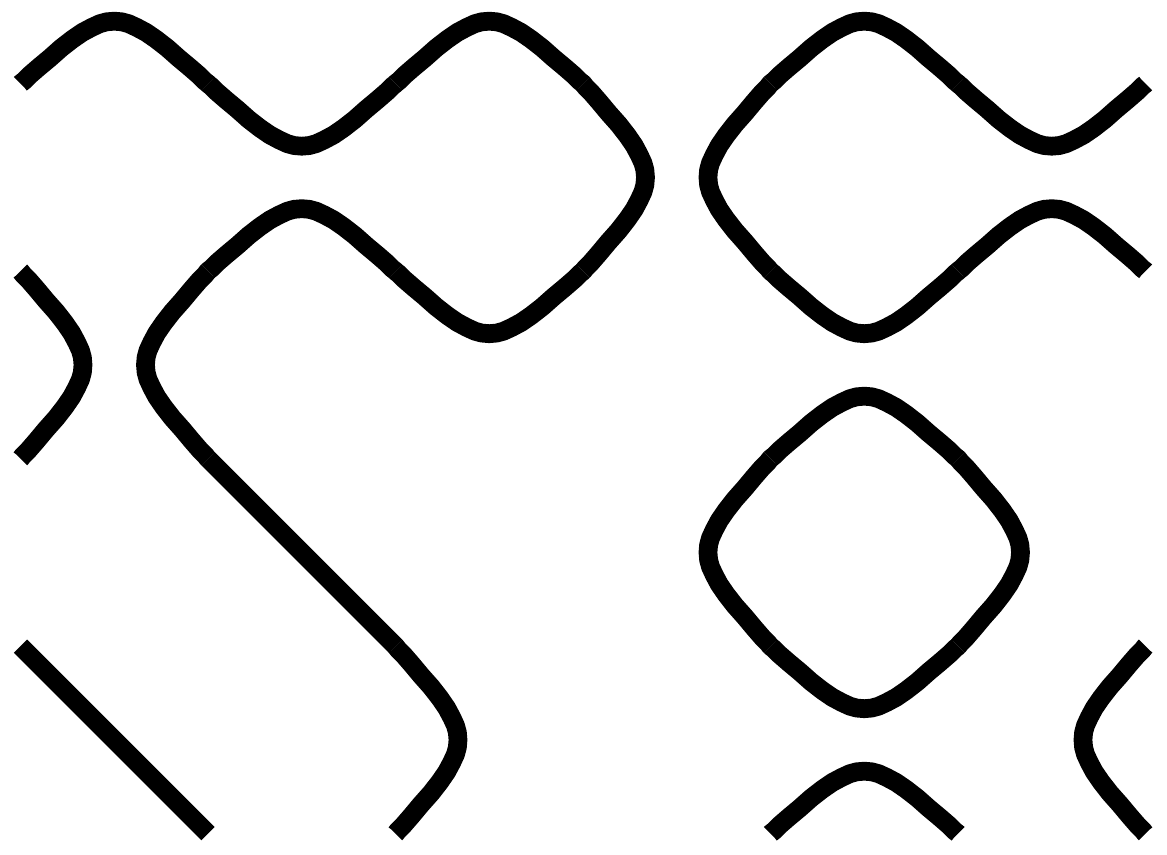}}\right \> $. 

\begin{figure}[tb]
\centerline{
\includegraphics[height=.7in]{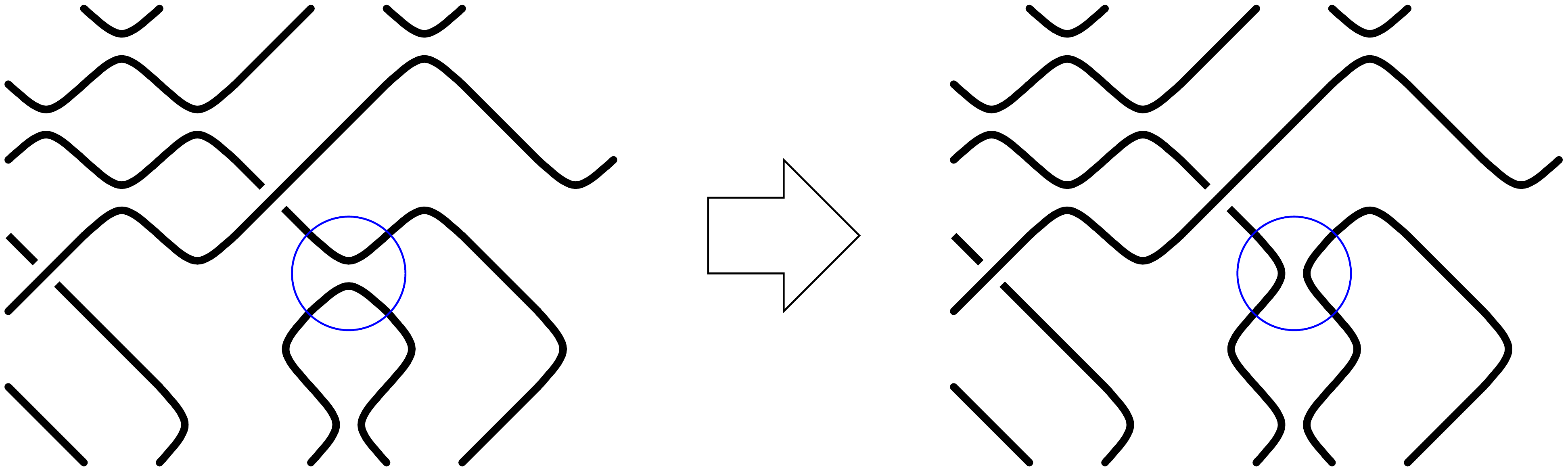}
}
\caption{
In string liquid, strings can move freely,
including reconnecting the strings.
}
\label{strnetSa}
\end{figure}

To obtain other topological orders, we may consider a different superposition
of strings. But those superpositions should all be determined by local rules,
so that there is a local Hamiltonian that can produce a given superposition.
What are those local rules that give rise to the string liquid
$|\Phi_{Z_2}\>=\sum_\text{all closed strings} \left
|\cfig{\includegraphics[height=0.2in]{strnetS}}\right \> $?  The first rule is
that, in the ground state, the down-spins are always connected with no open
ends.  To describe the second rule, we need to introduce the amplitudes of
close strings in the ground state: $\Phi \left(
\cfig{\includegraphics[height=0.2in]{strnetS}}\right) $.  The ground state is
given by
\begin{align}
\sum_\text{all closed strings} 
\Phi \left( \cfig{\includegraphics[height=0.2in]{strnetS}}\right)
 \left |\cfig{\includegraphics[height=0.2in]{strnetS}}\right \>.
\end{align}
Then the second rule relates the amplitudes of close strings in the ground state
as we change the strings locally:
\begin{align}
\label{Z2rl}
 \Phi
\bpm \includegraphics[height=0.2in]{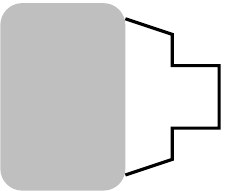} \epm  =&
\Phi
\bpm \includegraphics[height=0.2in]{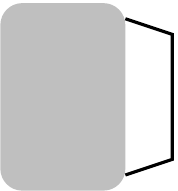} \epm ,
&
 \Phi
\bpm \includegraphics[height=0.2in]{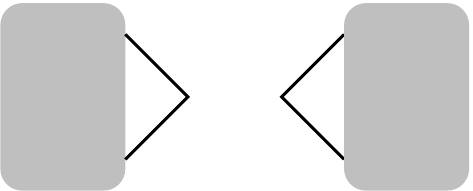} \epm  =&
\Phi
\bpm \includegraphics[height=0.2in]{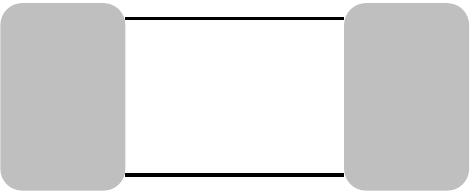} \epm,
\end{align}
In other words, if we locally deform/reconnect the strings as in Fig.
\ref{strnetSa}, the amplitude (or the ground state wave function) does not
change.

The first rule tells us that the amplitude of a string configuration only
depend on the topology of the string configuration.  Starting from a single
loop, using the local deformation and the local reconnection in Fig.
\ref{strnetSa}, we can generate all closed string configurations with any
number of loops.  So all those  closed string configurations have the same
amplitude.  Therefore, the local dancing rule fixes the wave function to be the
equal-weight superposition of all closed strings: 
\begin{align}
\label{Z2wav}
|\Phi_{Z_2}\>=\sum_\text{all closed strings} \left |\cfig{\includegraphics[height=0.2in]{strnetS}}\right \>
.  
\end{align}
In other words,  the local dancing rule fixes the global dancing pattern.

If we choose another local dancing rule, then we will get a different global
dancing pattern that corresponds to a different topological order.  One of the
new choices is obtained by just modifying the sign in \eqn{Z2rl}:
\begin{align}
\label{Semrl}
 \Phi
\bpm \includegraphics[height=0.2in]{Xi1} \epm  =&
\Phi
\bpm \includegraphics[height=0.2in]{Xi} \epm ,
&
 \Phi
\bpm \includegraphics[height=0.2in]{XijklX} \epm  =&
- \Phi
\bpm \includegraphics[height=0.2in]{XijX} \epm  .
\end{align}
We note that each local reconnection operation changes the number of loops by
1.  Thus the new local dancing rules gives rise to a wave function which has a
form 
\begin{align}
\label{Semwav}
|\Phi_\text{Semi}\>=\sum_\text{all closed strings} (-)^{N_\text{loops}}
\left |\cfig{\includegraphics[height=0.2in]{strnetS}}\right \> , 
\end{align}
where
$N_\text{loops}$ is the number of loops.  The wave function
$|\Phi_\text{Semi}\>$ corresponds to a different global dance and a different
topological order.

\subsubsection{Emergence of Fermi and fractional statistics}

Why the two  wave functions of non-oriented strings, $|\Phi_{Z_2}\>$ and
$|\Phi_\text{Semi}\>$ (see \eqn{Z2wav} and \eqn{Semwav}), have
topological orders?  This is because the two  wave functions give rise to
non-trivial topological properties.  The two  wave functions correspond to
different topological orders since they give rise to different topological
properties.  In this section, we will discuss two topological properties:
emergence of fractional statistics and, in next section, topological degeneracy
on torus.

The two topological states in two dimensions contain only closed strings, which
represent the ground states.  If the wave functions contain open strings (\ie
have non-zero amplitudes for open string states), then the ends of the open
strings will correspond to point-like topological excitations above the ground
states.  Although an open string is an extended object, its middle part merge
with the strings already in the ground states and is unobservable.  Only its
two ends carry energies and correspond to two point-like particles.  

We note that such a point-like particle from an end of string cannot be created
alone.  Thus an end of string correspond to a topological point defect, which
may carry fractional quantum numbers.  This is because an open string as a
whole always carry non-fractionalized quantum numbers.  But an open string
corresponds to \emph{two} topological point defects from the two ends.  So we
cannot say that each end of string also carries non-fractionalized quantum
numbers.  Some times, they do carry fractionalized quantum numbers.

Let us first consider the defects in the $|\Phi_{Z_2}\>$ state.  To understand
the fractionalization, let us first consider the \emph{spin} of such a defect,
to see if the \emph{spin} is fractionalized or not \cite{FFN0683}.  Note
that, here the \emph{spin} is not the spin of the spin-1/2 that form our model.
The \emph{spin} is the orbital angular momentum of an end.  We use different
fonts to distinguish them.  An end of string can be represented by 
\begin{align}
\big |\bmm \includegraphics[scale=0.33]{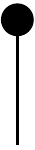}\emm \big \>_\text{def} 
= 
\big |\bmm \includegraphics[scale=0.33]{def1}\emm \big \>+
\big |\bmm \includegraphics[scale=0.33]{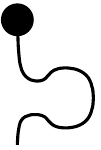}\emm \big \>+
\big |\bmm \includegraphics[scale=0.33]{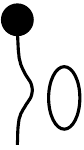}\emm \big \>+ ...
.
\end{align}
which is an equal-weight superposition of all string states obtained from the
deformations and the reconnections of $\bmm
\includegraphics[scale=0.33]{def1}\emm$.

Under a $360^\circ$ rotation, the end of string is changed to $\big |\bmm
\includegraphics[scale=0.33]{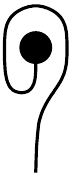}\emm \big \>_\text{def} $, which is an equal
weight superposition of all string states obtained from the
deformations and the reconnections of $\bmm \includegraphics[scale=0.33]{def3}\emm$.
Since $\big |\bmm \includegraphics[scale=0.33]{def1}\emm \big \>_\text{def} $
and $\big |\bmm \includegraphics[scale=0.33]{def3}\emm \big \>_\text{def} $ are
always different, $\big |\bmm \includegraphics[scale=0.33]{def1}\emm \big
\>_\text{def} $ is not an eigenstate of $360^\circ$ rotation and does not carry
a definite \emph{spin}.

To construct the  eigenstates of $360^\circ$ rotation, let us make a
$360^\circ$ rotation to $\big |\bmm \includegraphics[scale=0.33]{def3}\emm \big
\>_\text{def}$.  To do that, we first use the string reconnection move in Fig.
\ref{strnetSa}, to show that $\big |\bmm \includegraphics[scale=0.33]{def3}\emm
\big \>_\text{def} = \big |\bmm \includegraphics[scale=0.33]{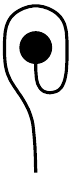}\emm \big \>_\text{def} $.  A
$360^\circ$ rotation on $\big |\bmm \includegraphics[scale=0.33]{def2}\emm \big
\>_\text{def} $ gives us $\big |\bmm \includegraphics[scale=0.33]{def1}\emm \big \>_\text{def} $.

We see that the $360^\circ$ rotation
exchanges $\big |\bmm \includegraphics[scale=0.33]{def1}\emm \big \>_\text{def} $
and $\big |\bmm \includegraphics[scale=0.33]{def3}\emm \big \>_\text{def} $.
Thus the  eigenstates of
$360^\circ$ rotation are given by
$\big |\bmm \includegraphics[scale=0.33]{def1}\emm \big \>_\text{def} + \big |\bmm
\includegraphics[scale=0.33]{def3}\emm \big \>_\text{def} $ with
eigenvalue 1, and by $\big |\bmm
\includegraphics[scale=0.33]{def1}\emm \big \>_\text{def} - \big |\bmm
\includegraphics[scale=0.33]{def3}\emm \big \>_\text{def} $ with eigenvalue $-1$.
So the particle $\big |\bmm \includegraphics[scale=0.33]{def1}\emm \big \>_\text{def} +
\big |\bmm \includegraphics[scale=0.33]{def3}\emm \big \>_\text{def} $ has a \emph{spin} 0 (mod
1), and the particle $\big |\bmm \includegraphics[scale=0.33]{def1}\emm \big \>_\text{def}
- \big |\bmm \includegraphics[scale=0.33]{def3}\emm \big \>_\text{def} $ has a \emph{spin} 1/2
(mod 1).

\begin{figure}[tb]
\centerline{
\includegraphics[height=.7in]{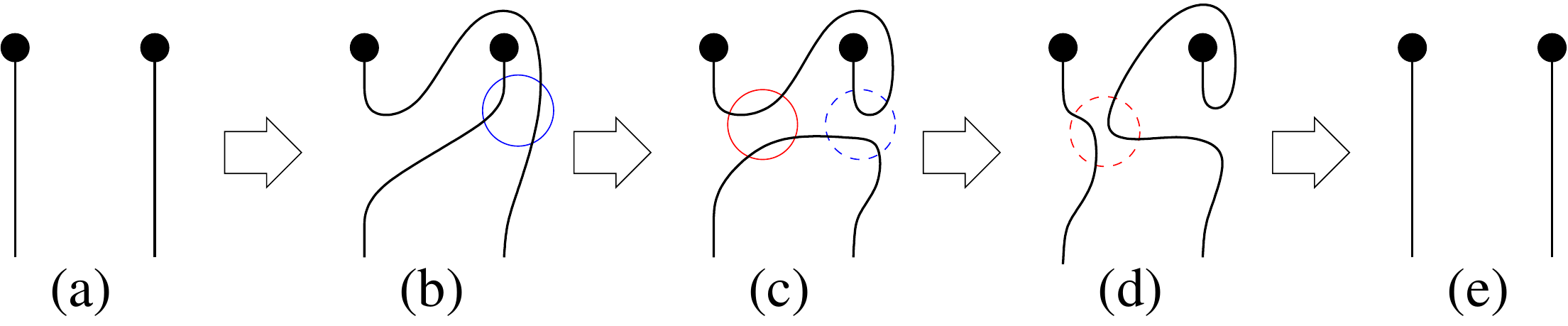}
}
\caption{
Deformation of strings and two reconnection moves, plus an exchange of two ends
of strings and a $360^\circ$ rotation of one of the end of string, change the
configuration (a) back to itself.  Note that from (a) to (b) we exchange the
two ends of strings, and from (d) to (e) we rotate of one of the end of string
by $360^\circ$.
The combination of those moves do not generate any phase.
}
\label{exch}
\end{figure}

\def\arraystretch{1.7} \setlength\tabcolsep{3pt}
\begin{table*}[t] 
\caption{\textbf{Topologically ordered states with long range entanglement}.  
Here 1B refers to 1-dimensional bosonic system,  2F 2-dimensional fermionic
system, \etc.  The second column  indicates the presence of fractionalized
point-like excitations.  The third column  indicates the presence of
non-abelian statistics.  The fourth column  indicates whether the boundary must
be gapless, or can be gapped, or for some must be gapless and for others can be
gapped.
} 
\label{top} 
\centering
\begin{tabular}{ |c|c|c|c|c|c| } 
\hline 
\textbf{Topological order}& \textbf{Frac. exc.}&\textbf{Non-ab. sta.}&\textbf{Boundary}&\textbf{Classification/comment} \\
\hline 
\hline 
1F Majorana chain \xcite{K0131} & No  & Not any & Maj. zero mode &  $\Z_2$ 
($Z_2^f$ symm. breaking)\\
\hline 
\hline 
2B bosonic $E_8$ state & No & No & Gapless &  Invertible topological order\\
\hline 
2B chiral spin liquid \xcite{KL8795,WWZcsp} & Semion & No & Gapless &  Spin quantum Hall state\\
\hline 
2B $Z_2$-spin liquid \xcite{RS9173,Wsrvb} & Fermion & No & Gapped & $Z_2$-gauge/toric-code \\
\hline 
2B double-semion state \xcite{FNS0428,LW0510} & Fermion & No & Gapped & $Z_2$-Dijkgraaf-Witten \xcite{DW9093}  \\
\hline 
2B string-net liquids \xcite{LW0510} & Yes & Yes & Gapped & Unitary fusion category \\
\hline 
2F $p+\ii p$ fermion paired state \xcite{SMF9945,RG0067} & No  & No  & Gapless & Invertible topological order \\
\hline 
\parbox[c]{9em}{2F integer quantum\\ Hall states \xcite{KDP8094} } & No & No & Gapless & $\Z$ (invertible topological order) \\
\hline 
\parbox[c]{10em}{2F Laughlin states \xcite{L8395} \\ 2F Halperin states \xcite{H8375} } & Yes & No &
Gapped/gapless & $K$-matrix (symmetric, integral)  \\
\hline 
2F $\chi_1\chi_2^2$ state \xcite{Wnab} & Yes & $SU(2)_2$ & Gapless &  Cannot do universal TQC \\
\hline 
2F $\chi_2^3$ state \xcite{Wnab} & Yes & $SU(3)_2$ & Gapless & Can do universal TQC \\
\hline 
2F Pfaffian state \xcite{MR9162} & Yes & $SU(2)_2$ & Gapless & Cannot do universal TQC \\
\hline 
2F $Z_3$ parafermion state \xcite{RR9984} & Yes & $SU(2)_3$ & Gapless & Can do universal TQC \\
\hline 
2F string-net liquids \xcite{GWW1017,BK160501640} & Yes & Yes & Gapped & Unitary super fusion category \\
\hline 
\hline 
3+1D superconductor \xcite{W9141,HOS0497} & Fermion & Not any & Gapped &  With dynamical $U(1)$ gauge field\\
\hline 
3B string-net liquids \xcite{LW0510} & Fermion & Not any & Gapped & Symmetric fusion category \\
\hline 
3B Walker-Wang model \xcite{WW1132} & Fermion & Not any & Gapped & Pre-modular tensor category \\
\hline 
3B all-boson topo. order \xcite{LW170404221}  & Boson & Not any & Gapped & 
Pointed fusion 2-category  \\
\hline 
\end{tabular}
\end{table*}

If one believes in the \emph{spin}-statistics theorem, one may guess that the particle
$\big |\bmm \includegraphics[scale=0.33]{def1}\emm \big \>_\text{def} + \big |\bmm
\includegraphics[scale=0.33]{def3}\emm \big \>_\text{def} $ is a boson and the particle
$\big |\bmm \includegraphics[scale=0.33]{def1}\emm \big \>_\text{def} - \big |\bmm
\includegraphics[scale=0.33]{def3}\emm \big \>_\text{def} $ is a fermion.  This guess is
indeed correct.  Form Fig. \ref{exch}, we see that we can use deformation of
strings and two reconnection moves to generate an exchange of two ends of
strings and a $360^\circ$ rotation of one of the end of string.  Such
operations allow us to show that Fig. \ref{exch}a and  Fig. \ref{exch}e have
the same amplitude, which means that an exchange of two ends of strings
followed by a $360^\circ$ rotation of one of the end of string do not generate
any phase.  This is nothing but the \emph{spin}-statistics theorem.  

The emergence of Fermi statistics in the $|\Phi_{Z_2}\>$ state of a purely
bosonic spin-1/2 model indicates that the state is a topologically ordered
state.  We also see that the $|\Phi_{Z_2}\>$ state has a bosonic quasi-particle
$\big |\bmm \includegraphics[scale=0.33]{def1}\emm \big \>_\text{def} + \big
|\bmm \includegraphics[scale=0.33]{def3}\emm \big \>_\text{def} $, and a
fermionic quasi-particle $\big |\bmm \includegraphics[scale=0.33]{def1}\emm
\big \>_\text{def} - \big |\bmm \includegraphics[scale=0.33]{def3}\emm \big
\>_\text{def} $.  The bound state of the above two particles is a boson (not a
fermion) due to their  mutual semion statistics.  Such quasi-particle content
agrees exactly with the $Z_2$ gauge theory which also has three type of
topological excitations, two bosons and one fermion.  In fact,
the low energy effective theory of the topologically ordered state
$|\Phi_{Z_2}\>$ is the $Z_2$ gauge theory and we will call  $|\Phi_{Z_2}\>$ a
$Z_2$-topologically ordered state \cite{RS9173,Wsrvb}.

Next, let us consider the defects in the $|\Phi_\text{Semi}\>$ state.  Now
\begin{align}
\big |\bmm \includegraphics[scale=0.33]{def1}\emm \big \>_\text{def} 
= 
\big |\bmm \includegraphics[scale=0.33]{def1}\emm \big \>+
\big |\bmm \includegraphics[scale=0.33]{def1a}\emm \big \>-
\big |\bmm \includegraphics[scale=0.33]{def1b}\emm \big \>+ ...
.
\end{align}
and a similar expression for $\big |\bmm \includegraphics[scale=0.33]{def3}\emm
\big \>_\text{def}$, due to a
change of the local rule
for reconnecting the strings (see \eqn{Semrl}).  
Using the string reconnection move in Fig.
\ref{strnetSa}, we find that $\big |\bmm \includegraphics[scale=0.33]{def3}\emm
\big \>_\text{def} = - \big |\bmm \includegraphics[scale=0.33]{def2}\emm \big
\>_\text{def} $.  So a $360^\circ$ rotation, changes $(\big |\bmm
\includegraphics[scale=0.33]{def1}\emm \big \>_\text{def}, \big |\bmm
\includegraphics[scale=0.33]{def3}\emm \big \>_\text{def} )$ to $( \big |\bmm
\includegraphics[scale=0.33]{def3}\emm \big \>_\text{def}, -\big |\bmm
\includegraphics[scale=0.33]{def1}\emm \big \>_\text{def} )$.  We find that
$\big |\bmm \includegraphics[scale=0.33]{def1}\emm \big \>_\text{def} + \ii
\big |\bmm \includegraphics[scale=0.33]{def3}\emm \big \>_\text{def} $ is the
eigenstate of the $360^\circ$ rotation with eigenvalue $-\ii$, and $\big
|\bmm \includegraphics[scale=0.33]{def1}\emm \big \>_\text{def} - \ii \big
|\bmm \includegraphics[scale=0.33]{def3}\emm \big \>_\text{def} $ is the other
eigenstate of the $360^\circ$ rotation with eigenvalue $\ii$.  So the
particle $\big |\bmm \includegraphics[scale=0.33]{def1}\emm \big \>_\text{def}
+ \ii \big |\bmm \includegraphics[scale=0.33]{def3}\emm \big \>_\text{def} $
has a \emph{spin} $-1/4$, and the particle $\big |\bmm
\includegraphics[scale=0.33]{def1}\emm \big \>_\text{def} - \ii \big |\bmm
\includegraphics[scale=0.33]{def3}\emm \big \>_\text{def} $ has a \emph{spin} $1/4$.
The \emph{spin}-statistics theorem is still valid for $|\Phi_\text{Semi}\>_\text{def}$
state, as one can see form Fig. \ref{exch}.  So, the particle $\big |\bmm
\includegraphics[scale=0.33]{def1}\emm \big \>_\text{def} + \ii\big |\bmm
\includegraphics[scale=0.33]{def3}\emm \big \>_\text{def} $ and particle $\big
|\bmm \includegraphics[scale=0.33]{def1}\emm \big \>_\text{def} - \ii\big
|\bmm \includegraphics[scale=0.33]{def3}\emm \big \>_\text{def} $ have
fractional statistics with statistical angles of semion: $\pm \pi/2$.  Thus the
$|\Phi_\text{Semi}\>$ state contains a topological order.  We will
call such a topological order a \textbf{double-semion topological order} \cite{FNS0428,LW0510}.

It is amazing to see that the long-range quantum entanglement in string liquid
can gives rise to fractional \emph{spin} and fractional statistics, even from a purely
bosonic model.  Fractional \emph{spin} and Fermi statistics are two of most mysterious
phenomena in natural.  Now, we can understand them as merely a phenomenon of
long-range quantum entanglement.  They are no longer mysterious.

\subsubsection{Topological degeneracy}

The $Z_2$-topological order has another important topological property:
topological degeneracy \cite{RC8933,Wsrvb}.  Topological degeneracy is the
ground state degeneracy of a gapped many-body system that is robust against any
local perturbations as long as the system size is large \cite{WNtop}.
It implies the presence of topological order.

Topological degeneracy can be used as protected qubits which allows us to
perform topological quantum computation.\cite{K032} It is believed that the
appearance of topological degeneracy implies the topological order (or
long-range entanglement) in the ground state. Many-body states
with topological degeneracy are described by topological quantum field theory
at low energies.

The simplest topological  degeneracy appears when we put topologically ordered
states on compact spaces with no boundary.  We can use the global entanglement
pattern to understand the  topological  degeneracy.  We know that the local
rules determine the global entanglement pattern.  On a sphere, the  local rules determine a unique global entanglement pattern.  So the ground state
is non-degenerate.  However on other  compact spaces, there can be several
global entanglement patterns that all satisfy the same local rules. In this
case, the ground state is degenerate.

\begin{figure}[tb]
\centerline{
\includegraphics[height=0.6in]{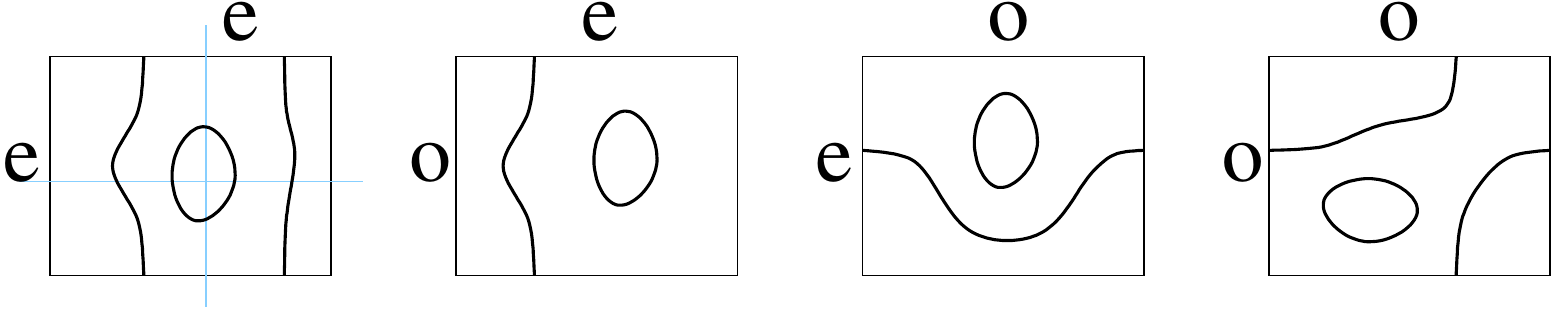}
}
\caption{
 On a torus, the closed string configurations
can be divided into four sectors, depending on even or
odd number of strings crossing the x- or y-axes.
}
\label{z2eo}
\end{figure}

For the $Z_2$-topological state on torus, the local rule relate the
amplitudes of the string configurations that differ by a string reconnection
operation in Fig. \ref{strnetSa}.  On a torus, the closed string configurations
can be divided into four sectors (see Fig. \ref{z2eo}), depending on even or
odd number of strings crossing the x- or y-axes.  The string reconnection
move only connect the string configurations among each sector.  So the
superposition of the string configurations in each sector represents a
different many-body wave functions.
Since those many-body wave functions are locally indistinguishable, they
correspond to different degenerate ground states.  Therefore, the local 
rule for the  $Z_2$-topological order gives rise to four fold degenerate ground
state on torus.

Similarly,  the double-semion topological order also gives rise to four fold
degenerate ground state on torus.

\subsection{Table of some topological orders}

In table \ref{top}, we list some topological orders in bosonic and fermionic
systems in various dimensions.  The simplest one in the table is the 2+1D
\textbf{IQH states} \cite{KDP8094}.  
%
Some entries in table \ref{top} have not been discussed above.  In particular,
the \textbf{string-net liquids} for bosonic systems \cite{LW0510} and fermionic
systems \cite{GWW1017,BK160501640} allow us to obtain all 2+1D topological
orders with gappable boundary \cite{KK1251,LW1384}. It reveals that 2+1D
bosonic topological orders are classified by \textbf{unitary fusion categories}
\cite{ENO0562}, while 2+1D fermionic topological orders are classified by
\textbf{unitary super fusion categories}.  For more general 2+1D bosonic
topological orders, it was conjectured \cite{Wrig},
and became more and more clear 
\cite{KW9327,K062,RSW0777,W150605768}, that they
are classified by the \textbf{modular matrices} $S,T$ (which
encode unitary modular tensor categories (MTC) \cite{MS8977})
plus the chiral central charge $c$ of the edge states.  Physically, the so
called \emph{MTC can be viewed as a set of topological excitations, together
with the data that describes the fusion and braiding of those excitations.} 


Many topological orders have fractionalized excitations (see the second column
of  table \ref{top}), some 2+1D  topological orders even have non-abelian
excitations (see the third column of table \ref{top}).  In 1+1D fermion systems
and 2+1D boson/fermion systems, there are even topological orders
that have no fractionalized excitations (the second column with an ``No''
entry). Those  topological orders are called invertible topological
orders \cite{KW1458,K1459,F1478}, and their non-trivialness is reflected in
their non-trivial boundary states which has a gravitational
anomaly \cite{W1313,KW1458}.

Regarding \textbf{Pt.8} in Sec. \ref{newworld}, we note that the fermions are
fractionalized topological excitations in bosonic systems. But they are local
non-topological excitations in fermionic systems.  For example \textbf{Majorana
fermions} are local non-topological excitations in fermionic superconductor
(with spin-orbital coupling and no dynamical $U(1)$ gauge field), since they
are antiparticles of themselves.  Therefore, Majorana fermions are indeed
fermions with Fermi statistics.  They are not particles with non-abelian
statistics.  In fact, Majorana fermions are the familiar Bogoliubov
quasiparticles in superconductors which were discovered long time ago.  So what
people are looking for, in the intensive experimental search, is not the
Majorana fermion first introduced by Majorana, but instead \textbf{Majorana
zero mode}, that can appear, for example, at the end of an 1D $p$-wave
superconductor \cite{K0131}, or at the center of a vortex in a 2D $p+\ii p$
fermion paired state \cite{SMF9945,RG0067}.  Majorana zero mode is not Majorana
fermion. In fact, it is not even a particle.  It is a \emph{property} of a
particle, just like the mass is a property of a particle.  If the mobile
particle carries a Majorana zero mode, then the particle will have a
non-abelian statistics \cite{I0168}.  So one should not mix Majorana zero mode
with Majorana fermion.

We also like to mention that the $SU(2)_2$-type of non-abelian statistics in
the $\chi_1\chi_2^2$ FQH state and the Pfaffian state contain a non-abelian
quasiparticle that carries an Majorana zero mode.  Such a particle has an
internal degrees of freedom of half of a qubit (\ie quantum dimension $d=\sqrt
2$).\footnote{An physical explanation of quantum dimension can be found in
\textcite{K062} and \textcite{W150605768}.}

Last, this paper only discusses topological phases at zero temperature.  Phases
beyond Landau symmetry breaking order also exist for $T\neq 0$, which are not
reviewed here since they requires a different theoretical framework.

\section{Even product states can be non-trivial, if there is a symmetry}
\label{secSPT}

One expects gapped product states that have neither symmetry breaking order nor
topological order to be trivial, in the sense that all those states belong to
one single phase. In this section, we will see that in fact those states can
belong to several different phases if there is a symmetry, and thus can be
non-trivial.

\subsection{Gapped integer-spin chain: Haldane phases}

The ground state of the $SO(3)$ symmetric anti-ferromagnetic spin-1/2
Heisenberg chain
\begin{align}
 H = \sum_i \v S_i \cdot \v S_{i+1}
\end{align}
cannot break the $SO(3)$ spin rotation symmetry due to quantum
fluctuations.\cite{MW6633} What is the nature of this symmetric ground state?
The Beth ansatz approach, bosonization, and Lie-Schultz-Mattis theorem
\cite{LSM6107} all indicate the ground state of spin-1/2 Heisenberg chain
behaves almost like a spontaneous $SO(3)$ symmetry breaking state: spin-spin
correlation has an slow algebraic decay (in contrast to exponential decay for a
typical disordered system) and the chain is gapless (as if having an Goldston
mode \cite{G6154}). This result led people to believe that all spin-$S$ chain
are also  gapless and have algebraic decaying spin-spin correlation, since for
$S>1/2$, the spins have even weaker quantum fluctuations than the spin-1/2
chain, 


In 1983, Haldane considered spin fluctuations in 1+1D space-time that have
non-trivial ``winding'' number in $\pi_2(S^2)$.
He realized that the spin configuration with ``winding'' number $\pm 1$ has a
phase factor $-1$ if the spin is half-integer and a phase factor $1$ if the
spin is integer. So the  half-integer spin chain and integer spin chain may
have different dynamics.  Haldane conjectured \cite{H8364} that the spin chain
is gapped if the spin is integer, despite it has weaker quantum fluctuations
than spin-1/2 chain.  If the spin is half-integer, then the spin chain is
gapless.  The gapped ground state of an integer spin chain is called a Haldane
phase.  At that time, people believed the Haldane phase to be a trivial
disordered phase, just like the product state formed by spin-0 on each site.

However, such an opinion was put in doubt by an exact soluble integer spin
chain. It was shown that, for the exactly soluble model \cite{AKL8877}, the
boundary of the integer spin-$S$ chain carries degenerate degrees of freedom of
spin-$S/2$.  Since the gapless edge excitations for 2+1D FQH states implies a
bulk topological order, people start to wonder that maybe the
similar picture applies to one lower dimensions: the gapped 1+1D ground states
of integer spin chains also have topological orders due to the gapless
spin-$S/2$ boundary.

But this point of view seems incorrect.  The gapless boundary of a 2+1D chiral
topological order is actually a bulk property, since gaplessness is robust
against any modifications on the boundary.  This is why the gapless boundary
reflects a bulk topological order.  However, gapless spin-$S/2$
boundary of spin-$S$ chain can be easily gapped by applying a Zeeman field at
the boundary.  This seems to suggest that the gapped ground state of integer
spin chain to be trivial.

\begin{figure}[t]
\begin{center}
\includegraphics[scale=0.5]{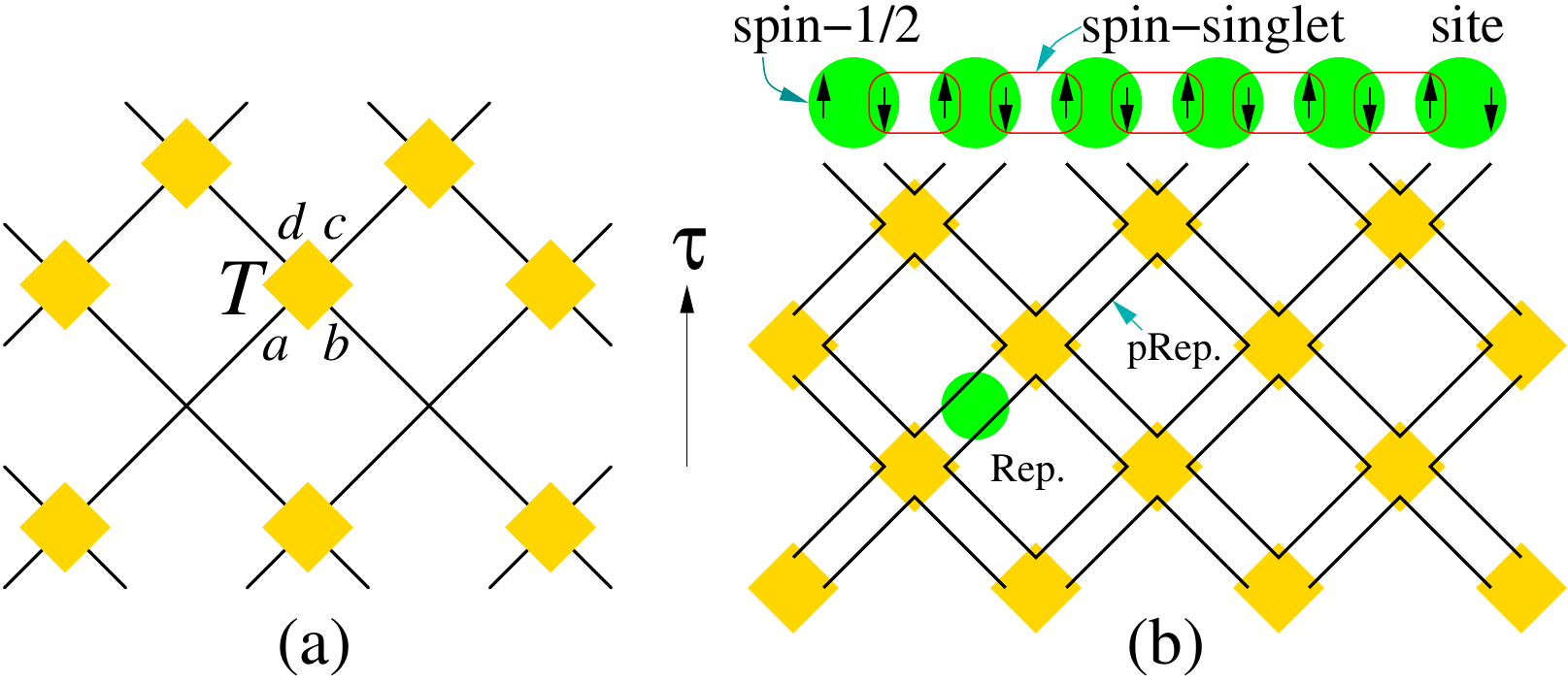}
\end{center}
\caption{
(a) A tensor network representation of the partition function
$Z=\text{Tr}\ee^{-\tau H}$ of a 1+1D quantum system obtained from path
integral.  Each vertex is a rank-4 tensor $T_{abcd}$ where each leg corresponds
to an index. The range of the index is the dimension of the tensor $T$.  The
partition function $Z$ is obtained as a product of all tensors, with the common
indices on the edges linking two vertices summed over (which corresponds to the
path integral).  We can combine several tensors $T$ to form a new tensor $T'$
and obtain a new coarse-grained tensor network that produces the same partition
function $Z$.  After many coarse-graining iterations, we obtain a fixed-point
tensor $T^\text{fix}$ that characterizes a quantum phase.  (b) The fixed-point
tensor of spin-1 Heisenberg chain has a corner-double-line structure.  It gives
rise to the fixed-point wave function of an ideal $SO(3)$-SPT state.
} 
\label{singlecdlT} 
\end{figure}

\subsection{Haldane phases are not topological for even-integer-spin, while
topological for odd-integer-spin }

What is the nature of the Haldane phase for integer spin-$S$ chain?
Topological or not topological? This question bothered me for 15 years, until
we used tensor-entanglement-filtering renormalization (TEFR) approach (see Fig.
\ref{singlecdlT}a) to study spin-1 XXZ chain \cite{GW0931}:
\begin{align}
 H  = \sum_{i} J \v S_{i}\cdot \v S_{i+1} + U (S^z_{i})^2
\end{align}
Unlike density matrix renormalization group (DMRG) approach \cite{W9263}, TEFR
approach gives us a simple fixed-point tensor.  We found that the fixed-point
tensor has a corner-double-line structure (with degenerate weights) when
$U\approx 0$ (see Fig.  \ref{singlecdlT}b), and the fixed-point tensor becomes
a dimension-1 trivial tensor when $U\gg J$ (see Fig.  \ref{singlecdlT}a where
the indices of $T$ are all equal to 1).

The ground state for $U\gg J$ is a product state of $|S^z_i=0\>$ which is
consistent with trivial dimension-1 fixed-point tensor.  The corner-double-line
fixed-point tensor for $U=0$ corresponds to a fixed-point wave function that
contains 4 states per site (increased from 3 states of spin-1, see Fig.
\ref{singlecdlT}b).  The 4 states form the $3\oplus 1$ dimensional
representation of $SO(3)$, which can be viewed as two spin-1/2 representations
(the projective representations of $SO(3)$) 
\begin{align} 
3\oplus 1 = 2\otimes 2 .  
\end{align} 
In such a fixed-point wave function, the two spin-1/2's on neighboring sites
form a spin singlet. The total fixed-point wave function is the product state
of those spin singlets (see Fig.  \ref{singlecdlT}b).  We discovered that, just
like the $U\gg J$ limit, \emph{the spin-1 Haldane phase is also a short-range
entangled state equivalent to a product state.  It is not topological despite
the fractionalized spin-$1/2$ boundary.}  

However, non-topological does not mean trivial.  We find that, for spin-1
chain, the corner-double-line structure even appear for the follow
generic Hamiltonian
\begin{align}
\label{Sp1xz}
 H & = \sum_{i} [ J \v S_{i}\cdot \v S_{i+1} + U (S^z_{i})^2 ]
\\
& \ \ \
+\sum_{i} B_x S^x_i + B_z S^z_i
+ B_x'
[S^x_i(S^z_{i+1})^2 +S^x_{i+1}(S^z_i)^2]
\nonumber 
\end{align}
when $U,B_{x,z},B_x' \approx 0$.  This suggests that the corner-double-line
structure is stable against any perturbations with time reversal symmetry $T^*$
(which is the usual time reversal plus a 180$^\circ$ spin-$S^y$ rotation) and
spacial reflection symmetry\footnote{In fact, the corner-double-line structure
is stable against any perturbations with time reversal symmetry $T^*$ \emph{or}
spacial reflection symmetry \cite{PBT1225}.}.  On the other hand, the
corner-double-line structure can be destroyed by perturbations that break those
symmetries.  This suggest that the spin-1 Haldane phase, characterized by the
corner-double-line tensor (or the dimmerized fixed-point wave function) is a
stable phase, distinct from the product state of $|S_z=0\>$'s, as long as we do
not break those symmetries.  We conclude that the Haldane phase of spin-1 chain is
non-trivial despite it is a product state that does not spontaneously break any
symmetry! This is a new state of matter and we propose the concept of
\textbf{symmetry protected trivial (SPT)} order to describe this new state of
matter. SPT orders is characterized by the corner-double-line fixed-point
tensors with degenerate weights (or the dimmerized fixed-point wave function).
Later, \textcite{PBT1039} also showed that SPT orders can be characterized
via the entanglement spectrum.  It is interesting to see that even product
states without spontaneous symmetry breaking can be non-trivial.  
%
%
However, the spin-1 Haldane phase at that time has already been widely referred
as a topological phase. So we gave the term ``SPT order'' another
representation ``\textbf{symmetry protected topological} order''\footnote{After
long debates, we eventually used the second less-accurate representation in our
paper.}.  

It is very important to regard SPT states as short-range entangled, not
topological (in the sense of orange-vs.-donut).  This correct way of thinking
leads to a complete classification  of all 1D gapped interacting phases
\cite{CGW1107,SPC1139}, in terms of projective representations of the symmetry
group\cite{PBT1039} one year later  and the systematic group cohomology theory
of SPT phases in higher dimensions two years later \cite{CGL1172}.  In
particular, the projective-representation classification of 1+1D SPT phases
indicates that \emph{only the odd-integer-spin Haldane phases are the $SO(3)$-SPT
phases, while the even-integer-spin Haldane phases are not the $SO(3)$-SPT phase just
like the product state of spin-0's} \cite{PBT1225}.  So Haldane phases can be
topological or non-topological depending on the spin to be odd or even integer.
This explains the \textbf{Pt.1} in Sec.  \ref{newworld}. 

\subsection{An $Z_2$-SPT state in 2+1D}

After realizing SPT states to be product states, it becomes easy to construct
SPT states in any dimension. We just need to write a product state in some
complicated form, and then try to find all the twisted way to implement the
symmetry.

First, we need to introduce the concept of on-site symmetry, which is usually
referred as global symmetry.  Relative to the tensor product decomposition $
\cH^\mathrm{tot}= \bigotimes_i \cH_i$ of the total Hilbert space, a symmetry
transformation is on-site if it has a tensor product decomposition $ U =
\prod_i U_i$, where $U_i$ is the  symmetry transformation acting on $\cH_i$.
The notion of on-site symmetry is stressed in \textcite{CLW1141,CGW1107}, which
is a key to understand SPT states.

\begin{figure}[t]
\begin{center}
\includegraphics[scale=0.5]{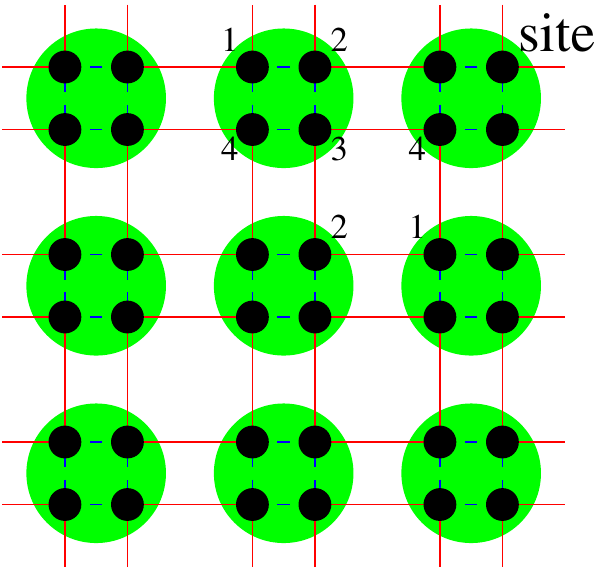}
\end{center}
\caption{
(Color on line) The filled dots are qubits (or spin-1/2's).  A circle (with
dots inside) represents a site.  The dash blue line connecting dots $i,j$
represents the phase factor $CZ_{ij}$ in the $Z_2$ global symmetry
transformation.  In the $Z_2$-SPT state, the four spins in a plaquette (connected
by a red square) is described by $ \frac{1}{\sqrt{2}} ( |\up\up\up\up\> +
|\down\down\down\down\>)$.
} 
\label{CZX} 
\end{figure}

The first lattice model that realizes \cite{CLW1141} a 2+1D SPT state has four
qubits (or spin-1/2 spins) on each site (see Fig. \ref{CZX}).  A complicated
product state is given by
\begin{align}
 |\Psi_0\> &
= \bigotimes_\text{plaquette}  \frac{1}{\sqrt{2}} ( |\up\up\up\up\> +   |\down\down\down\down\> )
\end{align}
where $ \frac{1}{\sqrt{2}} ( |\up\up\up\up\> + |\down\down\down\down\> )$ is
the wave function for the four spins in the plaquette (see Fig. \ref{CZX}).
Note that the four spins in $ \frac{1}{\sqrt{2}} ( |\up\up\up\up\> + |\down\down\down\down\> )$ are  on four different sites.  

One way to introduce a $Z_2$ symmetry is to define the transformation on each
site to be the spin flipping:
\begin{align}
\label{UX}
U_X= 
\si^x_1 \si^x_2 \si^x_3 \si^x_4 ,\ \ \ \ U_X^2=1.
\end{align}
Obviously, $|\Psi_0\>$ is invariant under such a spin flipping $Z_2$
transformation.  But for such a $Z_2$ symmetry, $|\Psi_0\>$ is not a SPT
state.

There is another way to define $Z_2$ symmetry (on each site, see Fig.
\ref{CZX}), but this time with a twist:
\begin{align}
U_{CZX}=U_{X}U_{CZ}, 
\end{align}
where the $\pm1$ phase twist, $U_{CZ}$, is a product of $CZ_{ij}$ that acts on
the two spins at $i$ and $j$: $CZ_{ij}=-1$ when acts on $|\down\down\>$ and
$CZ_{ij}=1$ otherwise. More specifically 
\begin{align}
U_{CZ}&=\prod_{j=1,2,3,4} CZ_{j,j+1}
\nonumber\\
&=\prod_{j=1,2,3,4} \frac{1+\si^z_{j+1}+\si^z_{j}  -\si^z_{j+1}\si^z_{j}}{2},
\end{align}
where $j=5$ is the same as $j= 1$.  It is a non-trivial exercise but one can
indeed check that $U_{CZX}^2=1$.  $|\Psi_0\>$ is invariant under such a twisted
spin flipping $Z_2$ transformation since all the $\pm 1$ $CZ_{ij}$ factors
cancel each other.  For the new $Z_2$ symmetry, $|\Psi_0\>$ is an
SPT state \cite{CLW1141}.  In fact, one can construct an exactly soluble
lattice Hamiltonian, which is symmetric under the new symmetry and has
$|\Psi_0\>$ as its unique gapped group state.  

The above construction has been generalized to higher dimensions and arbitrary
compact symmetry group via group cohomology theory: for each element in
$\cH^{d+1}(G;\R/\Z)$, we can construct an $d+1$D SPT state protected by
$G$-symmetry.  But one thing remain unclear: how to see those constructed state
to be a $G$-SPT state?

\subsection{Probing SPT orders}

\begin{figure}[tb]
\begin{center}
\includegraphics[scale=0.3]{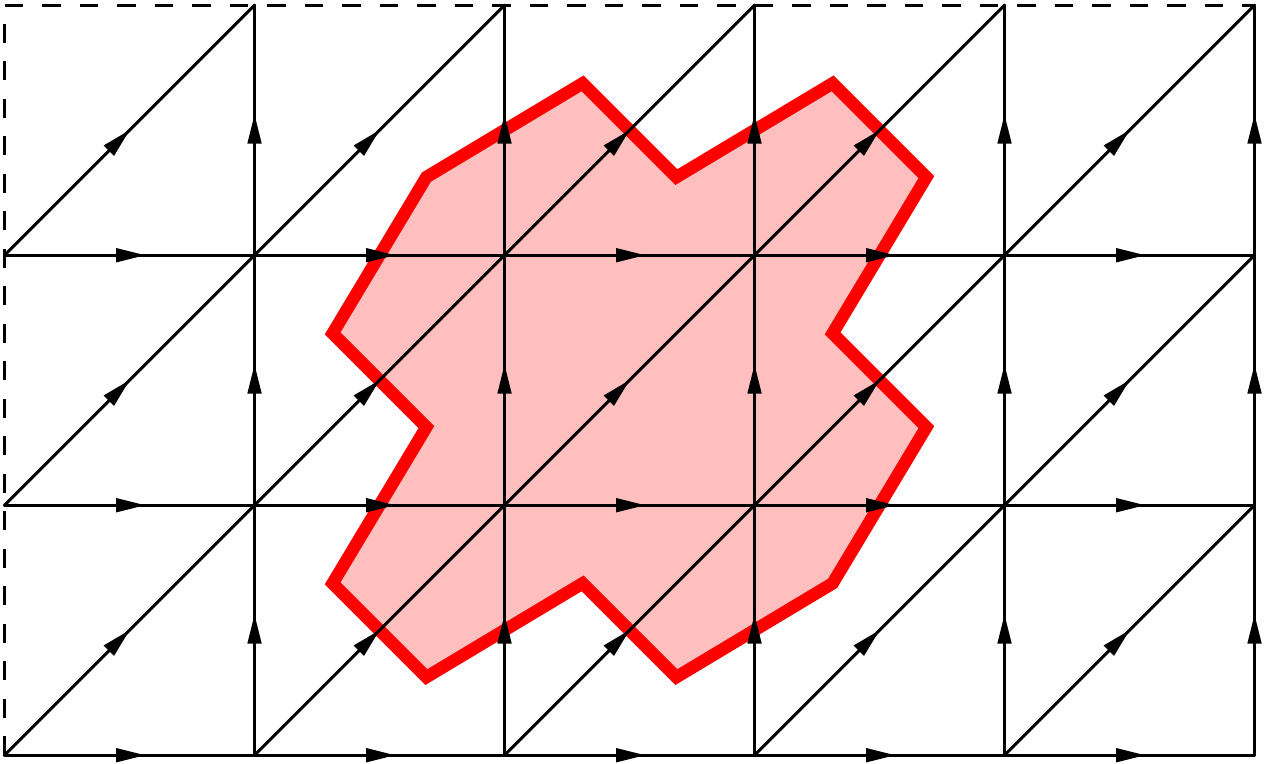} \end{center}
\caption{ (Color online)
A 2D lattice on a torus.  A $g\in G$ transformation is performed on the sites in
the shaded region.  The $g$ transformation changes the Hamiltonian term on
the triangle $(ijk)$ across the boundary from $H_{ijk}$ to $H^g_{ijk}$.
}
\label{ltrans}
\end{figure}

\begin{figure}[tb]
\begin{center}
\includegraphics[scale=0.25]{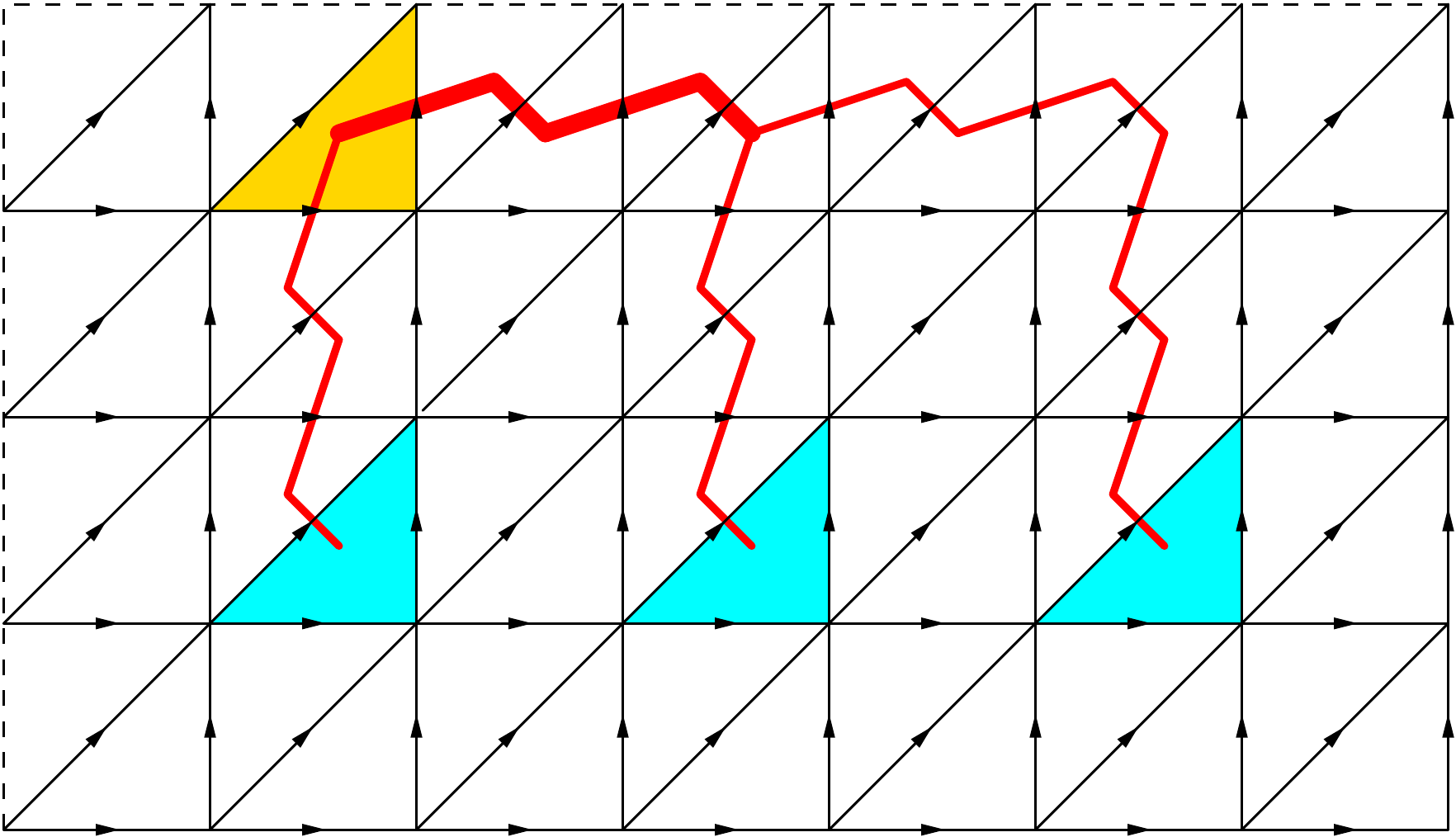} \end{center}
\caption{ (Color online)
Three identical monodromy defects (blue triangles) from $G=Z_3=\{0,1,2\}$
symmetry twist. The think red lines are $1$-cuts, and the thick  red line is a
$2$-cut. The $g$-cuts can be relocated by local $Z_3$ transformations as in Fig. \ref{ltrans}.
The yellow triangle can also be relocated by local $Z_3$ transformations.
Thus it is not a monodromy defect.
}
\label{ZnDefect}
\end{figure}

\def\arraystretch{1.7} \setlength\tabcolsep{3pt}
\begin{table*}[t] 
\caption{\textbf{SPT states with short-range entanglement}.  Here 1B refers to
1-dimensional bosonic system,  2F 2-dimensional fermionic system, \etc. Also
$T$ represents the time reversal symmetry, which generates the group $Z_2^T$
for bosonic systems, and $Z_4^T$ for electron systems.  This is because
$T^2=(-)^{N_F}$ is the fermion-number-parity operator for electron systems.
The last column describes the degenerate state at the end
of 1D SPT phases, or other SPT-probes for higher dimensions.
} 
\label{SPT} 
\centering
\begin{tabular}{ |c|c|c|c| } 
\hline 
\textbf{SPT order}& \textbf{Symm.}&\textbf{Classification}  & \textbf{Chain-end/SPT-probe}\\
\hline 
1B spin-1 Haldane phase \xcite{H8364} &  $SO(3)$ &  $\cH^2(SO(3),\R/\Z)=\Z_2$ &Spin-$1/2$ \\
\hline 
1B spin-1 Haldane phase \xcite{H8364} &  $Z_2^T$ &  $\cH^2(Z_2^T,\R/\Z)=\Z_2$ &Kramer doublet \\
\hline 
1B symm. gapped phases \xcite{PBT1039,CGW1107} &  $G$ &  $\cH^2(G,\R/\Z)$ & Proj. rep. of $G$ \\
\hline 
1F ins. w/ coplanar spin order \xcite{CGW1128} &  $U^f(1)\rtimes Z_2^{T} $ &  $\Z_2$ & Kramer doublet\\
\hline 
1F topo. superconductor \xcite{QZ09083550,K0986,W1103} &  $Z_4^{T} $ &  $\Z_2$ & charge-$0$ Kramer doublet\\
\hline 
1F $G^f$-SPT phases \xcite{CGW1128} &  $G^f$ &  $\cH^2(G^f,\R/\Z)$ & Proj. rep. of $G^f$ \\
\hline 
\hline 
2B $Z_n$-SPT states \xcite{CLW1141} &  $Z_n$ &  $\cH^3(Z_n,\R/\Z)=\Z_n$ & \parbox[c]{11em}{$Z_n$-dislocation has frac.\\ statistics/$Z_n$-charge} \\
\hline 
2B SPT insulator \xcite{CGL1172,LV1219} &  $U(1)$ &  $\cH^3(U(1),\R/\Z)=\Z$ & Even-int. Hall conductance \\
\hline 
2B $T$-symm. SPT insulator \xcite{CGL1172,LV1219} &  $U(1)\rtimes Z_2^T$ &  $\cH^3( U(1)\rtimes Z_2^T,\R/\Z)=\Z_2$ & $\pi$-flux has Kramer doub.\\
\hline 
2B spin quantum Hall states \xcite{CGL1172,LW1305} &  $SO(3)$ &  $\cH^3(SO(3),\R/\Z)=\Z$ & Quantized spin Hall cond. \\
\hline 
2B $T$-symm. SPT spin liquid \xcite{CGL1172} &  $Z_2^T\times SO(3)$ &  $\cH^3(Z_2^T\times SO(3),\R/\Z)=\Z_2$ & \\
\hline 
2B $G$-SPT states \xcite{CGL1172} &  $G$ & $\cH^3(G,\R/\Z)$ & \\
\hline 
2F quantum spin Hall states \xcite{KM0501,BZ0602} &  $U^f(1)\times U^f(1)$ &  $\Z$ & Spin-charge Hall cond. \\
\hline 
2F topological insulator \xcite{KM0502} &  $[U^f(1)\rtimes Z_4^T]/Z_2$ &  $\Z_2$ &  \parbox[c]{11em}{$\pi$-flux carries\\ charge-0 Kramer doublet} \\
\hline 
2F topo. superconductor \xcite{R0664,QHR0901,SF0904} &  $Z_4^{T} $ &  $\Z_2$ & \parbox[c]{12em}{$\pi$-flux carries\\charge-even Kramer doub.} \\
\hline 
2F $G^f$-SPT states \xcite{LW150704673,LW160205946} &  $G^f$ without $T$& \parbox[c]{16em}{Chiral central charge $c=0$ \\modular extensions of sRep$(G^f)$}  & \\
\hline 
\hline 
3B $T$-symm. SPT states \xcite{CGL1172,WS1334} &  $Z_2^T$ &  $\cH^4(Z_2^T,\R/\Z)\oplus \Z_2=\Z_2^2$ &  \\
\hline 
3B $T$-symm. SPT insulator \xcite{CGL1172,WS1334} &  $U(1)\rtimes Z_2^T$ &  $\cH^4(U(1)\rtimes Z_2^T,\R/\Z)\oplus \Z_2=\Z_2^3$ & A monople is a fermion \xcite{MKF1331}\\
\hline 
3B $T$-symm. SPT spin liquid \xcite{CGL1172} &  $Z_2^T\times SO(3)$ &  $\cH^4(Z_2^T\times SO(3),\R/\Z)\oplus \Z_2=\Z_2^4$ & \\
\hline 
3B $G$-SPT states \xcite{CGL1172} &  $G$ without $T$ &  $\cH^{4}(G,\R/\Z)$ & \\
\hline
3B $G$-SPT states \xcite{CGL1172,VS1306} &  $G$ with $T$ &  $\cH^{4}(G,\R/\Z)\oplus \Z_2$ & \\
\hline 
3F topological insulator \xcite{MB0706,R0922,FKM0703,QHZ0824} &  $[U^f(1)\rtimes Z_4^T]/Z_2$ &  $\Z_2$ & \parbox[c]{11em}{A monople carries\\ half-integer charge}\\
\hline 
3F topo. superconductor \xcite{R0664,QHR0901,SF0904} &  $Z_4^{T} $ &  $\Z_{16}$ \xcite{WS14011142,KTT1429} & \\
\hline 
\end{tabular}
\end{table*}

An SPT state is almost trivial. For example, all the correlations are short
ranged and featureless, as well as all the bulk excitations are local
excitations without fractionalization.  So, it is not easy to see the
non-trivialness of a SPT state.  One way to reveal the non-trivialness is to
probe the boundary \cite{CLW1141}:\\
\emph{The boundary of a SPT state cannot be gapped and non-degenerate if the
symmetry is not broken explicitly}.\\
This because the effective symmetry on the low energy boundary degrees of
freedom must be non-on-site, and the non-on-site property for the boundary
theory exactly corresponds to and classify the anomaly in global symmetry
\cite{W1313}.  This implies the boundary of a SPT state to be either symmetry
breaking, gapless, and/or topologically ordered.

Another way to detect the non-trivialness of a SPT state is to twist the
symmetry and measure the ground state response under the twisted symmetry
\cite{LG1220}.  To understand how to twist the symmetry, let us assume that a
2D lattice Hamiltonian for a SPT state with symmetry $G$ to have a form (see
Fig.  \ref{ltrans}) $ H=\sum_{(ijk)} H_{ijk} $, where $\sum_{(ijk)}$ sums over
all the triangles $(ijk)$ in Fig. \ref{ltrans} and $H_{ijk}$ acts on the states
on site-$i$, site-$j$, and site-$k$.  $H$ and $H_{ijk}$ are invariant under the
global $G$ transformations.

Let us perform a \emph{local} $g\in G$ transformation which only acts on the
sites in the shaded region in Fig.  \ref{ltrans}. Such a local transformation
will change $H$ to $\t H$.  However, only the Hamiltonian terms on the
triangles $(ijk)$ across the boundary of the shaded region are changed from
$H_{ijk}$ to $H^g_{ijk}$.  Since the $G$ transformation is an unitary
transformation, $H$ and $\t H$ have the same energy spectrum.  In other words
the boundary (called the $g$-cut) in Fig.  \ref{ltrans} (described by
$H^g_{ijk}$'s) does not cost any energy.

Now let us consider a Hamiltonian on a lattice with some $g$-cuts (see Fig.
\ref{ZnDefect}) $\t H= { \sum_{(ijk)}}' H_{ijk} +\sum_{(ijk)}^{g\text{-cut}}
H^g_{ijk}$, where $ \sum'_{(ijk)}$ sums over the triangles not on the cut and
$\sum^{g\text{-cut}}_{(ijk)}$ sums over the triangles that are divided into
disconnected pieces by the $g$-cut.  The triangles at the ends of the cut have
no Hamiltonian terms.  We note that the cut carries no energy. Only the ends of
cut cost energies. So the Fig. \ref{ZnDefect}  corresponds to three monodromy
defects.  If the $g$ is a generator of $G$, then the end of $g$-cut will be
called elementary monodromy defect.  We like to point out that dislocation in a
crystal is an example of monodromy defect of translation symmetry.  It has been
used to detect SPT phases protected by translation symmetry (the so called weak
topological phases) \cite{RZV08105121,TK10060690,SZ14014044}.  We also like to
point out that the above procedure to obtain $\t H$ is actually the ``gauging''
of the $G$ symmetry \cite{LG1220}.  $\t H$ is a gauged Hamiltonian that contain
three $G$ vortices at the ends of the cut.

Using the above  monodromy defects, we can detect the $Z_n$-SPT order
\cite{W161201418}:\\
\emph{$n$ \textbf{identical} elementary monodromy defects in a 2+1D $Z_n$-SPT
state on a torus always carry a total $Z_n$-charge $m$, if the $Z_n$ SPT state
is described by the $m^{th}$ cocycle in $\cH^3(Z_n,\R/\Z)$.}\\
The total $Z_n$-charge of $n$ identical monodromy
defects allows us to completely characterize the 2+1D $Z_n$ SPT states.
Another way to probe the $Z_n$-SPT order
is to use the statistics of the monodromy defects \cite{LG1220}:\\ \emph{The
statistical angle $\th_M$ of an elementary monodromy defect 
satisfies mod$(\frac{\th_M}{2\pi},\frac 1 n)=\frac{m}{n^2}$ for a $Z_n$-SPT
state characterized by $m\in \cH^3(Z_n,\R/\Z)=\Z_n$.}

This way of probing an SPT state is like using the modular extensions of Rep$(G)$
to probe the $G$-SPT order \cite{LW160205936,LW160205946}.  (The so called
\emph{modular extension} can be viewed as including all the  monodromy defects
and considering their statistics.) It has been shown that the modular
extensions of Rep$(G)$ one-to-one correspond to the elements in
$\cH^3(G,\R/\Z)$ \cite{DO07040195,LW160205946}.  So the modular extensions can
fully characterize $\cH^3(G,\R/\Z)$.  In other words, measuring the abelian
and/or non-abelian statistics among the  monodromy defects and the local
excitations described by Rep$(G)$, allows us to fully detect the $G$-SPT order
in 2+1D for any unitary symmetry $G$.  The similar idea also applies to 3+1D
SPT states \cite{WL1437,LW170404221}.  If the symmetry group contain $U(1)$,
one can also use the $U(1)$ monopoles to probe the 3+1D SPT states
\cite{MKF1331,YW1427,W1447}.  A systematic discussion to probe all SPT orders
in any dimensions can be found in \textcite{HW1339}.

\def\arraystretch{1.0} \setlength\tabcolsep{3pt}
\begin{table*}[t]
\caption{\textbf{Classification of the gapped phases of noninteracting
fermions} in $d$-dimensional space, for some symmetries.  The space of the
gapped states is given by $C_{p+d \text{ mod }2}$, where $p$ depends on the
symmetry group.  The distinct phases are given by $\pi_0(C_{p+d\text{ mod
}2})$.  ``0'' means that only trivial phase exist.  $\Z$ means that nontrivial
phases are labeled by nonzero integers and the trivial phase is labeled by 0.
$U^f(1)$ means that the $\pi$ rotation is $(-)^{N_F}$.
$Z_4^f$ is generated by $C$ satisfying $C^2=(-)^{N_F}$.
Adapted from \textcite{W1103}.
}
 \label{frfC}
 \centering
 \begin{tabular}{ |c||c|c|c|c|c|c|c|c|c|c|c|c| }
 \hline
Symm. group & $C_p|_{\text{for }d=0}$ & class & $p \backslash d$
  & 0 & 1 & 2 & 3 & 4 & 5 & 6 & 7 &
example
\\
\hline
\hline
{
\footnotesize
$\bmm
U^f(1)\\[1mm]
Z_4^f
\emm$
} & $\frac{U(l+m)}{U(l)\times U(m)}\times \Z$ & A & 0 
& $\Z$ & $0$ & $\Z$ & $0$ & $\Z$ & $0$ & $\Z$ & $0$ &
$\bmm
\text{(Chern)}\\
\text{insulator}
\emm$
~~
$\bmm
\text{supercond.}\\[-1mm]
\text{with collinear}\\[-1mm]
\text{spin order}
\emm$
\\
 \hline
{
\footnotesize
$\bmm
U^f(1)\times Z_2^T\\[1mm]
Z_4^f\times Z_2^T
\emm$
}  & $U(n)$ & AIII & 1
& $0$ & $\Z$ & $0$ & $\Z$ & $0$ & $\Z$ & $0$ & $\Z$ &
$\bmm
\text{supercond. w/ real pairing}\\[-1mm]
\text{and $S_z$ conserving}\\[-1mm]
\text{spin-orbital coupling}
\emm$
\\
\hline
 \end{tabular}
\end{table*}

\begin{table*}[tb]
 \caption{
\textbf{Classification of gapped phases of noninteracting fermions} in $d$
spatial dimensions, for some symmetries.  The space of the gapped states is
$R_{p-d \text{ mod }8}$, where $p$ depends on the symmetry.  The
phases are classified by $\pi_0(R_{p-d\text{ mod }8})$.  $\Z_2$ means that
there is one nontrivial and one trivial phases labeled by 1 and 0.  Note that $
\frac{U^f(1)\rtimes Z_4^T \times Z_4^f}{Z_2^2}$ is the symmetry group generated
by time reversal $T$, charge conjugation $c \to \ii\si^y c^\dag$ and charge
conservation.  Adapted from \textcite{W1103}.  
}
 \label{frfR}
{
 \begin{tabular}{ |c||c|c|c|c|c|c|c|c| }
 \hline
$\bmm
\text{Symm.}\\
\text{group}
\emm$
&
{\scriptsize
$\bmm
U^f(1)\rtimes Z_2^T
\emm$
}%
& 
{\scriptsize
$\bmm
\Z_2^T\times Z_2^f
\emm$
}%
&
{\scriptsize
$\bmm
Z_2^f\\[1mm]
Z_2\times Z_2^f
\emm$
}%
&
{\scriptsize
$\bmm
Z_4^T\\[1mm]
Z_4^T \times Z_2
\emm$
}%
&
{\scriptsize
$\bmm
[U^f(1)\rtimes Z_4^T]/Z_2\\[1mm]
[Z_4^f\rtimes Z_4^T]/Z_2 
\emm$
}%
&
{\scriptsize
$\bmm
\frac{U^f(1)\rtimes Z_4^T \times Z_4^f}{Z_2^2}
\emm$
}%
&
{\scriptsize
$\bmm
SU^f(2)
\emm$
}
&
{\scriptsize
$\bmm
\frac{SU^f(2)\times Z_4^T}{Z_2}
\emm$
}%
\\
 \hline
$R_p|_{\text{for }d=0}$ &
{\footnotesize
$\frac{O(l+m)}{O(l)\times O(m)}\times \Z$
}
& 
 $O(n)$ & 
 $\frac{O(2n)}{U(n)}$ & 
 $\frac{U(2n)}{Sp(n)}$ &
{\footnotesize
$\frac{Sp(l+m)}{Sp(l)\times Sp(m)}\times \Z$
}
& 
 $Sp(n)$ & 
 $\frac{Sp(n)}{U(n)}$ & 
 $\frac{U(n)}{O(n)}$  \\ 
 \hline
 &  $p=0$  & $p=1$ & $p=2$ & $p=3$ & $p=4$ & $p=5$ & $p=6$ & $p=7$ \\ 
\hline
class &  AI &  BDI & D &  DIII & AII & CII & C &   CI \\ 
\hline
$d=0$ & $\Z$ & $\Z_2$ & $\Z_2$ & $0$ & $\Z$ & $0$ & $0$ & $0$ \\ 
$d=1$ & $0$ & $\Z$ & $\Z_2$ & $\Z_2$ & $0$ & $\Z$ & $0$ & $0$ \\
$d=2$ & $0$ & $0$ & $\Z$ & $\Z_2$ & $\Z_2$ & $0$ & $\Z$ & $0$ \\
$d=3$ & $0$ & $0$ & $0$ & $\Z$ & $\Z_2$ & $\Z_2$ & $0$ & $\Z$ \\
\hline
$d=4$ & $\Z$ & $0$ & $0$ & $0$ & $\Z$ & $\Z_2$ & $\Z_2$ & $0$ \\
$d=5$ & $0$ & $\Z$ & $0$ & $0$ & $0$ & $\Z$ & $\Z_2$ & $\Z_2$ \\
$d=6$ & $\Z_2$ & $0$ & $\Z$ & $0$ & $0$ & $0$ & $\Z$ & $\Z_2$ \\
$d=7$ & $\Z_2$ & $\Z_2$ & $0$ & $\Z$ & $0$ & $0$ & $0$ & $\Z$ \\
\hline
{\footnotesize
Example 
}
&
{\footnotesize
$\bmm
\text{insulator}\\
\text{w/ coplanar}\\
\text{spin order}
\emm$
}
&
{\footnotesize
$\bmm
\text{supercond.}\\
\text{w/ coplanar}\\
\text{spin order}
\emm$
}
&
{\footnotesize
$\bmm
\text{supercond.}
\emm$
}
&
{\footnotesize
$\bmm
\text{supercond.}\\
\text{w/ time}\\
\text{reversal}\\
\emm$
}
&
{\footnotesize
$\bmm
\text{insulator}\\
\text{w/ time}\\
\text{reversal}\\
\emm$
}
&
{\footnotesize
$\bmm
\text{insulator}\\[-1mm]
\text{w/ time}\\[-1mm]
\text{reversal and}\\[-1mm]
\text{intersublattice}\\[-1mm]
\text{hopping}
\emm$
}
&
{\footnotesize
$\bmm
\text{spin}\\
\text{singlet}\\
\text{supercond.}\\
\emm$
}
&
{\footnotesize
$\bmm
\text{spin}\\[-1mm]
\text{singlet}\\[-1mm]
\text{supercond.}\\[-1mm]
\text{w/ time}\\[-1mm]
\text{reversal}
\emm$
}
\\
\hline
 \end{tabular}
}
\end{table*}

\subsection{Table of some SPT states}

In table \ref{SPT}, we list bosonic/fermionic SPT states for various symmetries
and in various dimensions.  For bosonic SPT states with on-site symmetry $G$, a
partial classification was first given by the group cohomology of the symmetry
group $\cH^{d+1}(G,\R/\Z)$ where $d$ is the space dimension \cite{CGL1172}.
Later, it was pointed out the group cohomology description is incomplete when
$d=3$ and when $G$ contains time reversal symmetry \cite{VS1306,WS1334}.  Then,
it was realized that bosonic SPT states can all be classified by generalized
group cohomology $\cH^{d+1}(G\times SO_\infty,\R/\Z)/\Ga$.
This implies that in 1+1D and 2+1D, bosonic SPT states are classified by
$\cH^{2}(G,\R/\Z)$ and $\cH^{3}(G,\R/\Z)$  respectively.  In 3+1D, bosonic
SPT states are classified by $\cH^{4}(G,\R/\Z)$  if the on-site symmetry $G$
does not contain time reversal, and by $\cH^{4}(G,(\R/\Z)_T)\oplus \Z_2$  if
$G$ contains time reversal.  Recent work also generalizes the cohomology
classification of bosonic SPT states to translation and point-group symmetries
\cite{HL14036902,YC14090168,HC150800573,SH160408151,L160802736,TE161200846}. 

For non-interacting fermionic SPT states
\cite{KM0502,BZ0611399,MB0706,FKM0703,QHZ0824,R0664}, there is a related
classification of non-interacting gapped states
based on K-theory\cite{K0986} or nonlinear $\si$-model of disordered fermions
\cite{SRF0825} (see Tables \ref{frfC} and \ref{frfR}).  But such a classification does
not apply to interacting fermions.  For interacting fermionic SPT states
\cite{WS13063238,WS14011142}, there is a systematic understanding based on
group super cohomology theory
\cite{GW1441,GK150505856,KT170108264,WG170310937}, if the total symmetry group
has a form $G^f=G_b\times Z_2^f$.  Here $Z_2^f$ is the fermion-number-parity
symmetry which is always present for fermion systems.  Recently, a complete
classification for all 2+1D fermionic SPT states was found for generic on-site
symmetry $G^f$ which does not contain time reversal \cite{LW160205946}:
\emph{2+1D fermionic SPT phases are classified by the modular extensions of
sRep$(G^f)$}. Here sRep$(G^f)$ is the symmetric fusion category formed
representations of $G^f$ where the representations with non-trivial $Z_2^f$
action are fermions.  Last, we would like to mention that, in addition to the
cohomological and categorical approach, there is also a cobordism approach for
bosonic/fermionic SPT states, which can lead a classifying result for all
dimensions and for some simple symmetries \cite{K1459,K1467,KTT1429}.

Regarding to \textbf{Pt.3} in Sec.  \ref{newworld}, quantum spin Hall effect
refers to quantized transverse $S^z$-spin current induced by force acting on
electric charges (\ie a quantized mixed-electro-spin Hall conductance)
\cite{KM0502,BZ0602},
while spin quantum Hall effect refer to quantized transverse $S^z$-spin current
induced by force acting on ``$S^z$-charge'' (\ie a quantized spin-Hall
conductance).  They have a vanishing charge-Hall and thermo-Hall conductances.
Under such definitions, the quantum spin Hall states \cite{KM0501,BZ0602} and
topological insulators in 2+1D \cite{KM0502} (both appear in table \ref{SPT})
are different fermionic SPT states. They even have different symmetries:
quantum spin Hall states have $U_\up^f(1)\times U_\down^f(1)$ symmetry, while
topological insulators $[U^f(1) \rtimes Z_4^T]/Z_2$ symmetry\footnote{The
superscript $f$ means that the $U(1)$ groups contain $Z_2^f$ as a subgroup.
$U^f_{\up,\down}(1)$ is the symmetry of $\up,\down$-spin conservation, and
$U^f(1)$ is the symmetry of charge conservation.  $Z_4^T$ is the group
generated by time reversal transformation $T$, that satisfies $T^2=(-)^{N_F}$
and $(-)^{N_F}$ is the fermion-number-parity. After the discovery of the
$Z_2$-topological invariant and the 2+1D topological insulator \cite{KM0502},
quantum spin Hall state, some times, was also defined as 2+1D topological
insulator. Such a quantum spin Hall state has no quantum spin Hall effect nor
spin quantum Hall effect, since even the $S^z$ current is not conserved.}.  

Even though topological insulator arises from the studies of quantum spin Hall
effect, it is incorrect to think topological insulator to be due to quantum
spin Hall effect. In particular, \textcite{KM0502}, in ``$Z_2$ Topological
Order and the Quantum Spin Hall Effect'', concluded that even without quantum
spin Hall effect, an insulator can still be non-trivial. This led to the notion
of topological insulator. This is a very surprising discovery which started the
very active field of topological insulator.  Despite the term ``Topological
Order'' in the title, the topological insulator is a short-range entangled SPT
state.  It has no topological order as introduced in \Ref{Wrig,WNtop},
which involves long-range entanglement.  This explains the \textbf{Pt.2} in
Sec.  \ref{newworld}.  \textcite{KM0502} only deal with non-interacting
fermions in 2+1D. Soon, it was shown that the 2+1D topological insulator is
stable against weak interactions \cite{XMc0508291,WZc0508273}.

With regard to the second part of \textbf{Pt.2}, many popular articles
characterize topological insulator as an insulator with conducting surface.
Such a characterization is incorrect, since both trivial insulator and
topological insulator can some times have conducting surfaces, and other times
have insulating surfaces (for interacting electrons) \cite{CFV1350,WS13063223}.
Maybe it is more correct to say ``topological insulator is an insulator with
conducting surface when electrons interact weakly''. But even when electrons
interact weakly, both trivial insulator and topological insulator can have
conducting surfaces.  We need to measure the surface Fermi surface to be sure
\cite{HH09021356}, but it does not work for 2+1D topological insulator.  So a
more accurate characterization of 2+1D topological insulator is that the
charge-$0$ time-reversal symmetric $\pi$-flux must be a Kramer doublet
\cite{QZ08010252,RL08010627}.


\section{Towards a classification of all gapped phases}

Only for a few times in history, we have completely classified some large class
of matter states.  The first time is the classification of all spontaneous
symmetry breaking orders, can be classified by a pair of groups: 
\begin{align}
(G_\Psi\subset G_H), 
\end{align}
where  $G_H$ is the
symmetry group of the system and $G_\Psi$, a subgroup of $G_H$, is the
symmetry group of the ground state. This includes the classification of all 230
crystal orders in 3-dimensions. 

The second time is the classification of all gapped 1-dimensional quantum
states:  gapped 1-dimensional quantum states 
with on-site symmetry $G_H$ can be classified by a
triple:\cite{CGW1107,SPC1139} 
\begin{align}
[G_\Psi\subset G_H;\ {\rm pRep}(G_\Psi)], 
\end{align}
where {\rm pRep}$(G_\Psi)$ is a projective
representation of $G_\Psi$. 

Recently, the 1+1D classification of all gapped quantum phases was generalized
to 2+1D, which is a major progress.  We find that all 2+1D bosonic topological
orders are classified by a pair \cite{Wrig,K062,RSW0777,W150605768}:
\begin{align}
(\text{MTC},c), 
\end{align}
where MTC is a unitary \textbf{modular tensor category} and
$c$ is  the chiral central charge $c$ of the edge states.  On the other hand,
2+1D fermionic topological orders are classified by a triple:\cite{LW150704673}
\begin{align}
[\text{sRep}(Z_2^f) \subset \text{BFC};\ c], 
\end{align}
where $\text{sRep}(Z_2^f)$ is
the \textbf{symmetric fusion category} (SFC) formed by the representations of
the fermion-number-parity symmetry $Z_2^f$ where the non-trivial representation
is assigned Fermi statistics, and BFC is a unitary \textbf{braided fusion
category}.  

In the presence of finite unitary on-site symmetry $G_H$, 2+1D gapped bosonic
phases are classified by \cite{BBC1440,LW160205946} 
\begin{align}
[G_\Psi\subset G_H;\ 
\text{Rep}(G_\Psi) \subset \text{BFC} \subset \text{MTC};\ c],
\end{align}
where $\text{Rep}(G_\Psi)$ is the SFC formed by the representations of $G_\Psi$
where all representations are assigned Bose statistics, and MTC is a minimal
modular extension of the BFC.  The above classification include symmetry
breaking orders, SPT orders, topological orders, and \emph{symmetry-enriched
topological orders} (SET) described by \emph{projective symmetry group}
\cite{W0213}.  SET orders of time-reversal/reflection symmetry are classified
by \textcite{BW161207792}. Some more discussions on SET orders can be found in
\cite{LV1334,HW1307,MR1315,HW1351,X13078131,CY14126589}.

We have a similar result for fermion systems: all 2+1D gapped fermionic phases
with unitary finite on-site symmetry $G_H^f$ are classified by
\cite{LW160205946} 
\begin{align}
[G_\Psi^f\subset G_H^f;\
\text{sRep}(G_\Psi^f) \subset \text{BFC} \subset \text{MTC};\ c], 
\end{align}
where
$\text{sRep}(G_\Psi^f)$ is the SFC formed by the representations of
$G_\Psi^f$ where some representations are assigned Fermi statistics.
But we are still struggling to obtain a systematic theory of topological order
in 3+1D, 28 years after the introduction of the concept.

Those results imply that the long-range
entanglement in 2+1D is described by an unfamiliar mathematics -- tensor
category theory.  This is the mathematics for the quantum topology, and it is
the quantum topology (instead of classical topology) that forms the
mathematical foundation of topological order (\ie long-range entanglement).
This explains the title of this paper ``quantum topological phases of matter'',
which really means ``highly entangled phases of matter''.

I would like to thank Cenke Xu and Xiao-Liang Qi for many comments.
This research was supported by NSF Grant No.  DMR-1506475 and NSFC 11274192.

\bibliography{../../bib/wencross,../../bib/all,../../bib/publst} 

\begin{thebibliography}{189}%
\makeatletter
\providecommand \@ifxundefined [1]{%
 \@ifx{#1\undefined}
}%
\providecommand \@ifnum [1]{%
 \ifnum #1\expandafter \@firstoftwo
 \else \expandafter \@secondoftwo
 \fi
}%
\providecommand \@ifx [1]{%
 \ifx #1\expandafter \@firstoftwo
 \else \expandafter \@secondoftwo
 \fi
}%
\providecommand \natexlab [1]{#1}%
\providecommand \enquote  [1]{``#1''}%
\providecommand \bibnamefont  [1]{#1}%
\providecommand \bibfnamefont [1]{#1}%
\providecommand \citenamefont [1]{#1}%
\providecommand \href@noop [0]{\@secondoftwo}%
\providecommand \href [0]{\begingroup \@sanitize@url \@href}%
\providecommand \@href[1]{\@@startlink{#1}\@@href}%
\providecommand \@@href[1]{\endgroup#1\@@endlink}%
\providecommand \@sanitize@url [0]{\catcode `\\12\catcode `\$12\catcode
  `\&12\catcode `\#12\catcode `\^12\catcode `\_12\catcode `\%12\relax}%
\providecommand \@@startlink[1]{}%
\providecommand \@@endlink[0]{}%
\providecommand \url  [0]{\begingroup\@sanitize@url \@url }%
\providecommand \@url [1]{\endgroup\@href {#1}{\urlprefix }}%
\providecommand \urlprefix  [0]{URL }%
\providecommand \Eprint [0]{\href }%
\providecommand \doibase [0]{http://dx.doi.org/}%
\providecommand \selectlanguage [0]{\@gobble}%
\providecommand \bibinfo  [0]{\@secondoftwo}%
\providecommand \bibfield  [0]{\@secondoftwo}%
\providecommand \translation [1]{[#1]}%
\providecommand \BibitemOpen [0]{}%
\providecommand \bibitemStop [0]{}%
\providecommand \bibitemNoStop [0]{.\EOS\space}%
\providecommand \EOS [0]{\spacefactor3000\relax}%
\providecommand \BibitemShut  [1]{\csname bibitem#1\endcsname}%
\let\auto@bib@innerbib\@empty
\bibitem [{\citenamefont {Landau}(1937)}]{L3726}%
  \BibitemOpen
  \bibfield  {author} {\bibinfo {author} {\bibfnamefont {L.~D.}\ \bibnamefont
  {Landau}},\ }\href@noop {} {\bibfield  {journal} {\bibinfo  {journal} {Phys.
  Z. Sowjetunion}\ }\textbf {\bibinfo {volume} {11}},\ \bibinfo {pages} {26}
  (\bibinfo {year} {1937})}\BibitemShut {NoStop}%
\bibitem [{\citenamefont {Ginzburg}\ and\ \citenamefont
  {Landau}(1950)}]{GL5064}%
  \BibitemOpen
  \bibfield  {author} {\bibinfo {author} {\bibfnamefont {V.~L.}\ \bibnamefont
  {Ginzburg}}\ and\ \bibinfo {author} {\bibfnamefont {L.~D.}\ \bibnamefont
  {Landau}},\ }\href@noop {} {\bibfield  {journal} {\bibinfo  {journal} {Zh.
  Eksp. Teor. Fiz.}\ }\textbf {\bibinfo {volume} {20}},\ \bibinfo {pages}
  {1064} (\bibinfo {year} {1950})}\BibitemShut {NoStop}%
\bibitem [{\citenamefont {Landau}\ and\ \citenamefont
  {Lifschitz}(1958)}]{LanL58}%
  \BibitemOpen
  \bibfield  {author} {\bibinfo {author} {\bibfnamefont {L.~D.}\ \bibnamefont
  {Landau}}\ and\ \bibinfo {author} {\bibfnamefont {E.~M.}\ \bibnamefont
  {Lifschitz}},\ }\href@noop {} {\emph {\bibinfo {title} {Statistical Physics -
  Course of Theoretical Physics Vol 5}}}\ (\bibinfo  {publisher} {Pergamon},\
  \bibinfo {address} {London},\ \bibinfo {year} {1958})\BibitemShut {NoStop}%
\bibitem [{\citenamefont {Nambu}(1960)}]{N6080}%
  \BibitemOpen
  \bibfield  {author} {\bibinfo {author} {\bibfnamefont {Y.}~\bibnamefont
  {Nambu}},\ }\href@noop {} {\bibfield  {journal} {\bibinfo  {journal} {Phys.
  Rev. Lett.}\ }\textbf {\bibinfo {volume} {4}},\ \bibinfo {pages} {380}
  (\bibinfo {year} {1960})}\BibitemShut {NoStop}%
\bibitem [{\citenamefont {Goldstone}(1961)}]{G6154}%
  \BibitemOpen
  \bibfield  {author} {\bibinfo {author} {\bibfnamefont {J.}~\bibnamefont
  {Goldstone}},\ }\href@noop {} {\bibfield  {journal} {\bibinfo  {journal}
  {Nuovo Cimento}\ }\textbf {\bibinfo {volume} {19}},\ \bibinfo {pages} {154}
  (\bibinfo {year} {1961})}\BibitemShut {NoStop}%
\bibitem [{\citenamefont {Tsui}\ \emph {et~al.}(1982)\citenamefont {Tsui},
  \citenamefont {Stormer},\ and\ \citenamefont {Gossard}}]{TSG8259}%
  \BibitemOpen
  \bibfield  {author} {\bibinfo {author} {\bibfnamefont {D.~C.}\ \bibnamefont
  {Tsui}}, \bibinfo {author} {\bibfnamefont {H.~L.}\ \bibnamefont {Stormer}}, \
  and\ \bibinfo {author} {\bibfnamefont {A.~C.}\ \bibnamefont {Gossard}},\
  }\href@noop {} {\bibfield  {journal} {\bibinfo  {journal} {Phys. Rev. Lett.}\
  }\textbf {\bibinfo {volume} {48}},\ \bibinfo {pages} {1559} (\bibinfo {year}
  {1982})}\BibitemShut {NoStop}%
\bibitem [{\citenamefont {Bednorz}\ and\ \citenamefont
  {Mueller}(1986)}]{BM8689}%
  \BibitemOpen
  \bibfield  {author} {\bibinfo {author} {\bibfnamefont {J.~G.}\ \bibnamefont
  {Bednorz}}\ and\ \bibinfo {author} {\bibfnamefont {K.~A.}\ \bibnamefont
  {Mueller}},\ }\href@noop {} {\bibfield  {journal} {\bibinfo  {journal} {Z.
  Phys. B}\ }\textbf {\bibinfo {volume} {64}},\ \bibinfo {pages} {189}
  (\bibinfo {year} {1986})}\BibitemShut {NoStop}%
\bibitem [{\citenamefont {Foerster}\ \emph {et~al.}(1980)\citenamefont
  {Foerster}, \citenamefont {Nielsen},\ and\ \citenamefont
  {Ninomiya}}]{FNN8035}%
  \BibitemOpen
  \bibfield  {author} {\bibinfo {author} {\bibfnamefont {D.}~\bibnamefont
  {Foerster}}, \bibinfo {author} {\bibfnamefont {H.~B.}\ \bibnamefont
  {Nielsen}}, \ and\ \bibinfo {author} {\bibfnamefont {M.}~\bibnamefont
  {Ninomiya}},\ }\href@noop {} {\bibfield  {journal} {\bibinfo  {journal}
  {Phys. Lett. B}\ }\textbf {\bibinfo {volume} {94}},\ \bibinfo {pages} {135}
  (\bibinfo {year} {1980})}\BibitemShut {NoStop}%
\bibitem [{\citenamefont {Baskaran}\ and\ \citenamefont
  {Anderson}(1988)}]{BA8880}%
  \BibitemOpen
  \bibfield  {author} {\bibinfo {author} {\bibfnamefont {G.}~\bibnamefont
  {Baskaran}}\ and\ \bibinfo {author} {\bibfnamefont {P.~W.}\ \bibnamefont
  {Anderson}},\ }\href@noop {} {\bibfield  {journal} {\bibinfo  {journal}
  {Phys. Rev. B}\ }\textbf {\bibinfo {volume} {37}},\ \bibinfo {pages} {580}
  (\bibinfo {year} {1988})}\BibitemShut {NoStop}%
\bibitem [{\citenamefont {Wen}(2002{\natexlab{a}})}]{W0202}%
  \BibitemOpen
  \bibfield  {author} {\bibinfo {author} {\bibfnamefont {X.-G.}\ \bibnamefont
  {Wen}},\ }\href@noop {} {\bibfield  {journal} {\bibinfo  {journal} {Phys.
  Rev. Lett.}\ }\textbf {\bibinfo {volume} {88}},\ \bibinfo {pages} {11602}
  (\bibinfo {year} {2002}{\natexlab{a}})},\ \Eprint
  {http://arxiv.org/abs/hep-th/01090120} {hep-th/01090120} \BibitemShut
  {NoStop}%
\bibitem [{\citenamefont {Wen}(2003)}]{W0303a}%
  \BibitemOpen
  \bibfield  {author} {\bibinfo {author} {\bibfnamefont {X.-G.}\ \bibnamefont
  {Wen}},\ }\href@noop {} {\bibfield  {journal} {\bibinfo  {journal} {Phys.
  Rev. D}\ }\textbf {\bibinfo {volume} {68}},\ \bibinfo {pages} {065003}
  (\bibinfo {year} {2003})},\ \Eprint {http://arxiv.org/abs/hep-th/0302201}
  {hep-th/0302201} \BibitemShut {NoStop}%
\bibitem [{\citenamefont {Levin}\ and\ \citenamefont
  {Wen}(2006{\natexlab{a}})}]{LWqed}%
  \BibitemOpen
  \bibfield  {author} {\bibinfo {author} {\bibfnamefont {M.}~\bibnamefont
  {Levin}}\ and\ \bibinfo {author} {\bibfnamefont {X.-G.}\ \bibnamefont
  {Wen}},\ }\href@noop {} {\bibfield  {journal} {\bibinfo  {journal} {Phys.
  Rev. B}\ }\textbf {\bibinfo {volume} {73}},\ \bibinfo {pages} {035122}
  (\bibinfo {year} {2006}{\natexlab{a}})},\ \Eprint
  {http://arxiv.org/abs/hep-th/0507118} {hep-th/0507118} \BibitemShut {NoStop}%
\bibitem [{\citenamefont {Wen}(2013{\natexlab{a}})}]{W1301}%
  \BibitemOpen
  \bibfield  {author} {\bibinfo {author} {\bibfnamefont {X.-G.}\ \bibnamefont
  {Wen}},\ }\href@noop {} {\bibfield  {journal} {\bibinfo  {journal} {Chin.
  Phys. Lett.}\ }\textbf {\bibinfo {volume} {30}},\ \bibinfo {pages} {111101}
  (\bibinfo {year} {2013}{\natexlab{a}})},\ \Eprint
  {http://arxiv.org/abs/arXiv:1305.1045} {arXiv:1305.1045} \BibitemShut
  {NoStop}%
\bibitem [{\citenamefont {{You}}\ \emph {et~al.}(2014)\citenamefont {{You}},
  \citenamefont {{BenTov}},\ and\ \citenamefont {{Xu}}}]{YBX1451}%
  \BibitemOpen
  \bibfield  {author} {\bibinfo {author} {\bibfnamefont {Y.-Z.}\ \bibnamefont
  {{You}}}, \bibinfo {author} {\bibfnamefont {Y.}~\bibnamefont {{BenTov}}}, \
  and\ \bibinfo {author} {\bibfnamefont {C.}~\bibnamefont {{Xu}}},\ }\href@noop
  {} {\  (\bibinfo {year} {2014})},\ \Eprint
  {http://arxiv.org/abs/arXiv:1402.4151} {arXiv:1402.4151} \BibitemShut
  {NoStop}%
\bibitem [{\citenamefont {{You}}\ and\ \citenamefont
  {{Xu}}(2015)}]{YX14124784}%
  \BibitemOpen
  \bibfield  {author} {\bibinfo {author} {\bibfnamefont {Y.-Z.}\ \bibnamefont
  {{You}}}\ and\ \bibinfo {author} {\bibfnamefont {C.}~\bibnamefont {{Xu}}},\
  }\href {\doibase 10.1103/PhysRevB.91.125147} {\bibfield  {journal} {\bibinfo
  {journal} {\prb}\ }\textbf {\bibinfo {volume} {91}},\ \bibinfo {pages}
  {125147} (\bibinfo {year} {2015})},\ \Eprint
  {http://arxiv.org/abs/arXiv:1412.4784} {arXiv:1412.4784} \BibitemShut
  {NoStop}%
\bibitem [{\citenamefont {Anderson}(1987)}]{A8796}%
  \BibitemOpen
  \bibfield  {author} {\bibinfo {author} {\bibfnamefont {P.~W.}\ \bibnamefont
  {Anderson}},\ }\href@noop {} {\bibfield  {journal} {\bibinfo  {journal}
  {Science}\ }\textbf {\bibinfo {volume} {235}},\ \bibinfo {pages} {1196}
  (\bibinfo {year} {1987})}\BibitemShut {NoStop}%
\bibitem [{\citenamefont {Baskaran}\ \emph {et~al.}(1987)\citenamefont
  {Baskaran}, \citenamefont {Zou},\ and\ \citenamefont {Anderson}}]{BZA8773}%
  \BibitemOpen
  \bibfield  {author} {\bibinfo {author} {\bibfnamefont {G.}~\bibnamefont
  {Baskaran}}, \bibinfo {author} {\bibfnamefont {Z.}~\bibnamefont {Zou}}, \
  and\ \bibinfo {author} {\bibfnamefont {P.~W.}\ \bibnamefont {Anderson}},\
  }\href@noop {} {\bibfield  {journal} {\bibinfo  {journal} {Solid State
  Comm.}\ }\textbf {\bibinfo {volume} {63}},\ \bibinfo {pages} {973} (\bibinfo
  {year} {1987})}\BibitemShut {NoStop}%
\bibitem [{\citenamefont {Affleck}\ and\ \citenamefont
  {Marston}(1988)}]{AM8874}%
  \BibitemOpen
  \bibfield  {author} {\bibinfo {author} {\bibfnamefont {I.}~\bibnamefont
  {Affleck}}\ and\ \bibinfo {author} {\bibfnamefont {J.~B.}\ \bibnamefont
  {Marston}},\ }\href@noop {} {\bibfield  {journal} {\bibinfo  {journal} {Phys.
  Rev. B}\ }\textbf {\bibinfo {volume} {37}},\ \bibinfo {pages} {3774}
  (\bibinfo {year} {1988})}\BibitemShut {NoStop}%
\bibitem [{\citenamefont {Rokhsar}\ and\ \citenamefont
  {Kivelson}(1988)}]{RK8876}%
  \BibitemOpen
  \bibfield  {author} {\bibinfo {author} {\bibfnamefont {D.~S.}\ \bibnamefont
  {Rokhsar}}\ and\ \bibinfo {author} {\bibfnamefont {S.~A.}\ \bibnamefont
  {Kivelson}},\ }\href@noop {} {\bibfield  {journal} {\bibinfo  {journal}
  {Phys. Rev. Lett.}\ }\textbf {\bibinfo {volume} {61}},\ \bibinfo {pages}
  {2376} (\bibinfo {year} {1988})}\BibitemShut {NoStop}%
\bibitem [{\citenamefont {Affleck}\ \emph
  {et~al.}(1988{\natexlab{a}})\citenamefont {Affleck}, \citenamefont {Zou},
  \citenamefont {Hsu},\ and\ \citenamefont {Anderson}}]{AZH8845}%
  \BibitemOpen
  \bibfield  {author} {\bibinfo {author} {\bibfnamefont {I.}~\bibnamefont
  {Affleck}}, \bibinfo {author} {\bibfnamefont {Z.}~\bibnamefont {Zou}},
  \bibinfo {author} {\bibfnamefont {T.}~\bibnamefont {Hsu}}, \ and\ \bibinfo
  {author} {\bibfnamefont {P.~W.}\ \bibnamefont {Anderson}},\ }\href@noop {}
  {\bibfield  {journal} {\bibinfo  {journal} {Phys. Rev. B}\ }\textbf {\bibinfo
  {volume} {38}},\ \bibinfo {pages} {745} (\bibinfo {year}
  {1988}{\natexlab{a}})}\BibitemShut {NoStop}%
\bibitem [{\citenamefont {Dagotto}\ \emph {et~al.}(1988)\citenamefont
  {Dagotto}, \citenamefont {Fradkin},\ and\ \citenamefont {Moreo}}]{DFM8826}%
  \BibitemOpen
  \bibfield  {author} {\bibinfo {author} {\bibfnamefont {E.}~\bibnamefont
  {Dagotto}}, \bibinfo {author} {\bibfnamefont {E.}~\bibnamefont {Fradkin}}, \
  and\ \bibinfo {author} {\bibfnamefont {A.}~\bibnamefont {Moreo}},\
  }\href@noop {} {\bibfield  {journal} {\bibinfo  {journal} {Phys. Rev. B}\
  }\textbf {\bibinfo {volume} {38}},\ \bibinfo {pages} {2926} (\bibinfo {year}
  {1988})}\BibitemShut {NoStop}%
\bibitem [{\citenamefont {Chung}\ \emph {et~al.}(2001)\citenamefont {Chung},
  \citenamefont {Marston},\ and\ \citenamefont {McKenzie}}]{CMM0159}%
  \BibitemOpen
  \bibfield  {author} {\bibinfo {author} {\bibfnamefont {C.~H.}\ \bibnamefont
  {Chung}}, \bibinfo {author} {\bibfnamefont {J.~B.}\ \bibnamefont {Marston}},
  \ and\ \bibinfo {author} {\bibfnamefont {R.~H.}\ \bibnamefont {McKenzie}},\
  }\href@noop {} {\bibfield  {journal} {\bibinfo  {journal} {J. Phys.: Condens.
  Matter}\ }\textbf {\bibinfo {volume} {13}},\ \bibinfo {pages} {5159}
  (\bibinfo {year} {2001})}\BibitemShut {NoStop}%
\bibitem [{\citenamefont {Rantner}\ and\ \citenamefont {Wen}(2001)}]{RW0171}%
  \BibitemOpen
  \bibfield  {author} {\bibinfo {author} {\bibfnamefont {W.}~\bibnamefont
  {Rantner}}\ and\ \bibinfo {author} {\bibfnamefont {X.-G.}\ \bibnamefont
  {Wen}},\ }\href@noop {} {\bibfield  {journal} {\bibinfo  {journal} {Phys.
  Rev. Lett.}\ }\textbf {\bibinfo {volume} {86}},\ \bibinfo {pages} {3871}
  (\bibinfo {year} {2001})},\ \Eprint {http://arxiv.org/abs/cond-mat/0010378}
  {cond-mat/0010378} \BibitemShut {NoStop}%
\bibitem [{\citenamefont {Rantner}\ and\ \citenamefont {Wen}(2002)}]{RW0201}%
  \BibitemOpen
  \bibfield  {author} {\bibinfo {author} {\bibfnamefont {W.}~\bibnamefont
  {Rantner}}\ and\ \bibinfo {author} {\bibfnamefont {X.-G.}\ \bibnamefont
  {Wen}},\ }\href@noop {} {\bibfield  {journal} {\bibinfo  {journal} {Phys.
  Rev. B}\ }\textbf {\bibinfo {volume} {66}},\ \bibinfo {pages} {144501}
  (\bibinfo {year} {2002})},\ \Eprint {http://arxiv.org/abs/cond-mat/0201521}
  {cond-mat/0201521} \BibitemShut {NoStop}%
\bibitem [{\citenamefont {Fradkin}\ \emph {et~al.}(2003)\citenamefont
  {Fradkin}, \citenamefont {Huse}, \citenamefont {Moessner}, \citenamefont
  {Oganesyan},\ and\ \citenamefont {Sondhi}}]{FHM0353}%
  \BibitemOpen
  \bibfield  {author} {\bibinfo {author} {\bibfnamefont {E.}~\bibnamefont
  {Fradkin}}, \bibinfo {author} {\bibfnamefont {D.~A.}\ \bibnamefont {Huse}},
  \bibinfo {author} {\bibfnamefont {R.}~\bibnamefont {Moessner}}, \bibinfo
  {author} {\bibfnamefont {V.}~\bibnamefont {Oganesyan}}, \ and\ \bibinfo
  {author} {\bibfnamefont {S.~L.}\ \bibnamefont {Sondhi}},\ }\href@noop {} {\
  (\bibinfo {year} {2003})},\ \Eprint {http://arxiv.org/abs/cond-mat/0311353}
  {cond-mat/0311353} \BibitemShut {NoStop}%
\bibitem [{\citenamefont {Hermele}\ \emph {et~al.}(2004)\citenamefont
  {Hermele}, \citenamefont {Senthil}, \citenamefont {Fisher}, \citenamefont
  {Lee}, \citenamefont {Nagaosa},\ and\ \citenamefont {Wen}}]{HSF0437}%
  \BibitemOpen
  \bibfield  {author} {\bibinfo {author} {\bibfnamefont {M.}~\bibnamefont
  {Hermele}}, \bibinfo {author} {\bibfnamefont {T.}~\bibnamefont {Senthil}},
  \bibinfo {author} {\bibfnamefont {M.~P.~A.}\ \bibnamefont {Fisher}}, \bibinfo
  {author} {\bibfnamefont {P.~A.}\ \bibnamefont {Lee}}, \bibinfo {author}
  {\bibfnamefont {N.}~\bibnamefont {Nagaosa}}, \ and\ \bibinfo {author}
  {\bibfnamefont {X.-G.}\ \bibnamefont {Wen}},\ }\href@noop {} {\bibfield
  {journal} {\bibinfo  {journal} {Phys. Rev. B}\ }\textbf {\bibinfo {volume}
  {70}},\ \bibinfo {pages} {214437} (\bibinfo {year} {2004})},\ \Eprint
  {http://arxiv.org/abs/cond-mat/0404751} {cond-mat/0404751} \BibitemShut
  {NoStop}%
\bibitem [{\citenamefont {Senthil}\ \emph {et~al.}(2004)\citenamefont
  {Senthil}, \citenamefont {Balents}, \citenamefont {Sachdev}, \citenamefont
  {Vishwanath},\ and\ \citenamefont {Fisher}}]{SBS0407}%
  \BibitemOpen
  \bibfield  {author} {\bibinfo {author} {\bibfnamefont {T.}~\bibnamefont
  {Senthil}}, \bibinfo {author} {\bibfnamefont {L.}~\bibnamefont {Balents}},
  \bibinfo {author} {\bibfnamefont {S.}~\bibnamefont {Sachdev}}, \bibinfo
  {author} {\bibfnamefont {A.}~\bibnamefont {Vishwanath}}, \ and\ \bibinfo
  {author} {\bibfnamefont {M.~P.~A.}\ \bibnamefont {Fisher}},\ }\href@noop {}
  {\bibfield  {journal} {\bibinfo  {journal} {Physical Review B}\ }\textbf
  {\bibinfo {volume} {70}},\ \bibinfo {pages} {144407} (\bibinfo {year}
  {2004})}\BibitemShut {NoStop}%
\bibitem [{\citenamefont {Kalmeyer}\ and\ \citenamefont
  {Laughlin}(1987)}]{KL8795}%
  \BibitemOpen
  \bibfield  {author} {\bibinfo {author} {\bibfnamefont {V.}~\bibnamefont
  {Kalmeyer}}\ and\ \bibinfo {author} {\bibfnamefont {R.~B.}\ \bibnamefont
  {Laughlin}},\ }\href@noop {} {\bibfield  {journal} {\bibinfo  {journal}
  {Phys. Rev. Lett.}\ }\textbf {\bibinfo {volume} {59}},\ \bibinfo {pages}
  {2095} (\bibinfo {year} {1987})}\BibitemShut {NoStop}%
\bibitem [{\citenamefont {{He}}\ and\ \citenamefont
  {{Chen}}(2015)}]{HC14072740}%
  \BibitemOpen
  \bibfield  {author} {\bibinfo {author} {\bibfnamefont {Y.-C.}\ \bibnamefont
  {{He}}}\ and\ \bibinfo {author} {\bibfnamefont {Y.}~\bibnamefont {{Chen}}},\
  }\href {\doibase 10.1103/PhysRevLett.114.037201} {\bibfield  {journal}
  {\bibinfo  {journal} {Physical Review Letters}\ }\textbf {\bibinfo {volume}
  {114}},\ \bibinfo {pages} {037201} (\bibinfo {year} {2015})},\ \Eprint
  {http://arxiv.org/abs/arXiv:1407.2740} {arXiv:1407.2740} \BibitemShut
  {NoStop}%
\bibitem [{\citenamefont {{Gong}}\ \emph {et~al.}(2015)\citenamefont {{Gong}},
  \citenamefont {{Zhu}}, \citenamefont {{Balents}},\ and\ \citenamefont
  {{Sheng}}}]{GS14121571}%
  \BibitemOpen
  \bibfield  {author} {\bibinfo {author} {\bibfnamefont {S.-S.}\ \bibnamefont
  {{Gong}}}, \bibinfo {author} {\bibfnamefont {W.}~\bibnamefont {{Zhu}}},
  \bibinfo {author} {\bibfnamefont {L.}~\bibnamefont {{Balents}}}, \ and\
  \bibinfo {author} {\bibfnamefont {D.~N.}\ \bibnamefont {{Sheng}}},\ }\href
  {\doibase 10.1103/PhysRevB.91.075112} {\bibfield  {journal} {\bibinfo
  {journal} {\prb}\ }\textbf {\bibinfo {volume} {91}},\ \bibinfo {pages}
  {075112} (\bibinfo {year} {2015})},\ \Eprint
  {http://arxiv.org/abs/arXiv:1412.1571} {arXiv:1412.1571} \BibitemShut
  {NoStop}%
\bibitem [{\citenamefont {Wen}\ \emph {et~al.}(1989)\citenamefont {Wen},
  \citenamefont {Wilczek},\ and\ \citenamefont {Zee}}]{WWZcsp}%
  \BibitemOpen
  \bibfield  {author} {\bibinfo {author} {\bibfnamefont {X.-G.}\ \bibnamefont
  {Wen}}, \bibinfo {author} {\bibfnamefont {F.}~\bibnamefont {Wilczek}}, \ and\
  \bibinfo {author} {\bibfnamefont {A.}~\bibnamefont {Zee}},\ }\href@noop {}
  {\bibfield  {journal} {\bibinfo  {journal} {Phys. Rev. B}\ }\textbf {\bibinfo
  {volume} {39}},\ \bibinfo {pages} {11413} (\bibinfo {year}
  {1989})}\BibitemShut {NoStop}%
\bibitem [{\citenamefont {Wen}(1989)}]{Wtop}%
  \BibitemOpen
  \bibfield  {author} {\bibinfo {author} {\bibfnamefont {X.-G.}\ \bibnamefont
  {Wen}},\ }\href@noop {} {\bibfield  {journal} {\bibinfo  {journal} {Phys.
  Rev. B}\ }\textbf {\bibinfo {volume} {40}},\ \bibinfo {pages} {7387}
  (\bibinfo {year} {1989})}\BibitemShut {NoStop}%
\bibitem [{\citenamefont {Wen}(1990{\natexlab{a}})}]{Wrig}%
  \BibitemOpen
  \bibfield  {author} {\bibinfo {author} {\bibfnamefont {X.-G.}\ \bibnamefont
  {Wen}},\ }\href@noop {} {\bibfield  {journal} {\bibinfo  {journal} {Int. J.
  Mod. Phys. B}\ }\textbf {\bibinfo {volume} {4}},\ \bibinfo {pages} {239}
  (\bibinfo {year} {1990}{\natexlab{a}})}\BibitemShut {NoStop}%
\bibitem [{\citenamefont {{Abanov}}\ and\ \citenamefont
  {{Gromov}}(2014)}]{AG14013703}%
  \BibitemOpen
  \bibfield  {author} {\bibinfo {author} {\bibfnamefont {A.~G.}\ \bibnamefont
  {{Abanov}}}\ and\ \bibinfo {author} {\bibfnamefont {A.}~\bibnamefont
  {{Gromov}}},\ }\href {\doibase 10.1103/PhysRevB.90.014435} {\bibfield
  {journal} {\bibinfo  {journal} {\prb}\ }\textbf {\bibinfo {volume} {90}},\
  \bibinfo {pages} {014435} (\bibinfo {year} {2014})},\ \Eprint
  {http://arxiv.org/abs/arXiv:1401.3703} {arXiv:1401.3703} \BibitemShut
  {NoStop}%
\bibitem [{\citenamefont {{Gromov}}\ \emph {et~al.}(2015)\citenamefont
  {{Gromov}}, \citenamefont {{Cho}}, \citenamefont {{You}}, \citenamefont
  {{Abanov}},\ and\ \citenamefont {{Fradkin}}}]{GF14106812}%
  \BibitemOpen
  \bibfield  {author} {\bibinfo {author} {\bibfnamefont {A.}~\bibnamefont
  {{Gromov}}}, \bibinfo {author} {\bibfnamefont {G.~Y.}\ \bibnamefont {{Cho}}},
  \bibinfo {author} {\bibfnamefont {Y.}~\bibnamefont {{You}}}, \bibinfo
  {author} {\bibfnamefont {A.~G.}\ \bibnamefont {{Abanov}}}, \ and\ \bibinfo
  {author} {\bibfnamefont {E.}~\bibnamefont {{Fradkin}}},\ }\href {\doibase
  10.1103/PhysRevLett.114.016805} {\bibfield  {journal} {\bibinfo  {journal}
  {Physical Review Letters}\ }\textbf {\bibinfo {volume} {114}},\ \bibinfo
  {pages} {016805} (\bibinfo {year} {2015})},\ \Eprint
  {http://arxiv.org/abs/arXiv:1410.6812} {arXiv:1410.6812} \BibitemShut
  {NoStop}%
\bibitem [{\citenamefont {{Bradlyn}}\ and\ \citenamefont
  {{Read}}(2015)}]{BR150204126}%
  \BibitemOpen
  \bibfield  {author} {\bibinfo {author} {\bibfnamefont {B.}~\bibnamefont
  {{Bradlyn}}}\ and\ \bibinfo {author} {\bibfnamefont {N.}~\bibnamefont
  {{Read}}},\ }\href {\doibase 10.1103/PhysRevB.91.165306} {\bibfield
  {journal} {\bibinfo  {journal} {\prb}\ }\textbf {\bibinfo {volume} {91}},\
  \bibinfo {pages} {165306} (\bibinfo {year} {2015})},\ \Eprint
  {http://arxiv.org/abs/arXiv:1502.04126} {arXiv:1502.04126
  [cond-mat.mes-hall]} \BibitemShut {NoStop}%
\bibitem [{\citenamefont {Kane}\ and\ \citenamefont {Fisher}(1997)}]{KF9732}%
  \BibitemOpen
  \bibfield  {author} {\bibinfo {author} {\bibfnamefont {C.~L.}\ \bibnamefont
  {Kane}}\ and\ \bibinfo {author} {\bibfnamefont {M.~P.~A.}\ \bibnamefont
  {Fisher}},\ }\href {\doibase 10.1103/PhysRevB.55.15832} {\bibfield  {journal}
  {\bibinfo  {journal} {Phys. Rev. B}\ }\textbf {\bibinfo {volume} {55}},\
  \bibinfo {pages} {15832} (\bibinfo {year} {1997})},\ \Eprint
  {http://arxiv.org/abs/cond-mat/9603118} {cond-mat/9603118} \BibitemShut
  {NoStop}%
\bibitem [{\citenamefont {Zeng}\ and\ \citenamefont {Wen}(2015)}]{ZW1490}%
  \BibitemOpen
  \bibfield  {author} {\bibinfo {author} {\bibfnamefont {B.}~\bibnamefont
  {Zeng}}\ and\ \bibinfo {author} {\bibfnamefont {X.-G.}\ \bibnamefont {Wen}},\
  }\href {\doibase 10.1103/PhysRevB.91.125121} {\bibfield  {journal} {\bibinfo
  {journal} {Phys. Rev. B}\ }\textbf {\bibinfo {volume} {91}},\ \bibinfo
  {pages} {125121} (\bibinfo {year} {2015})},\ \Eprint
  {http://arxiv.org/abs/arXiv:1406.5090} {arXiv:1406.5090} \BibitemShut
  {NoStop}%
\bibitem [{\citenamefont {{Swingle}}\ and\ \citenamefont
  {{McGreevy}}(2016)}]{SM1403}%
  \BibitemOpen
  \bibfield  {author} {\bibinfo {author} {\bibfnamefont {B.}~\bibnamefont
  {{Swingle}}}\ and\ \bibinfo {author} {\bibfnamefont {J.}~\bibnamefont
  {{McGreevy}}},\ }\href {\doibase 10.1103/PhysRevB.93.045127} {\bibfield
  {journal} {\bibinfo  {journal} {Phys. Rev. B}\ }\textbf {\bibinfo {volume}
  {93}},\ \bibinfo {pages} {045127} (\bibinfo {year} {2016})},\ \Eprint
  {http://arxiv.org/abs/arXiv:1407.8203} {arXiv:1407.8203} \BibitemShut
  {NoStop}%
\bibitem [{\citenamefont {{Haah}}(2011)}]{H11011962}%
  \BibitemOpen
  \bibfield  {author} {\bibinfo {author} {\bibfnamefont {J.}~\bibnamefont
  {{Haah}}},\ }\href {\doibase 10.1103/PhysRevA.83.042330} {\bibfield
  {journal} {\bibinfo  {journal} {\pra}\ }\textbf {\bibinfo {volume} {83}},\
  \bibinfo {pages} {042330} (\bibinfo {year} {2011})},\ \Eprint
  {http://arxiv.org/abs/arXiv:1101.1962} {arXiv:1101.1962} \BibitemShut
  {NoStop}%
\bibitem [{\citenamefont {Kitaev}\ and\ \citenamefont
  {Preskill}(2006)}]{KP0604}%
  \BibitemOpen
  \bibfield  {author} {\bibinfo {author} {\bibfnamefont {A.}~\bibnamefont
  {Kitaev}}\ and\ \bibinfo {author} {\bibfnamefont {J.}~\bibnamefont
  {Preskill}},\ }\href@noop {} {\bibfield  {journal} {\bibinfo  {journal}
  {Phys. Rev. Lett.}\ }\textbf {\bibinfo {volume} {96}},\ \bibinfo {pages}
  {110404} (\bibinfo {year} {2006})}\BibitemShut {NoStop}%
\bibitem [{\citenamefont {Levin}\ and\ \citenamefont
  {Wen}(2006{\natexlab{b}})}]{LW0605}%
  \BibitemOpen
  \bibfield  {author} {\bibinfo {author} {\bibfnamefont {M.}~\bibnamefont
  {Levin}}\ and\ \bibinfo {author} {\bibfnamefont {X.-G.}\ \bibnamefont
  {Wen}},\ }\href@noop {} {\bibfield  {journal} {\bibinfo  {journal} {Phys.
  Rev. Lett.}\ }\textbf {\bibinfo {volume} {96}},\ \bibinfo {pages} {110405}
  (\bibinfo {year} {2006}{\natexlab{b}})},\ \Eprint
  {http://arxiv.org/abs/cond-mat/0510613} {cond-mat/0510613} \BibitemShut
  {NoStop}%
\bibitem [{\citenamefont {Chen}\ \emph {et~al.}(2010)\citenamefont {Chen},
  \citenamefont {Gu},\ and\ \citenamefont {Wen}}]{CGW1038}%
  \BibitemOpen
  \bibfield  {author} {\bibinfo {author} {\bibfnamefont {X.}~\bibnamefont
  {Chen}}, \bibinfo {author} {\bibfnamefont {Z.-C.}\ \bibnamefont {Gu}}, \ and\
  \bibinfo {author} {\bibfnamefont {X.-G.}\ \bibnamefont {Wen}},\ }\href@noop
  {} {\bibfield  {journal} {\bibinfo  {journal} {Phys. Rev. B}\ }\textbf
  {\bibinfo {volume} {82}},\ \bibinfo {pages} {155138} (\bibinfo {year}
  {2010})},\ \Eprint {http://arxiv.org/abs/arXiv:1004.3835} {arXiv:1004.3835}
  \BibitemShut {NoStop}%
\bibitem [{\citenamefont {Witten}(1989)}]{W8951}%
  \BibitemOpen
  \bibfield  {author} {\bibinfo {author} {\bibfnamefont {E.}~\bibnamefont
  {Witten}},\ }\href@noop {} {\bibfield  {journal} {\bibinfo  {journal} {Comm.
  Math. Phys.}\ }\textbf {\bibinfo {volume} {121}},\ \bibinfo {pages} {351}
  (\bibinfo {year} {1989})}\BibitemShut {NoStop}%
\bibitem [{\citenamefont {Lawrence}\ \emph {et~al.}(1992)\citenamefont
  {Lawrence}, \citenamefont {Sz\"oke},\ and\ \citenamefont
  {Laughlin}}]{LSL9239}%
  \BibitemOpen
  \bibfield  {author} {\bibinfo {author} {\bibfnamefont {T.~W.}\ \bibnamefont
  {Lawrence}}, \bibinfo {author} {\bibfnamefont {A.}~\bibnamefont {Sz\"oke}}, \
  and\ \bibinfo {author} {\bibfnamefont {R.~B.}\ \bibnamefont {Laughlin}},\
  }\href {\doibase 10.1103/PhysRevLett.69.1439} {\bibfield  {journal} {\bibinfo
   {journal} {Phys. Rev. Lett.}\ }\textbf {\bibinfo {volume} {69}},\ \bibinfo
  {pages} {1439} (\bibinfo {year} {1992})}\BibitemShut {NoStop}%
\bibitem [{\citenamefont {Laughlin}(1983)}]{L8395}%
  \BibitemOpen
  \bibfield  {author} {\bibinfo {author} {\bibfnamefont {R.~B.}\ \bibnamefont
  {Laughlin}},\ }\href@noop {} {\bibfield  {journal} {\bibinfo  {journal}
  {Phys. Rev. Lett.}\ }\textbf {\bibinfo {volume} {50}},\ \bibinfo {pages}
  {1395} (\bibinfo {year} {1983})}\BibitemShut {NoStop}%
\bibitem [{\citenamefont {Girvin}\ and\ \citenamefont
  {MacDonald}(1987)}]{GM8752}%
  \BibitemOpen
  \bibfield  {author} {\bibinfo {author} {\bibfnamefont {S.~M.}\ \bibnamefont
  {Girvin}}\ and\ \bibinfo {author} {\bibfnamefont {A.~H.}\ \bibnamefont
  {MacDonald}},\ }\href@noop {} {\bibfield  {journal} {\bibinfo  {journal}
  {Phys. Rev. Lett.}\ }\textbf {\bibinfo {volume} {58}},\ \bibinfo {pages}
  {1252} (\bibinfo {year} {1987})}\BibitemShut {NoStop}%
\bibitem [{\citenamefont {Zhang}\ \emph {et~al.}(1989)\citenamefont {Zhang},
  \citenamefont {Hansson},\ and\ \citenamefont {Kivelson}}]{ZHK8982}%
  \BibitemOpen
  \bibfield  {author} {\bibinfo {author} {\bibfnamefont {S.~C.}\ \bibnamefont
  {Zhang}}, \bibinfo {author} {\bibfnamefont {T.~H.}\ \bibnamefont {Hansson}},
  \ and\ \bibinfo {author} {\bibfnamefont {S.}~\bibnamefont {Kivelson}},\
  }\href@noop {} {\bibfield  {journal} {\bibinfo  {journal} {Phys. Rev. Lett.}\
  }\textbf {\bibinfo {volume} {62}},\ \bibinfo {pages} {82} (\bibinfo {year}
  {1989})}\BibitemShut {NoStop}%
\bibitem [{\citenamefont {Read}(1989)}]{R8986}%
  \BibitemOpen
  \bibfield  {author} {\bibinfo {author} {\bibfnamefont {N.}~\bibnamefont
  {Read}},\ }\href@noop {} {\bibfield  {journal} {\bibinfo  {journal} {Phys.
  Rev. Lett.}\ }\textbf {\bibinfo {volume} {62}},\ \bibinfo {pages} {86}
  (\bibinfo {year} {1989})}\BibitemShut {NoStop}%
\bibitem [{\citenamefont {Wen}\ and\ \citenamefont {Niu}(1990)}]{WNtop}%
  \BibitemOpen
  \bibfield  {author} {\bibinfo {author} {\bibfnamefont {X.-G.}\ \bibnamefont
  {Wen}}\ and\ \bibinfo {author} {\bibfnamefont {Q.}~\bibnamefont {Niu}},\
  }\href@noop {} {\bibfield  {journal} {\bibinfo  {journal} {Phys. Rev. B}\
  }\textbf {\bibinfo {volume} {41}},\ \bibinfo {pages} {9377} (\bibinfo {year}
  {1990})}\BibitemShut {NoStop}%
\bibitem [{\citenamefont {Halperin}(1982)}]{H8285}%
  \BibitemOpen
  \bibfield  {author} {\bibinfo {author} {\bibfnamefont {B.~I.}\ \bibnamefont
  {Halperin}},\ }\href@noop {} {\bibfield  {journal} {\bibinfo  {journal}
  {Phys. Rev. B}\ }\textbf {\bibinfo {volume} {25}},\ \bibinfo {pages} {2185}
  (\bibinfo {year} {1982})}\BibitemShut {NoStop}%
\bibitem [{\citenamefont {Wen}(1990{\natexlab{b}})}]{Wcll}%
  \BibitemOpen
  \bibfield  {author} {\bibinfo {author} {\bibfnamefont {X.-G.}\ \bibnamefont
  {Wen}},\ }\href@noop {} {\bibfield  {journal} {\bibinfo  {journal} {Phys.
  Rev. B}\ }\textbf {\bibinfo {volume} {41}},\ \bibinfo {pages} {12838}
  (\bibinfo {year} {1990}{\natexlab{b}})}\BibitemShut {NoStop}%
\bibitem [{\citenamefont {Halperin}(1984)}]{H8483}%
  \BibitemOpen
  \bibfield  {author} {\bibinfo {author} {\bibfnamefont {B.~I.}\ \bibnamefont
  {Halperin}},\ }\href@noop {} {\bibfield  {journal} {\bibinfo  {journal}
  {Phys. Rev. Lett.}\ }\textbf {\bibinfo {volume} {52}},\ \bibinfo {pages}
  {1583} (\bibinfo {year} {1984})}\BibitemShut {NoStop}%
\bibitem [{\citenamefont {Arovas}\ \emph {et~al.}(1984)\citenamefont {Arovas},
  \citenamefont {Schrieffer},\ and\ \citenamefont {Wilczek}}]{ASW8422}%
  \BibitemOpen
  \bibfield  {author} {\bibinfo {author} {\bibfnamefont {D.}~\bibnamefont
  {Arovas}}, \bibinfo {author} {\bibfnamefont {J.~R.}\ \bibnamefont
  {Schrieffer}}, \ and\ \bibinfo {author} {\bibfnamefont {F.}~\bibnamefont
  {Wilczek}},\ }\href@noop {} {\bibfield  {journal} {\bibinfo  {journal} {Phys.
  Rev. Lett.}\ }\textbf {\bibinfo {volume} {53}},\ \bibinfo {pages} {722}
  (\bibinfo {year} {1984})}\BibitemShut {NoStop}%
\bibitem [{\citenamefont {Leinaas}\ and\ \citenamefont
  {Myrheim}(1977)}]{LM7701}%
  \BibitemOpen
  \bibfield  {author} {\bibinfo {author} {\bibfnamefont {J.~M.}\ \bibnamefont
  {Leinaas}}\ and\ \bibinfo {author} {\bibfnamefont {J.}~\bibnamefont
  {Myrheim}},\ }\href@noop {} {\bibfield  {journal} {\bibinfo  {journal} {Il
  Nuovo Cimento}\ }\textbf {\bibinfo {volume} {37B}},\ \bibinfo {pages} {1}
  (\bibinfo {year} {1977})}\BibitemShut {NoStop}%
\bibitem [{\citenamefont {Wilczek}(1982)}]{W8257}%
  \BibitemOpen
  \bibfield  {author} {\bibinfo {author} {\bibfnamefont {F.}~\bibnamefont
  {Wilczek}},\ }\href@noop {} {\bibfield  {journal} {\bibinfo  {journal} {Phys.
  Rev. Lett.}\ }\textbf {\bibinfo {volume} {49}},\ \bibinfo {pages} {957}
  (\bibinfo {year} {1982})}\BibitemShut {NoStop}%
\bibitem [{\citenamefont {Wu}(1984)}]{W8413}%
  \BibitemOpen
  \bibfield  {author} {\bibinfo {author} {\bibfnamefont {Y.-S.}\ \bibnamefont
  {Wu}},\ }\href {\doibase 10.1103/PhysRevLett.52.2103} {\bibfield  {journal}
  {\bibinfo  {journal} {Phys. Rev. Lett.}\ }\textbf {\bibinfo {volume} {52}},\
  \bibinfo {pages} {2013} (\bibinfo {year} {1984})}\BibitemShut {NoStop}%
\bibitem [{\citenamefont {de~Picciotto}\ \emph {et~al.}(1997)\citenamefont
  {de~Picciotto}, \citenamefont {Reznikov}, \citenamefont {Heiblum},
  \citenamefont {Umansky}, \citenamefont {Bunin},\ and\ \citenamefont
  {Mahalu}}]{dRH9762}%
  \BibitemOpen
  \bibfield  {author} {\bibinfo {author} {\bibfnamefont {R.}~\bibnamefont
  {de~Picciotto}}, \bibinfo {author} {\bibfnamefont {M.}~\bibnamefont
  {Reznikov}}, \bibinfo {author} {\bibfnamefont {M.}~\bibnamefont {Heiblum}},
  \bibinfo {author} {\bibfnamefont {V.}~\bibnamefont {Umansky}}, \bibinfo
  {author} {\bibfnamefont {G.}~\bibnamefont {Bunin}}, \ and\ \bibinfo {author}
  {\bibfnamefont {D.}~\bibnamefont {Mahalu}},\ }\href@noop {} {\bibfield
  {journal} {\bibinfo  {journal} {Nature}\ }\textbf {\bibinfo {volume} {389}},\
  \bibinfo {pages} {162} (\bibinfo {year} {1997})}\BibitemShut {NoStop}%
\bibitem [{\citenamefont {Kitaev}(2011)}]{K11sre}%
  \BibitemOpen
  \bibfield  {author} {\bibinfo {author} {\bibfnamefont {A.}~\bibnamefont
  {Kitaev}},\ }\href@noop {} {\bibfield  {journal} {\bibinfo  {journal}
  {http://online.kitp.ucsb.edu/online/topomat11}\ } (\bibinfo {year}
  {2011})}\BibitemShut {NoStop}%
\bibitem [{\citenamefont {von Klitzing}\ \emph {et~al.}(1980)\citenamefont {von
  Klitzing}, \citenamefont {Dorda},\ and\ \citenamefont {Pepper}}]{KDP8094}%
  \BibitemOpen
  \bibfield  {author} {\bibinfo {author} {\bibfnamefont {K.}~\bibnamefont {von
  Klitzing}}, \bibinfo {author} {\bibfnamefont {G.}~\bibnamefont {Dorda}}, \
  and\ \bibinfo {author} {\bibfnamefont {M.}~\bibnamefont {Pepper}},\
  }\href@noop {} {\bibfield  {journal} {\bibinfo  {journal} {Phys. Rev. Lett.}\
  }\textbf {\bibinfo {volume} {45}},\ \bibinfo {pages} {494} (\bibinfo {year}
  {1980})}\BibitemShut {NoStop}%
\bibitem [{\citenamefont {Hofstadter}(1976)}]{H7639}%
  \BibitemOpen
  \bibfield  {author} {\bibinfo {author} {\bibfnamefont {D.~R.}\ \bibnamefont
  {Hofstadter}},\ }\href {\doibase 10.1103/PhysRevB.14.2239} {\bibfield
  {journal} {\bibinfo  {journal} {\prb}\ }\textbf {\bibinfo {volume} {14}},\
  \bibinfo {pages} {2239–2249} (\bibinfo {year} {1976})}\BibitemShut
  {NoStop}%
\bibitem [{\citenamefont {Thouless}\ \emph {et~al.}(1982)\citenamefont
  {Thouless}, \citenamefont {Kohmoto}, \citenamefont {Nightingale},\ and\
  \citenamefont {den Nijs}}]{TKN8205}%
  \BibitemOpen
  \bibfield  {author} {\bibinfo {author} {\bibfnamefont {D.~J.}\ \bibnamefont
  {Thouless}}, \bibinfo {author} {\bibfnamefont {M.}~\bibnamefont {Kohmoto}},
  \bibinfo {author} {\bibfnamefont {M.~P.}\ \bibnamefont {Nightingale}}, \ and\
  \bibinfo {author} {\bibfnamefont {M.}~\bibnamefont {den Nijs}},\ }\href@noop
  {} {\bibfield  {journal} {\bibinfo  {journal} {Phys. Rev. Lett.}\ }\textbf
  {\bibinfo {volume} {49}},\ \bibinfo {pages} {405} (\bibinfo {year}
  {1982})}\BibitemShut {NoStop}%
\bibitem [{\citenamefont {Haldane}(1988)}]{H8815}%
  \BibitemOpen
  \bibfield  {author} {\bibinfo {author} {\bibfnamefont {F.}~\bibnamefont
  {Haldane}},\ }\href@noop {} {\bibfield  {journal} {\bibinfo  {journal}
  {\prl}\ }\textbf {\bibinfo {volume} {61}},\ \bibinfo {pages} {2015} (\bibinfo
  {year} {1988})}\BibitemShut {NoStop}%
\bibitem [{\citenamefont {Wen}(1991{\natexlab{a}})}]{Wnab}%
  \BibitemOpen
  \bibfield  {author} {\bibinfo {author} {\bibfnamefont {X.-G.}\ \bibnamefont
  {Wen}},\ }\href@noop {} {\bibfield  {journal} {\bibinfo  {journal} {Phys.
  Rev. Lett.}\ }\textbf {\bibinfo {volume} {66}},\ \bibinfo {pages} {802}
  (\bibinfo {year} {1991}{\natexlab{a}})}\BibitemShut {NoStop}%
\bibitem [{\citenamefont {Goldin}\ \emph {et~al.}(1985)\citenamefont {Goldin},
  \citenamefont {Menikoff},\ and\ \citenamefont {Sharp}}]{GMS8503}%
  \BibitemOpen
  \bibfield  {author} {\bibinfo {author} {\bibfnamefont {G.~A.}\ \bibnamefont
  {Goldin}}, \bibinfo {author} {\bibfnamefont {R.}~\bibnamefont {Menikoff}}, \
  and\ \bibinfo {author} {\bibfnamefont {D.~H.}\ \bibnamefont {Sharp}},\ }\href
  {\doibase 10.1103/PhysRevLett.54.603} {\bibfield  {journal} {\bibinfo
  {journal} {Phys. Rev. Lett.}\ }\textbf {\bibinfo {volume} {54}},\ \bibinfo
  {pages} {603} (\bibinfo {year} {1985})}\BibitemShut {NoStop}%
\bibitem [{\citenamefont {Kitaev}(2006)}]{K062}%
  \BibitemOpen
  \bibfield  {author} {\bibinfo {author} {\bibfnamefont {A.}~\bibnamefont
  {Kitaev}},\ }\href@noop {} {\bibfield  {journal} {\bibinfo  {journal} {Annals
  of Physics}\ }\textbf {\bibinfo {volume} {321}},\ \bibinfo {pages} {2}
  (\bibinfo {year} {2006})},\ \Eprint {http://arxiv.org/abs/cond-mat/0506438}
  {cond-mat/0506438} \BibitemShut {NoStop}%
\bibitem [{\citenamefont {Lan}\ and\ \citenamefont {Wen}(2017)}]{LW170107820}%
  \BibitemOpen
  \bibfield  {author} {\bibinfo {author} {\bibfnamefont {T.}~\bibnamefont
  {Lan}}\ and\ \bibinfo {author} {\bibfnamefont {X.-G.}\ \bibnamefont {Wen}},\
  }\href@noop {} {\  (\bibinfo {year} {2017})},\ \Eprint
  {http://arxiv.org/abs/arXiv:1701.07820} {arXiv:1701.07820} \BibitemShut
  {NoStop}%
\bibitem [{\citenamefont {Jain}(1991)}]{J9153}%
  \BibitemOpen
  \bibfield  {author} {\bibinfo {author} {\bibfnamefont {J.~K.}\ \bibnamefont
  {Jain}},\ }\href@noop {} {\bibfield  {journal} {\bibinfo  {journal} {Phys.
  Rev. B}\ }\textbf {\bibinfo {volume} {41}},\ \bibinfo {pages} {7653}
  (\bibinfo {year} {1991})}\BibitemShut {NoStop}%
\bibitem [{\citenamefont {Blok}\ and\ \citenamefont {Wen}(1992)}]{BW9215}%
  \BibitemOpen
  \bibfield  {author} {\bibinfo {author} {\bibfnamefont {B.}~\bibnamefont
  {Blok}}\ and\ \bibinfo {author} {\bibfnamefont {X.-G.}\ \bibnamefont {Wen}},\
  }\href@noop {} {\bibfield  {journal} {\bibinfo  {journal} {Nucl. Phys. B}\
  }\textbf {\bibinfo {volume} {374}},\ \bibinfo {pages} {615} (\bibinfo {year}
  {1992})}\BibitemShut {NoStop}%
\bibitem [{\citenamefont {Moore}\ and\ \citenamefont {Read}(1991)}]{MR9162}%
  \BibitemOpen
  \bibfield  {author} {\bibinfo {author} {\bibfnamefont {G.}~\bibnamefont
  {Moore}}\ and\ \bibinfo {author} {\bibfnamefont {N.}~\bibnamefont {Read}},\
  }\href@noop {} {\bibfield  {journal} {\bibinfo  {journal} {Nucl. Phys. B}\
  }\textbf {\bibinfo {volume} {360}},\ \bibinfo {pages} {362} (\bibinfo {year}
  {1991})}\BibitemShut {NoStop}%
\bibitem [{\citenamefont {Wen}(1993)}]{Wnabhalf}%
  \BibitemOpen
  \bibfield  {author} {\bibinfo {author} {\bibfnamefont {X.-G.}\ \bibnamefont
  {Wen}},\ }\href@noop {} {\bibfield  {journal} {\bibinfo  {journal} {Phys.
  Rev. Lett.}\ }\textbf {\bibinfo {volume} {70}},\ \bibinfo {pages} {355}
  (\bibinfo {year} {1993})}\BibitemShut {NoStop}%
\bibitem [{\citenamefont {Wen}(1999)}]{W9927}%
  \BibitemOpen
  \bibfield  {author} {\bibinfo {author} {\bibfnamefont {X.-G.}\ \bibnamefont
  {Wen}},\ }\href@noop {} {\bibfield  {journal} {\bibinfo  {journal} {Phys.
  Rev. B}\ }\textbf {\bibinfo {volume} {60}},\ \bibinfo {pages} {8827}
  (\bibinfo {year} {1999})},\ \Eprint {http://arxiv.org/abs/cond-mat/9811111}
  {cond-mat/9811111} \BibitemShut {NoStop}%
\bibitem [{\citenamefont {{Bonderson}}\ \emph {et~al.}(2011)\citenamefont
  {{Bonderson}}, \citenamefont {{Gurarie}},\ and\ \citenamefont
  {{Nayak}}}]{BN10085194}%
  \BibitemOpen
  \bibfield  {author} {\bibinfo {author} {\bibfnamefont {P.}~\bibnamefont
  {{Bonderson}}}, \bibinfo {author} {\bibfnamefont {V.}~\bibnamefont
  {{Gurarie}}}, \ and\ \bibinfo {author} {\bibfnamefont {C.}~\bibnamefont
  {{Nayak}}},\ }\href {\doibase 10.1103/PhysRevB.83.075303} {\bibfield
  {journal} {\bibinfo  {journal} {\prb}\ }\textbf {\bibinfo {volume} {83}},\
  \bibinfo {pages} {075303} (\bibinfo {year} {2011})},\ \Eprint
  {http://arxiv.org/abs/arXiv:1008.5194} {arXiv:1008.5194} \BibitemShut
  {NoStop}%
\bibitem [{\citenamefont {Willett}\ \emph {et~al.}(1987)\citenamefont
  {Willett}, \citenamefont {Eisenstein}, \citenamefont {Str{\"o}rmer},
  \citenamefont {Tsui}, \citenamefont {Gossard},\ and\ \citenamefont
  {English}}]{WES8776}%
  \BibitemOpen
  \bibfield  {author} {\bibinfo {author} {\bibfnamefont {R.}~\bibnamefont
  {Willett}}, \bibinfo {author} {\bibfnamefont {J.~P.}\ \bibnamefont
  {Eisenstein}}, \bibinfo {author} {\bibfnamefont {H.~L.}\ \bibnamefont
  {Str{\"o}rmer}}, \bibinfo {author} {\bibfnamefont {D.~C.}\ \bibnamefont
  {Tsui}}, \bibinfo {author} {\bibfnamefont {A.~C.}\ \bibnamefont {Gossard}}, \
  and\ \bibinfo {author} {\bibfnamefont {J.~H.}\ \bibnamefont {English}},\
  }\href {\doibase 10.1103/PhysRevLett.59.1776} {\bibfield  {journal} {\bibinfo
   {journal} {Phys. Rev. Lett.}\ }\textbf {\bibinfo {volume} {59}},\ \bibinfo
  {pages} {1776} (\bibinfo {year} {1987})}\BibitemShut {NoStop}%
\bibitem [{\citenamefont {{Dolev}}\ \emph {et~al.}(2008)\citenamefont
  {{Dolev}}, \citenamefont {{Heiblum}}, \citenamefont {{Umansky}},
  \citenamefont {{Stern}},\ and\ \citenamefont {{Mahalu}}}]{DM08020930}%
  \BibitemOpen
  \bibfield  {author} {\bibinfo {author} {\bibfnamefont {M.}~\bibnamefont
  {{Dolev}}}, \bibinfo {author} {\bibfnamefont {M.}~\bibnamefont {{Heiblum}}},
  \bibinfo {author} {\bibfnamefont {V.}~\bibnamefont {{Umansky}}}, \bibinfo
  {author} {\bibfnamefont {A.}~\bibnamefont {{Stern}}}, \ and\ \bibinfo
  {author} {\bibfnamefont {D.}~\bibnamefont {{Mahalu}}},\ }\href {\doibase
  10.1038/nature06855} {\bibfield  {journal} {\bibinfo  {journal} {\nat}\
  }\textbf {\bibinfo {volume} {452}},\ \bibinfo {pages} {829} (\bibinfo {year}
  {2008})},\ \Eprint {http://arxiv.org/abs/arXiv:0802.0930} {arXiv:0802.0930}
  \BibitemShut {NoStop}%
\bibitem [{\citenamefont {Radu}\ \emph {et~al.}(2008)\citenamefont {Radu},
  \citenamefont {Miller}, \citenamefont {Marcus}, \citenamefont {Kastner},
  \citenamefont {Pfeiffer},\ and\ \citenamefont {West}}]{RMM0899}%
  \BibitemOpen
  \bibfield  {author} {\bibinfo {author} {\bibfnamefont {I.~P.}\ \bibnamefont
  {Radu}}, \bibinfo {author} {\bibfnamefont {J.~B.}\ \bibnamefont {Miller}},
  \bibinfo {author} {\bibfnamefont {C.~M.}\ \bibnamefont {Marcus}}, \bibinfo
  {author} {\bibfnamefont {M.~A.}\ \bibnamefont {Kastner}}, \bibinfo {author}
  {\bibfnamefont {L.~N.}\ \bibnamefont {Pfeiffer}}, \ and\ \bibinfo {author}
  {\bibfnamefont {K.~W.}\ \bibnamefont {West}},\ }\href@noop {} {\bibfield
  {journal} {\bibinfo  {journal} {Science}\ }\textbf {\bibinfo {volume}
  {320}},\ \bibinfo {pages} {899} (\bibinfo {year} {2008})}\BibitemShut
  {NoStop}%
\bibitem [{\citenamefont {Onnes}(1911)}]{O1122}%
  \BibitemOpen
  \bibfield  {author} {\bibinfo {author} {\bibfnamefont {H.~K.}\ \bibnamefont
  {Onnes}},\ }\href@noop {} {\bibfield  {journal} {\bibinfo  {journal} {Comm.
  Phys. Lab. Univ. Leiden, Nos 119}\ }\textbf {\bibinfo {volume} {120}},\
  \bibinfo {pages} {122} (\bibinfo {year} {1911})}\BibitemShut {NoStop}%
\bibitem [{\citenamefont {Wen}(1991{\natexlab{b}})}]{W9141}%
  \BibitemOpen
  \bibfield  {author} {\bibinfo {author} {\bibfnamefont {X.-G.}\ \bibnamefont
  {Wen}},\ }\href@noop {} {\bibfield  {journal} {\bibinfo  {journal} {Int. J.
  Mod. Phys. B}\ }\textbf {\bibinfo {volume} {5}},\ \bibinfo {pages} {1641}
  (\bibinfo {year} {1991}{\natexlab{b}})}\BibitemShut {NoStop}%
\bibitem [{\citenamefont {Hansson}\ \emph {et~al.}(2004)\citenamefont
  {Hansson}, \citenamefont {Oganesyan},\ and\ \citenamefont
  {Sondhi}}]{HOS0497}%
  \BibitemOpen
  \bibfield  {author} {\bibinfo {author} {\bibfnamefont {T.~H.}\ \bibnamefont
  {Hansson}}, \bibinfo {author} {\bibfnamefont {V.}~\bibnamefont {Oganesyan}},
  \ and\ \bibinfo {author} {\bibfnamefont {S.~L.}\ \bibnamefont {Sondhi}},\
  }\href@noop {} {\bibfield  {journal} {\bibinfo  {journal} {Annals of
  Physics}\ }\textbf {\bibinfo {volume} {313}},\ \bibinfo {pages} {497}
  (\bibinfo {year} {2004})},\ \Eprint {http://arxiv.org/abs/cond-mat/0404327}
  {cond-mat/0404327} \BibitemShut {NoStop}%
\bibitem [{\citenamefont {Read}\ and\ \citenamefont {Sachdev}(1991)}]{RS9173}%
  \BibitemOpen
  \bibfield  {author} {\bibinfo {author} {\bibfnamefont {N.}~\bibnamefont
  {Read}}\ and\ \bibinfo {author} {\bibfnamefont {S.}~\bibnamefont {Sachdev}},\
  }\href@noop {} {\bibfield  {journal} {\bibinfo  {journal} {Phys. Rev. Lett.}\
  }\textbf {\bibinfo {volume} {66}},\ \bibinfo {pages} {1773} (\bibinfo {year}
  {1991})}\BibitemShut {NoStop}%
\bibitem [{\citenamefont {Wen}(1991{\natexlab{c}})}]{Wsrvb}%
  \BibitemOpen
  \bibfield  {author} {\bibinfo {author} {\bibfnamefont {X.-G.}\ \bibnamefont
  {Wen}},\ }\href@noop {} {\bibfield  {journal} {\bibinfo  {journal} {Phys.
  Rev. B}\ }\textbf {\bibinfo {volume} {44}},\ \bibinfo {pages} {2664}
  (\bibinfo {year} {1991}{\natexlab{c}})}\BibitemShut {NoStop}%
\bibitem [{\citenamefont {Kitaev}(2003)}]{K032}%
  \BibitemOpen
  \bibfield  {author} {\bibinfo {author} {\bibfnamefont {A.~Y.}\ \bibnamefont
  {Kitaev}},\ }\href@noop {} {\bibfield  {journal} {\bibinfo  {journal} {Ann.
  Phys. (N.Y.)}\ }\textbf {\bibinfo {volume} {303}},\ \bibinfo {pages} {2}
  (\bibinfo {year} {2003})}\BibitemShut {NoStop}%
\bibitem [{\citenamefont {Helton}\ \emph {et~al.}(2007)\citenamefont {Helton},
  \citenamefont {Matan}, \citenamefont {Shores}, \citenamefont {Nytko},
  \citenamefont {Bartlett}, \citenamefont {Yoshida}, \citenamefont {Takano},
  \citenamefont {Suslov}, \citenamefont {Qiu}, \citenamefont {Chung},
  \citenamefont {Nocera},\ and\ \citenamefont {Lee}}]{HMS0704}%
  \BibitemOpen
  \bibfield  {author} {\bibinfo {author} {\bibfnamefont {J.~S.}\ \bibnamefont
  {Helton}}, \bibinfo {author} {\bibfnamefont {K.}~\bibnamefont {Matan}},
  \bibinfo {author} {\bibfnamefont {M.~P.}\ \bibnamefont {Shores}}, \bibinfo
  {author} {\bibfnamefont {E.~A.}\ \bibnamefont {Nytko}}, \bibinfo {author}
  {\bibfnamefont {B.~M.}\ \bibnamefont {Bartlett}}, \bibinfo {author}
  {\bibfnamefont {Y.}~\bibnamefont {Yoshida}}, \bibinfo {author} {\bibfnamefont
  {Y.}~\bibnamefont {Takano}}, \bibinfo {author} {\bibfnamefont
  {A.}~\bibnamefont {Suslov}}, \bibinfo {author} {\bibfnamefont
  {Y.}~\bibnamefont {Qiu}}, \bibinfo {author} {\bibfnamefont {J.-H.}\
  \bibnamefont {Chung}}, \bibinfo {author} {\bibfnamefont {D.~G.}\ \bibnamefont
  {Nocera}}, \ and\ \bibinfo {author} {\bibfnamefont {Y.~S.}\ \bibnamefont
  {Lee}},\ }\href {\doibase 10.1103/PhysRevLett.98.107204} {\bibfield
  {journal} {\bibinfo  {journal} {Phys. Rev. Lett.}\ }\textbf {\bibinfo
  {volume} {98}},\ \bibinfo {pages} {107204} (\bibinfo {year} {2007})},\
  \Eprint {http://arxiv.org/abs/cond-mat/0610539} {cond-mat/0610539}
  \BibitemShut {NoStop}%
\bibitem [{\citenamefont {Fu}\ \emph {et~al.}(2015)\citenamefont {Fu},
  \citenamefont {Imai}, \citenamefont {Han},\ and\ \citenamefont
  {Lee}}]{FL151102174}%
  \BibitemOpen
  \bibfield  {author} {\bibinfo {author} {\bibfnamefont {M.}~\bibnamefont
  {Fu}}, \bibinfo {author} {\bibfnamefont {T.}~\bibnamefont {Imai}}, \bibinfo
  {author} {\bibfnamefont {T.-H.}\ \bibnamefont {Han}}, \ and\ \bibinfo
  {author} {\bibfnamefont {Y.~S.}\ \bibnamefont {Lee}},\ }\href {\doibase
  10.1126/science.aab2120} {\bibfield  {journal} {\bibinfo  {journal}
  {Science}\ }\textbf {\bibinfo {volume} {350}},\ \bibinfo {pages} {655}
  (\bibinfo {year} {2015})},\ \Eprint {http://arxiv.org/abs/arXiv:1511.02174}
  {arXiv:1511.02174} \BibitemShut {NoStop}%
\bibitem [{\citenamefont {Han}\ \emph {et~al.}(2016)\citenamefont {Han},
  \citenamefont {Norman}, \citenamefont {Wen}, \citenamefont
  {Rodriguez-Rivera}, \citenamefont {Helton}, \citenamefont {Broholm},\ and\
  \citenamefont {Lee}}]{HL151206807}%
  \BibitemOpen
  \bibfield  {author} {\bibinfo {author} {\bibfnamefont {T.-H.}\ \bibnamefont
  {Han}}, \bibinfo {author} {\bibfnamefont {M.~R.}\ \bibnamefont {Norman}},
  \bibinfo {author} {\bibfnamefont {J.-J.}\ \bibnamefont {Wen}}, \bibinfo
  {author} {\bibfnamefont {J.~A.}\ \bibnamefont {Rodriguez-Rivera}}, \bibinfo
  {author} {\bibfnamefont {J.~S.}\ \bibnamefont {Helton}}, \bibinfo {author}
  {\bibfnamefont {C.}~\bibnamefont {Broholm}}, \ and\ \bibinfo {author}
  {\bibfnamefont {Y.~S.}\ \bibnamefont {Lee}},\ }\href {\doibase
  10.1103/PhysRevB.94.060409} {\bibfield  {journal} {\bibinfo  {journal} {Phys.
  Rev. B}\ }\textbf {\bibinfo {volume} {94}},\ \bibinfo {pages} {060409}
  (\bibinfo {year} {2016})},\ \Eprint {http://arxiv.org/abs/arXiv:1512.06807}
  {arXiv:1512.06807} \BibitemShut {NoStop}%
\bibitem [{\citenamefont {Yan}\ \emph {et~al.}(2011)\citenamefont {Yan},
  \citenamefont {Huse},\ and\ \citenamefont {White}}]{YHW1173}%
  \BibitemOpen
  \bibfield  {author} {\bibinfo {author} {\bibfnamefont {S.}~\bibnamefont
  {Yan}}, \bibinfo {author} {\bibfnamefont {D.~A.}\ \bibnamefont {Huse}}, \
  and\ \bibinfo {author} {\bibfnamefont {S.~R.}\ \bibnamefont {White}},\ }\href
  {\doibase 10.1126/science.1201080} {\bibfield  {journal} {\bibinfo  {journal}
  {Science}\ }\textbf {\bibinfo {volume} {332}},\ \bibinfo {pages} {1173 }
  (\bibinfo {year} {2011})},\ \Eprint {http://arxiv.org/abs/arXiv:1011.6114}
  {arXiv:1011.6114} \BibitemShut {NoStop}%
\bibitem [{\citenamefont {Mei}\ \emph {et~al.}(2016)\citenamefont {Mei},
  \citenamefont {Chen}, \citenamefont {He},\ and\ \citenamefont
  {Wen}}]{MW160609639}%
  \BibitemOpen
  \bibfield  {author} {\bibinfo {author} {\bibfnamefont {J.-W.}\ \bibnamefont
  {Mei}}, \bibinfo {author} {\bibfnamefont {J.-Y.}\ \bibnamefont {Chen}},
  \bibinfo {author} {\bibfnamefont {H.}~\bibnamefont {He}}, \ and\ \bibinfo
  {author} {\bibfnamefont {X.-G.}\ \bibnamefont {Wen}},\ }\href@noop {} {\
  (\bibinfo {year} {2016})},\ \Eprint {http://arxiv.org/abs/arXiv:160609639}
  {arXiv:160609639} \BibitemShut {NoStop}%
\bibitem [{\citenamefont {{Jiang}}\ \emph {et~al.}(2016)\citenamefont
  {{Jiang}}, \citenamefont {{Kim}}, \citenamefont {{Han}},\ and\ \citenamefont
  {{Ran}}}]{JR161002024}%
  \BibitemOpen
  \bibfield  {author} {\bibinfo {author} {\bibfnamefont {S.}~\bibnamefont
  {{Jiang}}}, \bibinfo {author} {\bibfnamefont {P.}~\bibnamefont {{Kim}}},
  \bibinfo {author} {\bibfnamefont {J.~H.}\ \bibnamefont {{Han}}}, \ and\
  \bibinfo {author} {\bibfnamefont {Y.}~\bibnamefont {{Ran}}},\ }\href@noop {}
  {\  (\bibinfo {year} {2016})},\ \Eprint
  {http://arxiv.org/abs/arXiv:1610.02024} {arXiv:1610.02024} \BibitemShut
  {NoStop}%
\bibitem [{\citenamefont {{Liao}}\ \emph {et~al.}(2016)\citenamefont {{Liao}},
  \citenamefont {{Xie}}, \citenamefont {{Chen}}, \citenamefont {{Liu}},
  \citenamefont {{Xie}}, \citenamefont {{Huang}}, \citenamefont {{Normand}},\
  and\ \citenamefont {{Xiang}}}]{LX161004727}%
  \BibitemOpen
  \bibfield  {author} {\bibinfo {author} {\bibfnamefont {H.~J.}\ \bibnamefont
  {{Liao}}}, \bibinfo {author} {\bibfnamefont {Z.~Y.}\ \bibnamefont {{Xie}}},
  \bibinfo {author} {\bibfnamefont {J.}~\bibnamefont {{Chen}}}, \bibinfo
  {author} {\bibfnamefont {Z.~Y.}\ \bibnamefont {{Liu}}}, \bibinfo {author}
  {\bibfnamefont {H.~D.}\ \bibnamefont {{Xie}}}, \bibinfo {author}
  {\bibfnamefont {R.~Z.}\ \bibnamefont {{Huang}}}, \bibinfo {author}
  {\bibfnamefont {B.}~\bibnamefont {{Normand}}}, \ and\ \bibinfo {author}
  {\bibfnamefont {T.}~\bibnamefont {{Xiang}}},\ }\href@noop {} {\  (\bibinfo
  {year} {2016})},\ \Eprint {http://arxiv.org/abs/arXiv:1610.04727}
  {arXiv:1610.04727} \BibitemShut {NoStop}%
\bibitem [{\citenamefont {{He}}\ \emph {et~al.}(2016)\citenamefont {{He}},
  \citenamefont {{Zaletel}}, \citenamefont {{Oshikawa}},\ and\ \citenamefont
  {{Pollmann}}}]{HP161106238}%
  \BibitemOpen
  \bibfield  {author} {\bibinfo {author} {\bibfnamefont {Y.-C.}\ \bibnamefont
  {{He}}}, \bibinfo {author} {\bibfnamefont {M.~P.}\ \bibnamefont {{Zaletel}}},
  \bibinfo {author} {\bibfnamefont {M.}~\bibnamefont {{Oshikawa}}}, \ and\
  \bibinfo {author} {\bibfnamefont {F.}~\bibnamefont {{Pollmann}}},\
  }\href@noop {} {\  (\bibinfo {year} {2016})},\ \Eprint
  {http://arxiv.org/abs/arXiv:1611.06238} {arXiv:1611.06238} \BibitemShut
  {NoStop}%
\bibitem [{\citenamefont {Fidkowski}\ \emph {et~al.}(2009)\citenamefont
  {Fidkowski}, \citenamefont {Freedman}, \citenamefont {Nayak}, \citenamefont
  {Walker},\ and\ \citenamefont {Wang}}]{FFN0683}%
  \BibitemOpen
  \bibfield  {author} {\bibinfo {author} {\bibfnamefont {L.}~\bibnamefont
  {Fidkowski}}, \bibinfo {author} {\bibfnamefont {M.}~\bibnamefont {Freedman}},
  \bibinfo {author} {\bibfnamefont {C.}~\bibnamefont {Nayak}}, \bibinfo
  {author} {\bibfnamefont {K.}~\bibnamefont {Walker}}, \ and\ \bibinfo {author}
  {\bibfnamefont {Z.}~\bibnamefont {Wang}},\ }\href@noop {} {\bibfield
  {journal} {\bibinfo  {journal} {Comm. Math. Phys.}\ }\textbf {\bibinfo
  {volume} {287}},\ \bibinfo {pages} {805} (\bibinfo {year} {2009})},\ \Eprint
  {http://arxiv.org/abs/cond-mat/0610583} {cond-mat/0610583} \BibitemShut
  {NoStop}%
\bibitem [{\citenamefont {Kitaev}(2001)}]{K0131}%
  \BibitemOpen
  \bibfield  {author} {\bibinfo {author} {\bibfnamefont {A.~Y.}\ \bibnamefont
  {Kitaev}},\ }\href {\doibase 10.1070/1063-7869/44/10S/S29} {\bibfield
  {journal} {\bibinfo  {journal} {Phys.-Usp.}\ }\textbf {\bibinfo {volume}
  {44}},\ \bibinfo {pages} {131} (\bibinfo {year} {2001})},\ \Eprint
  {http://arxiv.org/abs/cond-mat/0010440} {cond-mat/0010440} \BibitemShut
  {NoStop}%
\bibitem [{\citenamefont {Freedman}\ \emph {et~al.}(2004)\citenamefont
  {Freedman}, \citenamefont {Nayak}, \citenamefont {Shtengel}, \citenamefont
  {Walker},\ and\ \citenamefont {Wang}}]{FNS0428}%
  \BibitemOpen
  \bibfield  {author} {\bibinfo {author} {\bibfnamefont {M.}~\bibnamefont
  {Freedman}}, \bibinfo {author} {\bibfnamefont {C.}~\bibnamefont {Nayak}},
  \bibinfo {author} {\bibfnamefont {K.}~\bibnamefont {Shtengel}}, \bibinfo
  {author} {\bibfnamefont {K.}~\bibnamefont {Walker}}, \ and\ \bibinfo {author}
  {\bibfnamefont {Z.}~\bibnamefont {Wang}},\ }\href@noop {} {\bibfield
  {journal} {\bibinfo  {journal} {Ann. Phys. (NY)}\ }\textbf {\bibinfo {volume}
  {310}},\ \bibinfo {pages} {428} (\bibinfo {year} {2004})},\ \Eprint
  {http://arxiv.org/abs/cond-mat/0307511} {cond-mat/0307511} \BibitemShut
  {NoStop}%
\bibitem [{\citenamefont {Levin}\ and\ \citenamefont {Wen}(2005)}]{LW0510}%
  \BibitemOpen
  \bibfield  {author} {\bibinfo {author} {\bibfnamefont {M.}~\bibnamefont
  {Levin}}\ and\ \bibinfo {author} {\bibfnamefont {X.-G.}\ \bibnamefont
  {Wen}},\ }\href@noop {} {\bibfield  {journal} {\bibinfo  {journal} {Phys.
  Rev. B}\ }\textbf {\bibinfo {volume} {71}},\ \bibinfo {pages} {045110}
  (\bibinfo {year} {2005})},\ \Eprint {http://arxiv.org/abs/cond-mat/0404617}
  {cond-mat/0404617} \BibitemShut {NoStop}%
\bibitem [{\citenamefont {Dijkgraaf}\ and\ \citenamefont
  {Witten}(1990)}]{DW9093}%
  \BibitemOpen
  \bibfield  {author} {\bibinfo {author} {\bibfnamefont {R.}~\bibnamefont
  {Dijkgraaf}}\ and\ \bibinfo {author} {\bibfnamefont {E.}~\bibnamefont
  {Witten}},\ }\href@noop {} {\bibfield  {journal} {\bibinfo  {journal} {Comm.
  Math. Phys.}\ }\textbf {\bibinfo {volume} {129}},\ \bibinfo {pages} {393}
  (\bibinfo {year} {1990})}\BibitemShut {NoStop}%
\bibitem [{\citenamefont {Senthil}\ \emph {et~al.}(1999)\citenamefont
  {Senthil}, \citenamefont {Marston},\ and\ \citenamefont {Fisher}}]{SMF9945}%
  \BibitemOpen
  \bibfield  {author} {\bibinfo {author} {\bibfnamefont {T.}~\bibnamefont
  {Senthil}}, \bibinfo {author} {\bibfnamefont {J.~B.}\ \bibnamefont
  {Marston}}, \ and\ \bibinfo {author} {\bibfnamefont {M.~P.~A.}\ \bibnamefont
  {Fisher}},\ }\href@noop {} {\bibfield  {journal} {\bibinfo  {journal} {Phys.
  Rev. B}\ }\textbf {\bibinfo {volume} {60}},\ \bibinfo {pages} {4245}
  (\bibinfo {year} {1999})}\BibitemShut {NoStop}%
\bibitem [{\citenamefont {Read}\ and\ \citenamefont {Green}(2000)}]{RG0067}%
  \BibitemOpen
  \bibfield  {author} {\bibinfo {author} {\bibfnamefont {N.}~\bibnamefont
  {Read}}\ and\ \bibinfo {author} {\bibfnamefont {D.}~\bibnamefont {Green}},\
  }\href@noop {} {\bibfield  {journal} {\bibinfo  {journal} {Phys. Rev. B}\
  }\textbf {\bibinfo {volume} {61}},\ \bibinfo {pages} {10267} (\bibinfo {year}
  {2000})}\BibitemShut {NoStop}%
\bibitem [{\citenamefont {Halperin}(1983)}]{H8375}%
  \BibitemOpen
  \bibfield  {author} {\bibinfo {author} {\bibfnamefont {B.~I.}\ \bibnamefont
  {Halperin}},\ }\href@noop {} {\bibfield  {journal} {\bibinfo  {journal}
  {Helv. Phys. Acta}\ }\textbf {\bibinfo {volume} {56}},\ \bibinfo {pages} {75}
  (\bibinfo {year} {1983})}\BibitemShut {NoStop}%
\bibitem [{\citenamefont {Read}\ and\ \citenamefont {Rezayi}(1999)}]{RR9984}%
  \BibitemOpen
  \bibfield  {author} {\bibinfo {author} {\bibfnamefont {N.}~\bibnamefont
  {Read}}\ and\ \bibinfo {author} {\bibfnamefont {E.}~\bibnamefont {Rezayi}},\
  }\href@noop {} {\bibfield  {journal} {\bibinfo  {journal} {Phys.Rev. B}\
  }\textbf {\bibinfo {volume} {59}},\ \bibinfo {pages} {8084} (\bibinfo {year}
  {1999})},\ \Eprint {http://arxiv.org/abs/cond-mat/9809384} {cond-mat/9809384}
  \BibitemShut {NoStop}%
\bibitem [{\citenamefont {Gu}\ \emph {et~al.}(2015)\citenamefont {Gu},
  \citenamefont {Wang},\ and\ \citenamefont {Wen}}]{GWW1017}%
  \BibitemOpen
  \bibfield  {author} {\bibinfo {author} {\bibfnamefont {Z.-C.}\ \bibnamefont
  {Gu}}, \bibinfo {author} {\bibfnamefont {Z.}~\bibnamefont {Wang}}, \ and\
  \bibinfo {author} {\bibfnamefont {X.-G.}\ \bibnamefont {Wen}},\ }\href
  {\doibase 10.1103/PhysRevB.91.125149} {\bibfield  {journal} {\bibinfo
  {journal} {Phys. Rev. B}\ }\textbf {\bibinfo {volume} {91}},\ \bibinfo
  {pages} {125149} (\bibinfo {year} {2015})},\ \Eprint
  {http://arxiv.org/abs/arXiv:1010.1517} {arXiv:1010.1517} \BibitemShut
  {NoStop}%
\bibitem [{\citenamefont {{Bhardwaj}}\ \emph {et~al.}(2016)\citenamefont
  {{Bhardwaj}}, \citenamefont {{Gaiotto}},\ and\ \citenamefont
  {{Kapustin}}}]{BK160501640}%
  \BibitemOpen
  \bibfield  {author} {\bibinfo {author} {\bibfnamefont {L.}~\bibnamefont
  {{Bhardwaj}}}, \bibinfo {author} {\bibfnamefont {D.}~\bibnamefont
  {{Gaiotto}}}, \ and\ \bibinfo {author} {\bibfnamefont {A.}~\bibnamefont
  {{Kapustin}}},\ }\href@noop {} {\  (\bibinfo {year} {2016})},\ \Eprint
  {http://arxiv.org/abs/arXiv:1605.01640} {arXiv:1605.01640} \BibitemShut
  {NoStop}%
\bibitem [{\citenamefont {Walker}\ and\ \citenamefont {Wang}(2011)}]{WW1132}%
  \BibitemOpen
  \bibfield  {author} {\bibinfo {author} {\bibfnamefont {K.}~\bibnamefont
  {Walker}}\ and\ \bibinfo {author} {\bibfnamefont {Z.}~\bibnamefont {Wang}},\
  }\href@noop {} {\  (\bibinfo {year} {2011})},\ \Eprint
  {http://arxiv.org/abs/arXiv:1104.2632} {arXiv:1104.2632} \BibitemShut
  {NoStop}%
\bibitem [{\citenamefont {Lan}\ \emph {et~al.}(2017{\natexlab{a}})\citenamefont
  {Lan}, \citenamefont {Kong},\ and\ \citenamefont {Wen}}]{LW170404221}%
  \BibitemOpen
  \bibfield  {author} {\bibinfo {author} {\bibfnamefont {T.}~\bibnamefont
  {Lan}}, \bibinfo {author} {\bibfnamefont {L.}~\bibnamefont {Kong}}, \ and\
  \bibinfo {author} {\bibfnamefont {X.-G.}\ \bibnamefont {Wen}},\ }\href@noop
  {} {\  (\bibinfo {year} {2017}{\natexlab{a}})},\ \Eprint
  {http://arxiv.org/abs/arXiv:1704.04221} {arXiv:1704.04221} \BibitemShut
  {NoStop}%
\bibitem [{\citenamefont {Read}\ and\ \citenamefont
  {Chakraborty}(1989)}]{RC8933}%
  \BibitemOpen
  \bibfield  {author} {\bibinfo {author} {\bibfnamefont {N.}~\bibnamefont
  {Read}}\ and\ \bibinfo {author} {\bibfnamefont {B.}~\bibnamefont
  {Chakraborty}},\ }\href@noop {} {\bibfield  {journal} {\bibinfo  {journal}
  {Phys. Rev. B}\ }\textbf {\bibinfo {volume} {40}},\ \bibinfo {pages} {7133}
  (\bibinfo {year} {1989})}\BibitemShut {NoStop}%
\bibitem [{\citenamefont {Kitaev}\ and\ \citenamefont {Kong}(2012)}]{KK1251}%
  \BibitemOpen
  \bibfield  {author} {\bibinfo {author} {\bibfnamefont {A.}~\bibnamefont
  {Kitaev}}\ and\ \bibinfo {author} {\bibfnamefont {L.}~\bibnamefont {Kong}},\
  }\href {\doibase 10.1007/s00220-012-1500-5} {\bibfield  {journal} {\bibinfo
  {journal} {Commun. Math. Phys.}\ }\textbf {\bibinfo {volume} {313}},\
  \bibinfo {pages} {351 } (\bibinfo {year} {2012})},\ \Eprint
  {http://arxiv.org/abs/arXiv:1104.5047} {arXiv:1104.5047} \BibitemShut
  {NoStop}%
\bibitem [{\citenamefont {Lan}\ and\ \citenamefont {Wen}(2014)}]{LW1384}%
  \BibitemOpen
  \bibfield  {author} {\bibinfo {author} {\bibfnamefont {T.}~\bibnamefont
  {Lan}}\ and\ \bibinfo {author} {\bibfnamefont {X.-G.}\ \bibnamefont {Wen}},\
  }\href@noop {} {\bibfield  {journal} {\bibinfo  {journal} {Phys. Rev. B}\
  }\textbf {\bibinfo {volume} {90}},\ \bibinfo {pages} {115119} (\bibinfo
  {year} {2014})},\ \Eprint {http://arxiv.org/abs/arXiv:1311.1784}
  {arXiv:1311.1784} \BibitemShut {NoStop}%
\bibitem [{\citenamefont {Etingof}\ \emph {et~al.}(2005)\citenamefont
  {Etingof}, \citenamefont {Nikshych},\ and\ \citenamefont {Ostrik}}]{ENO0562}%
  \BibitemOpen
  \bibfield  {author} {\bibinfo {author} {\bibfnamefont {P.}~\bibnamefont
  {Etingof}}, \bibinfo {author} {\bibfnamefont {D.}~\bibnamefont {Nikshych}}, \
  and\ \bibinfo {author} {\bibfnamefont {V.}~\bibnamefont {Ostrik}},\ }\href
  {\doibase 10.4007/annals.2005.162.581} {\bibfield  {journal} {\bibinfo
  {journal} {Annals of Mathematics}\ }\textbf {\bibinfo {volume} {162}},\
  \bibinfo {pages} {581} (\bibinfo {year} {2005})}\BibitemShut {NoStop}%
\bibitem [{\citenamefont {Keski-Vakkuri}\ and\ \citenamefont
  {Wen}(1993)}]{KW9327}%
  \BibitemOpen
  \bibfield  {author} {\bibinfo {author} {\bibfnamefont {E.}~\bibnamefont
  {Keski-Vakkuri}}\ and\ \bibinfo {author} {\bibfnamefont {X.-G.}\ \bibnamefont
  {Wen}},\ }\href@noop {} {\bibfield  {journal} {\bibinfo  {journal} {Int. J.
  Mod. Phys. B}\ }\textbf {\bibinfo {volume} {7}},\ \bibinfo {pages} {4227}
  (\bibinfo {year} {1993})}\BibitemShut {NoStop}%
\bibitem [{\citenamefont {{Rowell}}\ \emph {et~al.}(2009)\citenamefont
  {{Rowell}}, \citenamefont {{Stong}},\ and\ \citenamefont {{Wang}}}]{RSW0777}%
  \BibitemOpen
  \bibfield  {author} {\bibinfo {author} {\bibfnamefont {E.}~\bibnamefont
  {{Rowell}}}, \bibinfo {author} {\bibfnamefont {R.}~\bibnamefont {{Stong}}}, \
  and\ \bibinfo {author} {\bibfnamefont {Z.}~\bibnamefont {{Wang}}},\
  }\href@noop {} {\bibfield  {journal} {\bibinfo  {journal} {Comm. Math.
  Phys.}\ }\textbf {\bibinfo {volume} {292}},\ \bibinfo {pages} {343} (\bibinfo
  {year} {2009})},\ \Eprint {http://arxiv.org/abs/arXiv:0712.1377}
  {arXiv:0712.1377} \BibitemShut {NoStop}%
\bibitem [{\citenamefont {{Wen}}(2016)}]{W150605768}%
  \BibitemOpen
  \bibfield  {author} {\bibinfo {author} {\bibfnamefont {X.-G.}\ \bibnamefont
  {{Wen}}},\ }\href {\doibase 10.1093/nsr/nwv077} {\bibfield  {journal}
  {\bibinfo  {journal} {Natl. Sci. Rev.}\ }\textbf {\bibinfo {volume} {3}},\
  \bibinfo {pages} {68} (\bibinfo {year} {2016})},\ \Eprint
  {http://arxiv.org/abs/arXiv:1506.05768} {arXiv:1506.05768} \BibitemShut
  {NoStop}%
\bibitem [{\citenamefont {{Moore}}\ and\ \citenamefont
  {{Seiberg}}(1989)}]{MS8977}%
  \BibitemOpen
  \bibfield  {author} {\bibinfo {author} {\bibfnamefont {G.}~\bibnamefont
  {{Moore}}}\ and\ \bibinfo {author} {\bibfnamefont {N.}~\bibnamefont
  {{Seiberg}}},\ }\href {\doibase 10.1007/BF01238857} {\bibfield  {journal}
  {\bibinfo  {journal} {Communications in Mathematical Physics}\ }\textbf
  {\bibinfo {volume} {123}},\ \bibinfo {pages} {177} (\bibinfo {year}
  {1989})}\BibitemShut {NoStop}%
\bibitem [{\citenamefont {Kong}\ and\ \citenamefont {Wen}(2014)}]{KW1458}%
  \BibitemOpen
  \bibfield  {author} {\bibinfo {author} {\bibfnamefont {L.}~\bibnamefont
  {Kong}}\ and\ \bibinfo {author} {\bibfnamefont {X.-G.}\ \bibnamefont {Wen}},\
  }\href@noop {} {\  (\bibinfo {year} {2014})},\ \Eprint
  {http://arxiv.org/abs/arXiv:1405.5858} {arXiv:1405.5858} \BibitemShut
  {NoStop}%
\bibitem [{\citenamefont {Kapustin}(2014{\natexlab{a}})}]{K1459}%
  \BibitemOpen
  \bibfield  {author} {\bibinfo {author} {\bibfnamefont {A.}~\bibnamefont
  {Kapustin}},\ }\href@noop {} {\  (\bibinfo {year} {2014}{\natexlab{a}})},\
  \Eprint {http://arxiv.org/abs/arXiv:1404.6659} {arXiv:1404.6659} \BibitemShut
  {NoStop}%
\bibitem [{\citenamefont {{Freed}}(2014)}]{F1478}%
  \BibitemOpen
  \bibfield  {author} {\bibinfo {author} {\bibfnamefont {D.~S.}\ \bibnamefont
  {{Freed}}},\ }\href@noop {} {\  (\bibinfo {year} {2014})},\ \Eprint
  {http://arxiv.org/abs/arXiv:1406.7278} {arXiv:1406.7278} \BibitemShut
  {NoStop}%
\bibitem [{\citenamefont {Wen}(2013{\natexlab{b}})}]{W1313}%
  \BibitemOpen
  \bibfield  {author} {\bibinfo {author} {\bibfnamefont {X.-G.}\ \bibnamefont
  {Wen}},\ }\href@noop {} {\bibfield  {journal} {\bibinfo  {journal} {Phys.
  Rev. D}\ }\textbf {\bibinfo {volume} {88}},\ \bibinfo {pages} {045013}
  (\bibinfo {year} {2013}{\natexlab{b}})},\ \Eprint
  {http://arxiv.org/abs/arXiv:1303.1803} {arXiv:1303.1803} \BibitemShut
  {NoStop}%
\bibitem [{\citenamefont {Ivanov}(2001)}]{I0168}%
  \BibitemOpen
  \bibfield  {author} {\bibinfo {author} {\bibfnamefont {D.~A.}\ \bibnamefont
  {Ivanov}},\ }\href {\doibase 10.1103/PhysRevLett.86.268} {\bibfield
  {journal} {\bibinfo  {journal} {Phys. Rev. Lett.}\ }\textbf {\bibinfo
  {volume} {86}},\ \bibinfo {pages} {268} (\bibinfo {year} {2001})},\ \Eprint
  {http://arxiv.org/abs/cond-mat/0005069} {cond-mat/0005069} \BibitemShut
  {NoStop}%
\bibitem [{\citenamefont {Mermin}\ and\ \citenamefont {Wagner}(1966)}]{MW6633}%
  \BibitemOpen
  \bibfield  {author} {\bibinfo {author} {\bibfnamefont {N.~D.}\ \bibnamefont
  {Mermin}}\ and\ \bibinfo {author} {\bibfnamefont {H.}~\bibnamefont
  {Wagner}},\ }\href {\doibase 10.1103/PhysRevLett.17.1133} {\bibfield
  {journal} {\bibinfo  {journal} {\prl}\ }\textbf {\bibinfo {volume} {17}},\
  \bibinfo {pages} {1133–1136} (\bibinfo {year} {1966})}\BibitemShut
  {NoStop}%
\bibitem [{\citenamefont {Lieb}\ \emph {et~al.}(1961)\citenamefont {Lieb},
  \citenamefont {Schultz},\ and\ \citenamefont {Mattis}}]{LSM6107}%
  \BibitemOpen
  \bibfield  {author} {\bibinfo {author} {\bibfnamefont {E.~H.}\ \bibnamefont
  {Lieb}}, \bibinfo {author} {\bibfnamefont {T.~D.}\ \bibnamefont {Schultz}}, \
  and\ \bibinfo {author} {\bibfnamefont {D.~C.}\ \bibnamefont {Mattis}},\
  }\href@noop {} {\bibfield  {journal} {\bibinfo  {journal} {Ann. Phys. (N.Y)}\
  }\textbf {\bibinfo {volume} {16}},\ \bibinfo {pages} {407} (\bibinfo {year}
  {1961})}\BibitemShut {NoStop}%
\bibitem [{\citenamefont {Haldane}(1983)}]{H8364}%
  \BibitemOpen
  \bibfield  {author} {\bibinfo {author} {\bibfnamefont {F.~D.~M.}\
  \bibnamefont {Haldane}},\ }\href@noop {} {\bibfield  {journal} {\bibinfo
  {journal} {Physics Letters A}\ }\textbf {\bibinfo {volume} {93}},\ \bibinfo
  {pages} {464} (\bibinfo {year} {1983})}\BibitemShut {NoStop}%
\bibitem [{\citenamefont {Affleck}\ \emph
  {et~al.}(1988{\natexlab{b}})\citenamefont {Affleck}, \citenamefont {Kennedy},
  \citenamefont {Lieb},\ and\ \citenamefont {Tasaki}}]{AKL8877}%
  \BibitemOpen
  \bibfield  {author} {\bibinfo {author} {\bibfnamefont {I.}~\bibnamefont
  {Affleck}}, \bibinfo {author} {\bibfnamefont {T.}~\bibnamefont {Kennedy}},
  \bibinfo {author} {\bibfnamefont {E.~H.}\ \bibnamefont {Lieb}}, \ and\
  \bibinfo {author} {\bibfnamefont {H.}~\bibnamefont {Tasaki}},\ }\href@noop {}
  {\bibfield  {journal} {\bibinfo  {journal} {Commun. Math. Phys.}\ }\textbf
  {\bibinfo {volume} {115}},\ \bibinfo {pages} {477} (\bibinfo {year}
  {1988}{\natexlab{b}})}\BibitemShut {NoStop}%
\bibitem [{\citenamefont {Gu}\ and\ \citenamefont {Wen}(2009)}]{GW0931}%
  \BibitemOpen
  \bibfield  {author} {\bibinfo {author} {\bibfnamefont {Z.-C.}\ \bibnamefont
  {Gu}}\ and\ \bibinfo {author} {\bibfnamefont {X.-G.}\ \bibnamefont {Wen}},\
  }\href@noop {} {\bibfield  {journal} {\bibinfo  {journal} {Phys. Rev. B}\
  }\textbf {\bibinfo {volume} {80}},\ \bibinfo {pages} {155131} (\bibinfo
  {year} {2009})},\ \Eprint {http://arxiv.org/abs/arXiv:0903.1069}
  {arXiv:0903.1069} \BibitemShut {NoStop}%
\bibitem [{\citenamefont {White}(1992)}]{W9263}%
  \BibitemOpen
  \bibfield  {author} {\bibinfo {author} {\bibfnamefont {S.~R.}\ \bibnamefont
  {White}},\ }\href@noop {} {\bibfield  {journal} {\bibinfo  {journal} {Phys.
  Rev. Lett.}\ }\textbf {\bibinfo {volume} {69}},\ \bibinfo {pages} {2863}
  (\bibinfo {year} {1992})}\BibitemShut {NoStop}%
\bibitem [{\citenamefont {Pollmann}\ \emph {et~al.}(2012)\citenamefont
  {Pollmann}, \citenamefont {Berg}, \citenamefont {Turner},\ and\ \citenamefont
  {Oshikawa}}]{PBT1225}%
  \BibitemOpen
  \bibfield  {author} {\bibinfo {author} {\bibfnamefont {F.}~\bibnamefont
  {Pollmann}}, \bibinfo {author} {\bibfnamefont {E.}~\bibnamefont {Berg}},
  \bibinfo {author} {\bibfnamefont {A.~M.}\ \bibnamefont {Turner}}, \ and\
  \bibinfo {author} {\bibfnamefont {M.}~\bibnamefont {Oshikawa}},\ }\href
  {\doibase 10.1103/PhysRevB.85.075125} {\bibfield  {journal} {\bibinfo
  {journal} {Phys. Rev. B}\ }\textbf {\bibinfo {volume} {85}},\ \bibinfo
  {pages} {075125} (\bibinfo {year} {2012})},\ \Eprint
  {http://arxiv.org/abs/arXiv:0909.4059} {arXiv:0909.4059} \BibitemShut
  {NoStop}%
\bibitem [{\citenamefont {Pollmann}\ \emph {et~al.}(2010)\citenamefont
  {Pollmann}, \citenamefont {Berg}, \citenamefont {Turner},\ and\ \citenamefont
  {Oshikawa}}]{PBT1039}%
  \BibitemOpen
  \bibfield  {author} {\bibinfo {author} {\bibfnamefont {F.}~\bibnamefont
  {Pollmann}}, \bibinfo {author} {\bibfnamefont {E.}~\bibnamefont {Berg}},
  \bibinfo {author} {\bibfnamefont {A.~M.}\ \bibnamefont {Turner}}, \ and\
  \bibinfo {author} {\bibfnamefont {M.}~\bibnamefont {Oshikawa}},\ }\href
  {\doibase 10.1103/PhysRevB.81.064439} {\bibfield  {journal} {\bibinfo
  {journal} {Phys. Rev. B}\ }\textbf {\bibinfo {volume} {81}},\ \bibinfo
  {pages} {064439} (\bibinfo {year} {2010})},\ \Eprint
  {http://arxiv.org/abs/arXiv:0910.1811} {arXiv:0910.1811} \BibitemShut
  {NoStop}%
\bibitem [{\citenamefont {Chen}\ \emph
  {et~al.}(2011{\natexlab{a}})\citenamefont {Chen}, \citenamefont {Gu},\ and\
  \citenamefont {Wen}}]{CGW1107}%
  \BibitemOpen
  \bibfield  {author} {\bibinfo {author} {\bibfnamefont {X.}~\bibnamefont
  {Chen}}, \bibinfo {author} {\bibfnamefont {Z.-C.}\ \bibnamefont {Gu}}, \ and\
  \bibinfo {author} {\bibfnamefont {X.-G.}\ \bibnamefont {Wen}},\ }\href@noop
  {} {\bibfield  {journal} {\bibinfo  {journal} {Phys. Rev. B}\ }\textbf
  {\bibinfo {volume} {83}},\ \bibinfo {pages} {035107} (\bibinfo {year}
  {2011}{\natexlab{a}})},\ \Eprint {http://arxiv.org/abs/arXiv:1008.3745}
  {arXiv:1008.3745} \BibitemShut {NoStop}%
\bibitem [{\citenamefont {Schuch}\ \emph {et~al.}(2011)\citenamefont {Schuch},
  \citenamefont {Perez-Garcia},\ and\ \citenamefont {Cirac}}]{SPC1139}%
  \BibitemOpen
  \bibfield  {author} {\bibinfo {author} {\bibfnamefont {N.}~\bibnamefont
  {Schuch}}, \bibinfo {author} {\bibfnamefont {D.}~\bibnamefont
  {Perez-Garcia}}, \ and\ \bibinfo {author} {\bibfnamefont {I.}~\bibnamefont
  {Cirac}},\ }\href@noop {} {\bibfield  {journal} {\bibinfo  {journal} {Phys.
  Rev. B}\ }\textbf {\bibinfo {volume} {84}},\ \bibinfo {pages} {165139}
  (\bibinfo {year} {2011})},\ \Eprint {http://arxiv.org/abs/arXiv:1010.3732}
  {arXiv:1010.3732} \BibitemShut {NoStop}%
\bibitem [{\citenamefont {Chen}\ \emph
  {et~al.}(2013{\natexlab{a}})\citenamefont {Chen}, \citenamefont {Gu},
  \citenamefont {Liu},\ and\ \citenamefont {Wen}}]{CGL1172}%
  \BibitemOpen
  \bibfield  {author} {\bibinfo {author} {\bibfnamefont {X.}~\bibnamefont
  {Chen}}, \bibinfo {author} {\bibfnamefont {Z.-C.}\ \bibnamefont {Gu}},
  \bibinfo {author} {\bibfnamefont {Z.-X.}\ \bibnamefont {Liu}}, \ and\
  \bibinfo {author} {\bibfnamefont {X.-G.}\ \bibnamefont {Wen}},\ }\href@noop
  {} {\bibfield  {journal} {\bibinfo  {journal} {Phys. Rev. B}\ }\textbf
  {\bibinfo {volume} {87}},\ \bibinfo {pages} {155114} (\bibinfo {year}
  {2013}{\natexlab{a}})},\ \Eprint {http://arxiv.org/abs/arXiv:1106.4772}
  {arXiv:1106.4772} \BibitemShut {NoStop}%
\bibitem [{\citenamefont {Chen}\ \emph
  {et~al.}(2011{\natexlab{b}})\citenamefont {Chen}, \citenamefont {Liu},\ and\
  \citenamefont {Wen}}]{CLW1141}%
  \BibitemOpen
  \bibfield  {author} {\bibinfo {author} {\bibfnamefont {X.}~\bibnamefont
  {Chen}}, \bibinfo {author} {\bibfnamefont {Z.-X.}\ \bibnamefont {Liu}}, \
  and\ \bibinfo {author} {\bibfnamefont {X.-G.}\ \bibnamefont {Wen}},\
  }\href@noop {} {\bibfield  {journal} {\bibinfo  {journal} {Phys. Rev. B}\
  }\textbf {\bibinfo {volume} {84}},\ \bibinfo {pages} {235141} (\bibinfo
  {year} {2011}{\natexlab{b}})},\ \Eprint
  {http://arxiv.org/abs/arXiv:1106.4752} {arXiv:1106.4752} \BibitemShut
  {NoStop}%
\bibitem [{\citenamefont {{Qi}}\ \emph {et~al.}(2010)\citenamefont {{Qi}},
  \citenamefont {{Hughes}},\ and\ \citenamefont {{Zhang}}}]{QZ09083550}%
  \BibitemOpen
  \bibfield  {author} {\bibinfo {author} {\bibfnamefont {X.-L.}\ \bibnamefont
  {{Qi}}}, \bibinfo {author} {\bibfnamefont {T.~L.}\ \bibnamefont {{Hughes}}},
  \ and\ \bibinfo {author} {\bibfnamefont {S.-C.}\ \bibnamefont {{Zhang}}},\
  }\href {\doibase 10.1103/PhysRevB.81.134508} {\bibfield  {journal} {\bibinfo
  {journal} {Phys. Rev. B}\ }\textbf {\bibinfo {volume} {81}},\ \bibinfo
  {pages} {134508} (\bibinfo {year} {2010})},\ \Eprint
  {http://arxiv.org/abs/arXiv:0908.3550} {arXiv:0908.3550} \BibitemShut
  {NoStop}%
\bibitem [{\citenamefont {Kitaev}(2009)}]{K0986}%
  \BibitemOpen
  \bibfield  {author} {\bibinfo {author} {\bibfnamefont {A.}~\bibnamefont
  {Kitaev}},\ }in\ \href@noop {} {\emph {\bibinfo {booktitle} {Advances in
  Theoretical Physics: Landau Memorial Conference, Chernogolovka, Russia,
  2008}}},\ Vol.\ \bibinfo {volume} {AIP Conf. Proc. No. 1134},\ \bibinfo
  {editor} {edited by\ \bibinfo {editor} {\bibfnamefont {V.}~\bibnamefont
  {Lebedev}}\ and\ \bibinfo {editor} {\bibfnamefont {M.}~\bibnamefont
  {Feigel’man}}}\ (\bibinfo  {publisher} {AIP},\ \bibinfo {address}
  {Melville, NY},\ \bibinfo {year} {2009})\ p.~\bibinfo {pages} {22},\ \Eprint
  {http://arxiv.org/abs/arXiv:0901.2686} {arXiv:0901.2686} \BibitemShut
  {NoStop}%
\bibitem [{\citenamefont {Wen}(2011)}]{W1103}%
  \BibitemOpen
  \bibfield  {author} {\bibinfo {author} {\bibfnamefont {X.-G.}\ \bibnamefont
  {Wen}},\ }\href@noop {} {\bibfield  {journal} {\bibinfo  {journal} {Phys.
  Rev. B}\ }\textbf {\bibinfo {volume} {85}},\ \bibinfo {pages} {085103}
  (\bibinfo {year} {2011})},\ \Eprint {http://arxiv.org/abs/arXiv:1111.6341}
  {arXiv:1111.6341} \BibitemShut {NoStop}%
\bibitem [{\citenamefont {Chen}\ \emph
  {et~al.}(2011{\natexlab{c}})\citenamefont {Chen}, \citenamefont {Gu},\ and\
  \citenamefont {Wen}}]{CGW1128}%
  \BibitemOpen
  \bibfield  {author} {\bibinfo {author} {\bibfnamefont {X.}~\bibnamefont
  {Chen}}, \bibinfo {author} {\bibfnamefont {Z.-C.}\ \bibnamefont {Gu}}, \ and\
  \bibinfo {author} {\bibfnamefont {X.-G.}\ \bibnamefont {Wen}},\ }\href@noop
  {} {\bibfield  {journal} {\bibinfo  {journal} {Phys. Rev. B}\ }\textbf
  {\bibinfo {volume} {84}},\ \bibinfo {pages} {235128} (\bibinfo {year}
  {2011}{\natexlab{c}})},\ \Eprint {http://arxiv.org/abs/arXiv:1103.3323}
  {arXiv:1103.3323} \BibitemShut {NoStop}%
\bibitem [{\citenamefont {Lu}\ and\ \citenamefont {Vishwanath}(2012)}]{LV1219}%
  \BibitemOpen
  \bibfield  {author} {\bibinfo {author} {\bibfnamefont {Y.-M.}\ \bibnamefont
  {Lu}}\ and\ \bibinfo {author} {\bibfnamefont {A.}~\bibnamefont
  {Vishwanath}},\ }\href {\doibase 10.1103/PhysRevB.86.125119} {\bibfield
  {journal} {\bibinfo  {journal} {Phys. Rev. B}\ }\textbf {\bibinfo {volume}
  {86}},\ \bibinfo {pages} {125119} (\bibinfo {year} {2012})},\ \Eprint
  {http://arxiv.org/abs/arXiv:1205.3156} {arXiv:1205.3156} \BibitemShut
  {NoStop}%
\bibitem [{\citenamefont {Liu}\ and\ \citenamefont {Wen}(2013)}]{LW1305}%
  \BibitemOpen
  \bibfield  {author} {\bibinfo {author} {\bibfnamefont {Z.-X.}\ \bibnamefont
  {Liu}}\ and\ \bibinfo {author} {\bibfnamefont {X.-G.}\ \bibnamefont {Wen}},\
  }\href@noop {} {\bibfield  {journal} {\bibinfo  {journal} {Phys. Rev. Lett.}\
  }\textbf {\bibinfo {volume} {110}},\ \bibinfo {pages} {067205} (\bibinfo
  {year} {2013})},\ \Eprint {http://arxiv.org/abs/arXiv:1205.7024}
  {arXiv:1205.7024} \BibitemShut {NoStop}%
\bibitem [{\citenamefont {Kane}\ and\ \citenamefont
  {Mele}(2005{\natexlab{a}})}]{KM0501}%
  \BibitemOpen
  \bibfield  {author} {\bibinfo {author} {\bibfnamefont {C.~L.}\ \bibnamefont
  {Kane}}\ and\ \bibinfo {author} {\bibfnamefont {E.~J.}\ \bibnamefont
  {Mele}},\ }\href@noop {} {\bibfield  {journal} {\bibinfo  {journal} {Phys.
  Rev. Lett.}\ }\textbf {\bibinfo {volume} {95}},\ \bibinfo {pages} {226801}
  (\bibinfo {year} {2005}{\natexlab{a}})},\ \Eprint
  {http://arxiv.org/abs/cond-mat/0411737} {cond-mat/0411737} \BibitemShut
  {NoStop}%
\bibitem [{\citenamefont {Bernevig}\ and\ \citenamefont
  {Zhang}(2006)}]{BZ0602}%
  \BibitemOpen
  \bibfield  {author} {\bibinfo {author} {\bibfnamefont {B.~A.}\ \bibnamefont
  {Bernevig}}\ and\ \bibinfo {author} {\bibfnamefont {S.-C.}\ \bibnamefont
  {Zhang}},\ }\href@noop {} {\bibfield  {journal} {\bibinfo  {journal} {Phys.
  Rev. Lett.}\ }\textbf {\bibinfo {volume} {96}},\ \bibinfo {pages} {106802}
  (\bibinfo {year} {2006})},\ \Eprint {http://arxiv.org/abs/cond-mat/0504147}
  {cond-mat/0504147} \BibitemShut {NoStop}%
\bibitem [{\citenamefont {Kane}\ and\ \citenamefont
  {Mele}(2005{\natexlab{b}})}]{KM0502}%
  \BibitemOpen
  \bibfield  {author} {\bibinfo {author} {\bibfnamefont {C.~L.}\ \bibnamefont
  {Kane}}\ and\ \bibinfo {author} {\bibfnamefont {E.~J.}\ \bibnamefont
  {Mele}},\ }\href@noop {} {\bibfield  {journal} {\bibinfo  {journal} {Phys.
  Rev. Lett.}\ }\textbf {\bibinfo {volume} {95}},\ \bibinfo {pages} {146802}
  (\bibinfo {year} {2005}{\natexlab{b}})},\ \Eprint
  {http://arxiv.org/abs/cond-mat/0506581} {cond-mat/0506581} \BibitemShut
  {NoStop}%
\bibitem [{\citenamefont {Roy}(2006)}]{R0664}%
  \BibitemOpen
  \bibfield  {author} {\bibinfo {author} {\bibfnamefont {R.}~\bibnamefont
  {Roy}},\ }\href@noop {} {\  (\bibinfo {year} {2006})},\ \Eprint
  {http://arxiv.org/abs/cond-mat/0608064} {cond-mat/0608064} \BibitemShut
  {NoStop}%
\bibitem [{\citenamefont {Qi}\ \emph {et~al.}(2009)\citenamefont {Qi},
  \citenamefont {Hughes}, \citenamefont {Raghu},\ and\ \citenamefont
  {Zhang}}]{QHR0901}%
  \BibitemOpen
  \bibfield  {author} {\bibinfo {author} {\bibfnamefont {X.-L.}\ \bibnamefont
  {Qi}}, \bibinfo {author} {\bibfnamefont {T.~L.}\ \bibnamefont {Hughes}},
  \bibinfo {author} {\bibfnamefont {S.}~\bibnamefont {Raghu}}, \ and\ \bibinfo
  {author} {\bibfnamefont {S.-C.}\ \bibnamefont {Zhang}},\ }\href@noop {}
  {\bibfield  {journal} {\bibinfo  {journal} {Phys. Rev. Lett.}\ }\textbf
  {\bibinfo {volume} {102}},\ \bibinfo {pages} {187001} (\bibinfo {year}
  {2009})},\ \Eprint {http://arxiv.org/abs/arXiv:0803.3614} {arXiv:0803.3614}
  \BibitemShut {NoStop}%
\bibitem [{\citenamefont {Sato}\ and\ \citenamefont {Fujimoto}(2009)}]{SF0904}%
  \BibitemOpen
  \bibfield  {author} {\bibinfo {author} {\bibfnamefont {M.}~\bibnamefont
  {Sato}}\ and\ \bibinfo {author} {\bibfnamefont {S.}~\bibnamefont
  {Fujimoto}},\ }\href@noop {} {\bibfield  {journal} {\bibinfo  {journal}
  {Phys. Rev. B}\ }\textbf {\bibinfo {volume} {79}},\ \bibinfo {pages} {094504}
  (\bibinfo {year} {2009})},\ \Eprint {http://arxiv.org/abs/arXiv:0811.3864}
  {arXiv:0811.3864} \BibitemShut {NoStop}%
\bibitem [{\citenamefont {{Lan}}\ \emph {et~al.}(2016)\citenamefont {{Lan}},
  \citenamefont {{Kong}},\ and\ \citenamefont {{Wen}}}]{LW150704673}%
  \BibitemOpen
  \bibfield  {author} {\bibinfo {author} {\bibfnamefont {T.}~\bibnamefont
  {{Lan}}}, \bibinfo {author} {\bibfnamefont {L.}~\bibnamefont {{Kong}}}, \
  and\ \bibinfo {author} {\bibfnamefont {X.-G.}\ \bibnamefont {{Wen}}},\ }\href
  {\doibase 10.1103/PhysRevB.94.155113} {\bibfield  {journal} {\bibinfo
  {journal} {\prb}\ }\textbf {\bibinfo {volume} {94}},\ \bibinfo {pages}
  {155113} (\bibinfo {year} {2016})},\ \Eprint
  {http://arxiv.org/abs/arXiv:1507.04673} {arXiv:1507.04673} \BibitemShut
  {NoStop}%
\bibitem [{\citenamefont {Lan}\ \emph {et~al.}(2016)\citenamefont {Lan},
  \citenamefont {Kong},\ and\ \citenamefont {Wen}}]{LW160205946}%
  \BibitemOpen
  \bibfield  {author} {\bibinfo {author} {\bibfnamefont {T.}~\bibnamefont
  {Lan}}, \bibinfo {author} {\bibfnamefont {L.}~\bibnamefont {Kong}}, \ and\
  \bibinfo {author} {\bibfnamefont {X.-G.}\ \bibnamefont {Wen}},\ }\href@noop
  {} {\  (\bibinfo {year} {2016})},\ \Eprint
  {http://arxiv.org/abs/arXiv:1602.05946} {arXiv:1602.05946} \BibitemShut
  {NoStop}%
\bibitem [{\citenamefont {Wang}\ and\ \citenamefont {Senthil}(2013)}]{WS1334}%
  \BibitemOpen
  \bibfield  {author} {\bibinfo {author} {\bibfnamefont {C.}~\bibnamefont
  {Wang}}\ and\ \bibinfo {author} {\bibfnamefont {T.}~\bibnamefont {Senthil}},\
  }\href@noop {} {\bibfield  {journal} {\bibinfo  {journal} {Phys. Rev. B}\
  }\textbf {\bibinfo {volume} {87}},\ \bibinfo {pages} {235122} (\bibinfo
  {year} {2013})},\ \Eprint {http://arxiv.org/abs/arXiv:1302.6234}
  {arXiv:1302.6234} \BibitemShut {NoStop}%
\bibitem [{\citenamefont {Metlitski}\ \emph {et~al.}(2013)\citenamefont
  {Metlitski}, \citenamefont {Kane},\ and\ \citenamefont {Fisher}}]{MKF1331}%
  \BibitemOpen
  \bibfield  {author} {\bibinfo {author} {\bibfnamefont {M.~A.}\ \bibnamefont
  {Metlitski}}, \bibinfo {author} {\bibfnamefont {C.~L.}\ \bibnamefont {Kane}},
  \ and\ \bibinfo {author} {\bibfnamefont {M.~P.~A.}\ \bibnamefont {Fisher}},\
  }\href@noop {} {\bibfield  {journal} {\bibinfo  {journal} {Phys. Rev. B}\
  }\textbf {\bibinfo {volume} {88}},\ \bibinfo {pages} {035131} (\bibinfo
  {year} {2013})},\ \Eprint {http://arxiv.org/abs/arXiv:1302.6535}
  {arXiv:1302.6535} \BibitemShut {NoStop}%
\bibitem [{\citenamefont {Vishwanath}\ and\ \citenamefont
  {Senthil}(2013)}]{VS1306}%
  \BibitemOpen
  \bibfield  {author} {\bibinfo {author} {\bibfnamefont {A.}~\bibnamefont
  {Vishwanath}}\ and\ \bibinfo {author} {\bibfnamefont {T.}~\bibnamefont
  {Senthil}},\ }\href@noop {} {\bibfield  {journal} {\bibinfo  {journal} {Phys.
  Rev. X}\ }\textbf {\bibinfo {volume} {3}},\ \bibinfo {pages} {011016}
  (\bibinfo {year} {2013})},\ \Eprint {http://arxiv.org/abs/arXiv:1209.3058}
  {arXiv:1209.3058} \BibitemShut {NoStop}%
\bibitem [{\citenamefont {Moore}\ and\ \citenamefont {Balents}(2007)}]{MB0706}%
  \BibitemOpen
  \bibfield  {author} {\bibinfo {author} {\bibfnamefont {J.~E.}\ \bibnamefont
  {Moore}}\ and\ \bibinfo {author} {\bibfnamefont {L.}~\bibnamefont
  {Balents}},\ }\href@noop {} {\bibfield  {journal} {\bibinfo  {journal} {Phys.
  Rev. B}\ }\textbf {\bibinfo {volume} {75}},\ \bibinfo {pages} {121306}
  (\bibinfo {year} {2007})},\ \Eprint {http://arxiv.org/abs/cond-mat/0607314}
  {cond-mat/0607314} \BibitemShut {NoStop}%
\bibitem [{\citenamefont {Roy}(2009)}]{R0922}%
  \BibitemOpen
  \bibfield  {author} {\bibinfo {author} {\bibfnamefont {R.}~\bibnamefont
  {Roy}},\ }\href {\doibase 10.1103/PhysRevB.79.195322} {\bibfield  {journal}
  {\bibinfo  {journal} {Phys. Rev. B}\ }\textbf {\bibinfo {volume} {79}},\
  \bibinfo {pages} {195322} (\bibinfo {year} {2009})},\ \Eprint
  {http://arxiv.org/abs/cond-mat/0607531} {cond-mat/0607531} \BibitemShut
  {NoStop}%
\bibitem [{\citenamefont {Fu}\ \emph {et~al.}(2007)\citenamefont {Fu},
  \citenamefont {Kane},\ and\ \citenamefont {Mele}}]{FKM0703}%
  \BibitemOpen
  \bibfield  {author} {\bibinfo {author} {\bibfnamefont {L.}~\bibnamefont
  {Fu}}, \bibinfo {author} {\bibfnamefont {C.~L.}\ \bibnamefont {Kane}}, \ and\
  \bibinfo {author} {\bibfnamefont {E.~J.}\ \bibnamefont {Mele}},\ }\href@noop
  {} {\bibfield  {journal} {\bibinfo  {journal} {Phys. Rev. Lett.}\ }\textbf
  {\bibinfo {volume} {98}},\ \bibinfo {pages} {106803} (\bibinfo {year}
  {2007})},\ \Eprint {http://arxiv.org/abs/cond-mat/0607699} {cond-mat/0607699}
  \BibitemShut {NoStop}%
\bibitem [{\citenamefont {Qi}\ \emph {et~al.}(2008)\citenamefont {Qi},
  \citenamefont {Hughes},\ and\ \citenamefont {Zhang}}]{QHZ0824}%
  \BibitemOpen
  \bibfield  {author} {\bibinfo {author} {\bibfnamefont {X.-L.}\ \bibnamefont
  {Qi}}, \bibinfo {author} {\bibfnamefont {T.}~\bibnamefont {Hughes}}, \ and\
  \bibinfo {author} {\bibfnamefont {S.-C.}\ \bibnamefont {Zhang}},\ }\href@noop
  {} {\bibfield  {journal} {\bibinfo  {journal} {Phys. Rev. B}\ }\textbf
  {\bibinfo {volume} {78}},\ \bibinfo {pages} {195424} (\bibinfo {year}
  {2008})},\ \Eprint {http://arxiv.org/abs/arXiv:0802.3537} {arXiv:0802.3537}
  \BibitemShut {NoStop}%
\bibitem [{\citenamefont {{Wang}}\ and\ \citenamefont
  {{Senthil}}(2014)}]{WS14011142}%
  \BibitemOpen
  \bibfield  {author} {\bibinfo {author} {\bibfnamefont {C.}~\bibnamefont
  {{Wang}}}\ and\ \bibinfo {author} {\bibfnamefont {T.}~\bibnamefont
  {{Senthil}}},\ }\href {\doibase 10.1103/PhysRevB.89.195124} {\bibfield
  {journal} {\bibinfo  {journal} {\prb}\ }\textbf {\bibinfo {volume} {89}},\
  \bibinfo {pages} {195124} (\bibinfo {year} {2014})},\ \Eprint
  {http://arxiv.org/abs/arXiv:1401.1142} {arXiv:1401.1142} \BibitemShut
  {NoStop}%
\bibitem [{\citenamefont {{Kapustin}}\ \emph {et~al.}(2015)\citenamefont
  {{Kapustin}}, \citenamefont {{Thorngren}}, \citenamefont {{Turzillo}},\ and\
  \citenamefont {{Wang}}}]{KTT1429}%
  \BibitemOpen
  \bibfield  {author} {\bibinfo {author} {\bibfnamefont {A.}~\bibnamefont
  {{Kapustin}}}, \bibinfo {author} {\bibfnamefont {R.}~\bibnamefont
  {{Thorngren}}}, \bibinfo {author} {\bibfnamefont {A.}~\bibnamefont
  {{Turzillo}}}, \ and\ \bibinfo {author} {\bibfnamefont {Z.}~\bibnamefont
  {{Wang}}},\ }\href@noop {} {\bibfield  {journal} {\bibinfo  {journal}
  {Journal of High Energy Physics}\ }\textbf {\bibinfo {volume} {2015}},\
  \bibinfo {pages} {52} (\bibinfo {year} {2015})},\ \Eprint
  {http://arxiv.org/abs/arXiv:1406.7329} {arXiv:1406.7329} \BibitemShut
  {NoStop}%
\bibitem [{\citenamefont {Levin}\ and\ \citenamefont {Gu}(2012)}]{LG1220}%
  \BibitemOpen
  \bibfield  {author} {\bibinfo {author} {\bibfnamefont {M.}~\bibnamefont
  {Levin}}\ and\ \bibinfo {author} {\bibfnamefont {Z.-C.}\ \bibnamefont {Gu}},\
  }\href@noop {} {\bibfield  {journal} {\bibinfo  {journal} {Phys. Rev. B}\
  }\textbf {\bibinfo {volume} {86}},\ \bibinfo {pages} {115109} (\bibinfo
  {year} {2012})},\ \Eprint {http://arxiv.org/abs/arXiv:1202.3120}
  {arXiv:1202.3120} \BibitemShut {NoStop}%
\bibitem [{\citenamefont {Wen}(2016)}]{W161201418}%
  \BibitemOpen
  \bibfield  {author} {\bibinfo {author} {\bibfnamefont {X.-G.}\ \bibnamefont
  {Wen}},\ }\href@noop {} {\  (\bibinfo {year} {2016})},\ \Eprint
  {http://arxiv.org/abs/arXiv:1612.01418} {arXiv:1612.01418} \BibitemShut
  {NoStop}%
\bibitem [{\citenamefont {Lan}\ \emph {et~al.}(2017{\natexlab{b}})\citenamefont
  {Lan}, \citenamefont {Kong},\ and\ \citenamefont {Wen}}]{LW160205936}%
  \BibitemOpen
  \bibfield  {author} {\bibinfo {author} {\bibfnamefont {T.}~\bibnamefont
  {Lan}}, \bibinfo {author} {\bibfnamefont {L.}~\bibnamefont {Kong}}, \ and\
  \bibinfo {author} {\bibfnamefont {X.-G.}\ \bibnamefont {Wen}},\ }\href
  {\doibase doi:10.1007/s00220-016-2748-y} {\bibfield  {journal} {\bibinfo
  {journal} {Commun. Math. Phys.}\ }\textbf {\bibinfo {volume} {351}},\
  \bibinfo {pages} {709} (\bibinfo {year} {2017}{\natexlab{b}})},\ \Eprint
  {http://arxiv.org/abs/arXiv:1602.05936} {arXiv:1602.05936} \BibitemShut
  {NoStop}%
\bibitem [{\citenamefont {Drinfeld}\ \emph {et~al.}(2007)\citenamefont
  {Drinfeld}, \citenamefont {Gelaki}, \citenamefont {Nikshych},\ and\
  \citenamefont {Ostrik}}]{DO07040195}%
  \BibitemOpen
  \bibfield  {author} {\bibinfo {author} {\bibfnamefont {V.}~\bibnamefont
  {Drinfeld}}, \bibinfo {author} {\bibfnamefont {S.}~\bibnamefont {Gelaki}},
  \bibinfo {author} {\bibfnamefont {D.}~\bibnamefont {Nikshych}}, \ and\
  \bibinfo {author} {\bibfnamefont {V.}~\bibnamefont {Ostrik}},\ }\href@noop {}
  {\  (\bibinfo {year} {2007})},\ \Eprint
  {http://arxiv.org/abs/arXiv:0704.0195} {arXiv:0704.0195} \BibitemShut
  {NoStop}%
\bibitem [{\citenamefont {Wang}\ and\ \citenamefont {Levin}(2014)}]{WL1437}%
  \BibitemOpen
  \bibfield  {author} {\bibinfo {author} {\bibfnamefont {C.}~\bibnamefont
  {Wang}}\ and\ \bibinfo {author} {\bibfnamefont {M.}~\bibnamefont {Levin}},\
  }\href@noop {} {\bibfield  {journal} {\bibinfo  {journal} {Phys. Rev. Lett.}\
  }\textbf {\bibinfo {volume} {113}},\ \bibinfo {pages} {080403} (\bibinfo
  {year} {2014})},\ \Eprint {http://arxiv.org/abs/arXiv:1403.7437}
  {arXiv:1403.7437} \BibitemShut {NoStop}%
\bibitem [{\citenamefont {Ye}\ and\ \citenamefont {Wen}(2014)}]{YW1427}%
  \BibitemOpen
  \bibfield  {author} {\bibinfo {author} {\bibfnamefont {P.}~\bibnamefont
  {Ye}}\ and\ \bibinfo {author} {\bibfnamefont {X.-G.}\ \bibnamefont {Wen}},\
  }\href@noop {} {\bibfield  {journal} {\bibinfo  {journal} {Phys. Rev. B}\
  }\textbf {\bibinfo {volume} {89}},\ \bibinfo {pages} {045127} (\bibinfo
  {year} {2014})},\ \Eprint {http://arxiv.org/abs/arXiv:1303.3572}
  {arXiv:1303.3572} \BibitemShut {NoStop}%
\bibitem [{\citenamefont {Wen}(2014)}]{W1447}%
  \BibitemOpen
  \bibfield  {author} {\bibinfo {author} {\bibfnamefont {X.-G.}\ \bibnamefont
  {Wen}},\ }\href {\doibase 10.1103/PhysRevB.89.035147} {\bibfield  {journal}
  {\bibinfo  {journal} {Phys. Rev. B}\ }\textbf {\bibinfo {volume} {89}},\
  \bibinfo {pages} {035147} (\bibinfo {year} {2014})},\ \Eprint
  {http://arxiv.org/abs/arXiv:1301.7675} {arXiv:1301.7675} \BibitemShut
  {NoStop}%
\bibitem [{\citenamefont {Hung}\ and\ \citenamefont {Wen}(2014)}]{HW1339}%
  \BibitemOpen
  \bibfield  {author} {\bibinfo {author} {\bibfnamefont {L.-Y.}\ \bibnamefont
  {Hung}}\ and\ \bibinfo {author} {\bibfnamefont {X.-G.}\ \bibnamefont {Wen}},\
  }\href@noop {} {\bibfield  {journal} {\bibinfo  {journal} {Phys. Rev. B}\
  }\textbf {\bibinfo {volume} {89}},\ \bibinfo {pages} {075121} (\bibinfo
  {year} {2014})},\ \Eprint {http://arxiv.org/abs/arXiv:1311.5539}
  {arXiv:1311.5539} \BibitemShut {NoStop}%
\bibitem [{\citenamefont {{Hsieh}}\ \emph {et~al.}(2014)\citenamefont
  {{Hsieh}}, \citenamefont {{Sule}}, \citenamefont {{Cho}}, \citenamefont
  {{Ryu}},\ and\ \citenamefont {{Leigh}}}]{HL14036902}%
  \BibitemOpen
  \bibfield  {author} {\bibinfo {author} {\bibfnamefont {C.-T.}\ \bibnamefont
  {{Hsieh}}}, \bibinfo {author} {\bibfnamefont {O.~M.}\ \bibnamefont {{Sule}}},
  \bibinfo {author} {\bibfnamefont {G.~Y.}\ \bibnamefont {{Cho}}}, \bibinfo
  {author} {\bibfnamefont {S.}~\bibnamefont {{Ryu}}}, \ and\ \bibinfo {author}
  {\bibfnamefont {R.~G.}\ \bibnamefont {{Leigh}}},\ }\href {\doibase
  10.1103/PhysRevB.90.165134} {\bibfield  {journal} {\bibinfo  {journal}
  {\prb}\ }\textbf {\bibinfo {volume} {90}},\ \bibinfo {pages} {165134}
  (\bibinfo {year} {2014})},\ \Eprint {http://arxiv.org/abs/arXiv:1403.6902}
  {arXiv:1403.6902} \BibitemShut {NoStop}%
\bibitem [{\citenamefont {{You}}\ and\ \citenamefont
  {{Xu}}(2014)}]{YC14090168}%
  \BibitemOpen
  \bibfield  {author} {\bibinfo {author} {\bibfnamefont {Y.-Z.}\ \bibnamefont
  {{You}}}\ and\ \bibinfo {author} {\bibfnamefont {C.}~\bibnamefont {{Xu}}},\
  }\href {\doibase 10.1103/PhysRevB.90.245120} {\bibfield  {journal} {\bibinfo
  {journal} {\prb}\ }\textbf {\bibinfo {volume} {90}},\ \bibinfo {pages}
  {245120} (\bibinfo {year} {2014})},\ \Eprint
  {http://arxiv.org/abs/arXiv:1409.0168} {arXiv:1409.0168} \BibitemShut
  {NoStop}%
\bibitem [{\citenamefont {{Hermele}}\ and\ \citenamefont
  {{Chen}}(2016)}]{HC150800573}%
  \BibitemOpen
  \bibfield  {author} {\bibinfo {author} {\bibfnamefont {M.}~\bibnamefont
  {{Hermele}}}\ and\ \bibinfo {author} {\bibfnamefont {X.}~\bibnamefont
  {{Chen}}},\ }\href {\doibase 10.1103/PhysRevX.6.041006} {\bibfield  {journal}
  {\bibinfo  {journal} {Physical Review X}\ }\textbf {\bibinfo {volume} {6}},\
  \bibinfo {pages} {041006} (\bibinfo {year} {2016})},\ \Eprint
  {http://arxiv.org/abs/arXiv:1508.00573} {arXiv:1508.00573} \BibitemShut
  {NoStop}%
\bibitem [{\citenamefont {{Song}}\ \emph {et~al.}(2017)\citenamefont {{Song}},
  \citenamefont {{Huang}}, \citenamefont {{Fu}},\ and\ \citenamefont
  {{Hermele}}}]{SH160408151}%
  \BibitemOpen
  \bibfield  {author} {\bibinfo {author} {\bibfnamefont {H.}~\bibnamefont
  {{Song}}}, \bibinfo {author} {\bibfnamefont {S.-J.}\ \bibnamefont {{Huang}}},
  \bibinfo {author} {\bibfnamefont {L.}~\bibnamefont {{Fu}}}, \ and\ \bibinfo
  {author} {\bibfnamefont {M.}~\bibnamefont {{Hermele}}},\ }\href {\doibase
  10.1103/PhysRevX.7.011020} {\bibfield  {journal} {\bibinfo  {journal}
  {Physical Review X}\ }\textbf {\bibinfo {volume} {7}},\ \bibinfo {pages}
  {011020} (\bibinfo {year} {2017})},\ \Eprint
  {http://arxiv.org/abs/arXiv:1604.08151} {arXiv:1604.08151} \BibitemShut
  {NoStop}%
\bibitem [{\citenamefont {{Lake}}(2016)}]{L160802736}%
  \BibitemOpen
  \bibfield  {author} {\bibinfo {author} {\bibfnamefont {E.}~\bibnamefont
  {{Lake}}},\ }\href {\doibase 10.1103/PhysRevB.94.205149} {\bibfield
  {journal} {\bibinfo  {journal} {\prb}\ }\textbf {\bibinfo {volume} {94}},\
  \bibinfo {pages} {205149} (\bibinfo {year} {2016})},\ \Eprint
  {http://arxiv.org/abs/arXiv:1608.02736} {arXiv:1608.02736} \BibitemShut
  {NoStop}%
\bibitem [{\citenamefont {{Thorngren}}\ and\ \citenamefont
  {{Else}}(2016)}]{TE161200846}%
  \BibitemOpen
  \bibfield  {author} {\bibinfo {author} {\bibfnamefont {R.}~\bibnamefont
  {{Thorngren}}}\ and\ \bibinfo {author} {\bibfnamefont {D.~V.}\ \bibnamefont
  {{Else}}},\ }\href@noop {} {\  (\bibinfo {year} {2016})},\ \Eprint
  {http://arxiv.org/abs/arXiv:1612.00846} {arXiv:1612.00846} \BibitemShut
  {NoStop}%
\bibitem [{\citenamefont {{Bernevig}}\ \emph {et~al.}(2006)\citenamefont
  {{Bernevig}}, \citenamefont {{Hughes}},\ and\ \citenamefont
  {{Zhang}}}]{BZ0611399}%
  \BibitemOpen
  \bibfield  {author} {\bibinfo {author} {\bibfnamefont {B.~A.}\ \bibnamefont
  {{Bernevig}}}, \bibinfo {author} {\bibfnamefont {T.~L.}\ \bibnamefont
  {{Hughes}}}, \ and\ \bibinfo {author} {\bibfnamefont {S.-C.}\ \bibnamefont
  {{Zhang}}},\ }\href {\doibase 10.1126/science.1133734} {\bibfield  {journal}
  {\bibinfo  {journal} {Science}\ }\textbf {\bibinfo {volume} {314}},\ \bibinfo
  {pages} {1757} (\bibinfo {year} {2006})},\ \Eprint
  {http://arxiv.org/abs/cond-mat/0611399} {cond-mat/0611399} \BibitemShut
  {NoStop}%
\bibitem [{\citenamefont {Schnyder}\ \emph {et~al.}(2008)\citenamefont
  {Schnyder}, \citenamefont {Ryu}, \citenamefont {Furusaki},\ and\
  \citenamefont {Ludwig}}]{SRF0825}%
  \BibitemOpen
  \bibfield  {author} {\bibinfo {author} {\bibfnamefont {A.~P.}\ \bibnamefont
  {Schnyder}}, \bibinfo {author} {\bibfnamefont {S.}~\bibnamefont {Ryu}},
  \bibinfo {author} {\bibfnamefont {A.}~\bibnamefont {Furusaki}}, \ and\
  \bibinfo {author} {\bibfnamefont {A.~W.~W.}\ \bibnamefont {Ludwig}},\
  }\href@noop {} {\bibfield  {journal} {\bibinfo  {journal} {Phys. Rev. B}\
  }\textbf {\bibinfo {volume} {78}},\ \bibinfo {pages} {195125} (\bibinfo
  {year} {2008})},\ \Eprint {http://arxiv.org/abs/arXiv:0803.2786}
  {arXiv:0803.2786} \BibitemShut {NoStop}%
\bibitem [{\citenamefont {{Wang}}\ \emph {et~al.}(2014)\citenamefont {{Wang}},
  \citenamefont {{Potter}},\ and\ \citenamefont {{Senthil}}}]{WS13063238}%
  \BibitemOpen
  \bibfield  {author} {\bibinfo {author} {\bibfnamefont {C.}~\bibnamefont
  {{Wang}}}, \bibinfo {author} {\bibfnamefont {A.~C.}\ \bibnamefont
  {{Potter}}}, \ and\ \bibinfo {author} {\bibfnamefont {T.}~\bibnamefont
  {{Senthil}}},\ }\href {\doibase 10.1126/science.1243326} {\bibfield
  {journal} {\bibinfo  {journal} {Science}\ }\textbf {\bibinfo {volume}
  {343}},\ \bibinfo {pages} {629} (\bibinfo {year} {2014})},\ \Eprint
  {http://arxiv.org/abs/arXiv:1306.3238} {arXiv:1306.3238} \BibitemShut
  {NoStop}%
\bibitem [{\citenamefont {Gu}\ and\ \citenamefont {Wen}(2014)}]{GW1441}%
  \BibitemOpen
  \bibfield  {author} {\bibinfo {author} {\bibfnamefont {Z.-C.}\ \bibnamefont
  {Gu}}\ and\ \bibinfo {author} {\bibfnamefont {X.-G.}\ \bibnamefont {Wen}},\
  }\href@noop {} {\bibfield  {journal} {\bibinfo  {journal} {Phys. Rev. B}\
  }\textbf {\bibinfo {volume} {90}},\ \bibinfo {pages} {115141} (\bibinfo
  {year} {2014})},\ \Eprint {http://arxiv.org/abs/arXiv:1201.2648}
  {arXiv:1201.2648} \BibitemShut {NoStop}%
\bibitem [{\citenamefont {{Gaiotto}}\ and\ \citenamefont
  {{Kapustin}}(2015)}]{GK150505856}%
  \BibitemOpen
  \bibfield  {author} {\bibinfo {author} {\bibfnamefont {D.}~\bibnamefont
  {{Gaiotto}}}\ and\ \bibinfo {author} {\bibfnamefont {A.}~\bibnamefont
  {{Kapustin}}},\ }\href@noop {} {\  (\bibinfo {year} {2015})},\ \Eprint
  {http://arxiv.org/abs/arXiv:1505.05856} {arXiv:1505.05856} \BibitemShut
  {NoStop}%
\bibitem [{\citenamefont {{Kapustin}}\ and\ \citenamefont
  {{Thorngren}}(2017)}]{KT170108264}%
  \BibitemOpen
  \bibfield  {author} {\bibinfo {author} {\bibfnamefont {A.}~\bibnamefont
  {{Kapustin}}}\ and\ \bibinfo {author} {\bibfnamefont {R.}~\bibnamefont
  {{Thorngren}}},\ }\href@noop {} {\  (\bibinfo {year} {2017})},\ \Eprint
  {http://arxiv.org/abs/arXiv:1701.08264} {arXiv:1701.08264} \BibitemShut
  {NoStop}%
\bibitem [{\citenamefont {{Wang}}\ and\ \citenamefont
  {{Gu}}(2017)}]{WG170310937}%
  \BibitemOpen
  \bibfield  {author} {\bibinfo {author} {\bibfnamefont {Q.-R.}\ \bibnamefont
  {{Wang}}}\ and\ \bibinfo {author} {\bibfnamefont {Z.-C.}\ \bibnamefont
  {{Gu}}},\ }\href@noop {} {\  (\bibinfo {year} {2017})},\ \Eprint
  {http://arxiv.org/abs/arXiv:1703.10937} {arXiv:1703.10937} \BibitemShut
  {NoStop}%
\bibitem [{\citenamefont {Kapustin}(2014{\natexlab{b}})}]{K1467}%
  \BibitemOpen
  \bibfield  {author} {\bibinfo {author} {\bibfnamefont {A.}~\bibnamefont
  {Kapustin}},\ }\href@noop {} {\  (\bibinfo {year} {2014}{\natexlab{b}})},\
  \Eprint {http://arxiv.org/abs/arXiv:1403.1467} {arXiv:1403.1467} \BibitemShut
  {NoStop}%
\bibitem [{\citenamefont {{Xu}}\ and\ \citenamefont
  {{Moore}}(2006)}]{XMc0508291}%
  \BibitemOpen
  \bibfield  {author} {\bibinfo {author} {\bibfnamefont {C.}~\bibnamefont
  {{Xu}}}\ and\ \bibinfo {author} {\bibfnamefont {J.~E.}\ \bibnamefont
  {{Moore}}},\ }\href {\doibase 10.1103/PhysRevB.73.045322} {\bibfield
  {journal} {\bibinfo  {journal} {\prb}\ }\textbf {\bibinfo {volume} {73}},\
  \bibinfo {pages} {045322} (\bibinfo {year} {2006})},\ \Eprint
  {http://arxiv.org/abs/cond-mat/0508291} {cond-mat/0508291} \BibitemShut
  {NoStop}%
\bibitem [{\citenamefont {{Wu}}\ \emph {et~al.}(2006)\citenamefont {{Wu}},
  \citenamefont {{Bernevig}},\ and\ \citenamefont {{Zhang}}}]{WZc0508273}%
  \BibitemOpen
  \bibfield  {author} {\bibinfo {author} {\bibfnamefont {C.}~\bibnamefont
  {{Wu}}}, \bibinfo {author} {\bibfnamefont {B.~A.}\ \bibnamefont
  {{Bernevig}}}, \ and\ \bibinfo {author} {\bibfnamefont {S.-C.}\ \bibnamefont
  {{Zhang}}},\ }\href {\doibase 10.1103/PhysRevLett.96.106401} {\bibfield
  {journal} {\bibinfo  {journal} {Physical Review Letters}\ }\textbf {\bibinfo
  {volume} {96}},\ \bibinfo {pages} {106401} (\bibinfo {year} {2006})},\
  \Eprint {http://arxiv.org/abs/cond-mat/0508273} {cond-mat/0508273}
  \BibitemShut {NoStop}%
\bibitem [{\citenamefont {Chen}\ \emph
  {et~al.}(2013{\natexlab{b}})\citenamefont {Chen}, \citenamefont {Fidkowski},\
  and\ \citenamefont {Vishwanath}}]{CFV1350}%
  \BibitemOpen
  \bibfield  {author} {\bibinfo {author} {\bibfnamefont {X.}~\bibnamefont
  {Chen}}, \bibinfo {author} {\bibfnamefont {L.}~\bibnamefont {Fidkowski}}, \
  and\ \bibinfo {author} {\bibfnamefont {A.}~\bibnamefont {Vishwanath}},\
  }\href@noop {} {\  (\bibinfo {year} {2013}{\natexlab{b}})},\ \Eprint
  {http://arxiv.org/abs/arXiv:1306.3250} {arXiv:1306.3250} \BibitemShut
  {NoStop}%
\bibitem [{\citenamefont {{Wang}}\ \emph {et~al.}(2013)\citenamefont {{Wang}},
  \citenamefont {{Potter}},\ and\ \citenamefont {{Senthil}}}]{WS13063223}%
  \BibitemOpen
  \bibfield  {author} {\bibinfo {author} {\bibfnamefont {C.}~\bibnamefont
  {{Wang}}}, \bibinfo {author} {\bibfnamefont {A.~C.}\ \bibnamefont
  {{Potter}}}, \ and\ \bibinfo {author} {\bibfnamefont {T.}~\bibnamefont
  {{Senthil}}},\ }\href {\doibase 10.1103/PhysRevB.88.115137} {\bibfield
  {journal} {\bibinfo  {journal} {\prb}\ }\textbf {\bibinfo {volume} {88}},\
  \bibinfo {pages} {115137} (\bibinfo {year} {2013})},\ \Eprint
  {http://arxiv.org/abs/arXiv:1306.3223} {arXiv:1306.3223} \BibitemShut
  {NoStop}%
\bibitem [{\citenamefont {{Hsieh}}\ \emph {et~al.}(2008)\citenamefont
  {{Hsieh}}, \citenamefont {{Qian}}, \citenamefont {{Wray}}, \citenamefont
  {{Xia}}, \citenamefont {{Hor}}, \citenamefont {{Cava}},\ and\ \citenamefont
  {{Hasan}}}]{HH09021356}%
  \BibitemOpen
  \bibfield  {author} {\bibinfo {author} {\bibfnamefont {D.}~\bibnamefont
  {{Hsieh}}}, \bibinfo {author} {\bibfnamefont {D.}~\bibnamefont {{Qian}}},
  \bibinfo {author} {\bibfnamefont {L.}~\bibnamefont {{Wray}}}, \bibinfo
  {author} {\bibfnamefont {Y.}~\bibnamefont {{Xia}}}, \bibinfo {author}
  {\bibfnamefont {Y.~S.}\ \bibnamefont {{Hor}}}, \bibinfo {author}
  {\bibfnamefont {R.~J.}\ \bibnamefont {{Cava}}}, \ and\ \bibinfo {author}
  {\bibfnamefont {M.~Z.}\ \bibnamefont {{Hasan}}},\ }\href {\doibase
  10.1038/nature06843} {\bibfield  {journal} {\bibinfo  {journal} {\nat}\
  }\textbf {\bibinfo {volume} {452}},\ \bibinfo {pages} {970} (\bibinfo {year}
  {2008})},\ \Eprint {http://arxiv.org/abs/arXiv:0902.1356} {arXiv:0902.1356}
  \BibitemShut {NoStop}%
\bibitem [{\citenamefont {{Qi}}\ and\ \citenamefont
  {{Zhang}}(2008)}]{QZ08010252}%
  \BibitemOpen
  \bibfield  {author} {\bibinfo {author} {\bibfnamefont {X.-L.}\ \bibnamefont
  {{Qi}}}\ and\ \bibinfo {author} {\bibfnamefont {S.-C.}\ \bibnamefont
  {{Zhang}}},\ }\href {\doibase 10.1103/PhysRevLett.101.086802} {\bibfield
  {journal} {\bibinfo  {journal} {Physical Review Letters}\ }\textbf {\bibinfo
  {volume} {101}},\ \bibinfo {pages} {086802} (\bibinfo {year} {2008})},\
  \Eprint {http://arxiv.org/abs/arXiv:0801.0252} {arXiv:0801.0252} \BibitemShut
  {NoStop}%
\bibitem [{\citenamefont {{Ran}}\ \emph {et~al.}(2008)\citenamefont {{Ran}},
  \citenamefont {{Vishwanath}},\ and\ \citenamefont {{Lee}}}]{RL08010627}%
  \BibitemOpen
  \bibfield  {author} {\bibinfo {author} {\bibfnamefont {Y.}~\bibnamefont
  {{Ran}}}, \bibinfo {author} {\bibfnamefont {A.}~\bibnamefont {{Vishwanath}}},
  \ and\ \bibinfo {author} {\bibfnamefont {D.-H.}\ \bibnamefont {{Lee}}},\
  }\href {\doibase 10.1103/PhysRevLett.101.086801} {\bibfield  {journal}
  {\bibinfo  {journal} {Physical Review Letters}\ }\textbf {\bibinfo {volume}
  {101}},\ \bibinfo {pages} {086801} (\bibinfo {year} {2008})},\ \Eprint
  {http://arxiv.org/abs/arXiv:0801.0627} {arXiv:0801.0627} \BibitemShut
  {NoStop}%
\bibitem [{\citenamefont {{Barkeshli}}\ \emph {et~al.}(2014)\citenamefont
  {{Barkeshli}}, \citenamefont {{Bonderson}}, \citenamefont {{Cheng}},\ and\
  \citenamefont {{Wang}}}]{BBC1440}%
  \BibitemOpen
  \bibfield  {author} {\bibinfo {author} {\bibfnamefont {M.}~\bibnamefont
  {{Barkeshli}}}, \bibinfo {author} {\bibfnamefont {P.}~\bibnamefont
  {{Bonderson}}}, \bibinfo {author} {\bibfnamefont {M.}~\bibnamefont
  {{Cheng}}}, \ and\ \bibinfo {author} {\bibfnamefont {Z.}~\bibnamefont
  {{Wang}}},\ }\href@noop {} {\  (\bibinfo {year} {2014})},\ \Eprint
  {http://arxiv.org/abs/arXiv:1410.4540} {arXiv:1410.4540} \BibitemShut
  {NoStop}%
\bibitem [{\citenamefont {Wen}(2002{\natexlab{b}})}]{W0213}%
  \BibitemOpen
  \bibfield  {author} {\bibinfo {author} {\bibfnamefont {X.-G.}\ \bibnamefont
  {Wen}},\ }\href@noop {} {\bibfield  {journal} {\bibinfo  {journal} {Phys.
  Rev. B}\ }\textbf {\bibinfo {volume} {65}},\ \bibinfo {pages} {165113}
  (\bibinfo {year} {2002}{\natexlab{b}})},\ \Eprint
  {http://arxiv.org/abs/cond-mat/0107071} {cond-mat/0107071} \BibitemShut
  {NoStop}%
\bibitem [{\citenamefont {{Barkeshli}}\ \emph {et~al.}(2016)\citenamefont
  {{Barkeshli}}, \citenamefont {{Bonderson}}, \citenamefont {{Jian}},
  \citenamefont {{Cheng}},\ and\ \citenamefont {{Walker}}}]{BW161207792}%
  \BibitemOpen
  \bibfield  {author} {\bibinfo {author} {\bibfnamefont {M.}~\bibnamefont
  {{Barkeshli}}}, \bibinfo {author} {\bibfnamefont {P.}~\bibnamefont
  {{Bonderson}}}, \bibinfo {author} {\bibfnamefont {C.-M.}\ \bibnamefont
  {{Jian}}}, \bibinfo {author} {\bibfnamefont {M.}~\bibnamefont {{Cheng}}}, \
  and\ \bibinfo {author} {\bibfnamefont {K.}~\bibnamefont {{Walker}}},\
  }\href@noop {} {\  (\bibinfo {year} {2016})},\ \Eprint
  {http://arxiv.org/abs/arXiv:1612.07792} {arXiv:1612.07792} \BibitemShut
  {NoStop}%
\bibitem [{\citenamefont {Lu}\ and\ \citenamefont {Vishwanath}(2013)}]{LV1334}%
  \BibitemOpen
  \bibfield  {author} {\bibinfo {author} {\bibfnamefont {Y.-M.}\ \bibnamefont
  {Lu}}\ and\ \bibinfo {author} {\bibfnamefont {A.}~\bibnamefont
  {Vishwanath}},\ }\href@noop {} {\  (\bibinfo {year} {2013})},\ \Eprint
  {http://arxiv.org/abs/arXiv:1302.2634} {arXiv:1302.2634} \BibitemShut
  {NoStop}%
\bibitem [{\citenamefont {Hung}\ and\ \citenamefont {Wen}(2013)}]{HW1307}%
  \BibitemOpen
  \bibfield  {author} {\bibinfo {author} {\bibfnamefont {L.-Y.}\ \bibnamefont
  {Hung}}\ and\ \bibinfo {author} {\bibfnamefont {X.-G.}\ \bibnamefont {Wen}},\
  }\href@noop {} {\bibfield  {journal} {\bibinfo  {journal} {Phys. Rev. B}\
  }\textbf {\bibinfo {volume} {87}},\ \bibinfo {pages} {165107} (\bibinfo
  {year} {2013})},\ \Eprint {http://arxiv.org/abs/arXiv:1212.1827}
  {arXiv:1212.1827} \BibitemShut {NoStop}%
\bibitem [{\citenamefont {Mesaros}\ and\ \citenamefont {Ran}(2013)}]{MR1315}%
  \BibitemOpen
  \bibfield  {author} {\bibinfo {author} {\bibfnamefont {A.}~\bibnamefont
  {Mesaros}}\ and\ \bibinfo {author} {\bibfnamefont {Y.}~\bibnamefont {Ran}},\
  }\href@noop {} {\bibfield  {journal} {\bibinfo  {journal} {Phys. Rev. B}\
  }\textbf {\bibinfo {volume} {87}},\ \bibinfo {pages} {155115} (\bibinfo
  {year} {2013})},\ \Eprint {http://arxiv.org/abs/arXiv:1212.0835}
  {arXiv:1212.0835} \BibitemShut {NoStop}%
\bibitem [{\citenamefont {Hung}\ and\ \citenamefont {Wan}(2013)}]{HW1351}%
  \BibitemOpen
  \bibfield  {author} {\bibinfo {author} {\bibfnamefont {L.-Y.}\ \bibnamefont
  {Hung}}\ and\ \bibinfo {author} {\bibfnamefont {Y.}~\bibnamefont {Wan}},\
  }\href@noop {} {\  (\bibinfo {year} {2013})},\ \Eprint
  {http://arxiv.org/abs/arXiv:1302.2951} {arXiv:1302.2951} \BibitemShut
  {NoStop}%
\bibitem [{\citenamefont {{Xu}}(2013)}]{X13078131}%
  \BibitemOpen
  \bibfield  {author} {\bibinfo {author} {\bibfnamefont {C.}~\bibnamefont
  {{Xu}}},\ }\href {\doibase 10.1103/PhysRevB.88.205137} {\bibfield  {journal}
  {\bibinfo  {journal} {\prb}\ }\textbf {\bibinfo {volume} {88}},\ \bibinfo
  {pages} {205137} (\bibinfo {year} {2013})},\ \Eprint
  {http://arxiv.org/abs/arXiv:1307.8131} {arXiv:1307.8131} \BibitemShut
  {NoStop}%
\bibitem [{\citenamefont {Chang}\ \emph {et~al.}(2015)\citenamefont {Chang},
  \citenamefont {Cheng}, \citenamefont {Cui}, \citenamefont {Hu}, \citenamefont
  {Jin}, \citenamefont {Movassagh}, \citenamefont {Naaijkens}, \citenamefont
  {Wang},\ and\ \citenamefont {Young}}]{CY14126589}%
  \BibitemOpen
  \bibfield  {author} {\bibinfo {author} {\bibfnamefont {L.}~\bibnamefont
  {Chang}}, \bibinfo {author} {\bibfnamefont {M.}~\bibnamefont {Cheng}},
  \bibinfo {author} {\bibfnamefont {S.~X.}\ \bibnamefont {Cui}}, \bibinfo
  {author} {\bibfnamefont {Y.}~\bibnamefont {Hu}}, \bibinfo {author}
  {\bibfnamefont {W.}~\bibnamefont {Jin}}, \bibinfo {author} {\bibfnamefont
  {R.}~\bibnamefont {Movassagh}}, \bibinfo {author} {\bibfnamefont
  {P.}~\bibnamefont {Naaijkens}}, \bibinfo {author} {\bibfnamefont
  {Z.}~\bibnamefont {Wang}}, \ and\ \bibinfo {author} {\bibfnamefont
  {A.}~\bibnamefont {Young}},\ }\href@noop {} {\bibfield  {journal} {\bibinfo
  {journal} {Journal of Physics A: Mathematical and Theoretical}\ }\textbf
  {\bibinfo {volume} {48}},\ \bibinfo {pages} {12FT01} (\bibinfo {year}
  {2015})},\ \Eprint {http://arxiv.org/abs/arXiv:1412.6589} {arXiv:1412.6589}
  \BibitemShut {NoStop}%
\end{thebibliography}%

\end{document}